\documentclass[prd,reprint,amsmath,amssymb,superscriptaddress]{revtex4-2}

\newcommand{\be}{\begin{equation}}
\newcommand{\ee}{\end{equation}}
\newcommand{\ba}{\begin{align}}
\newcommand{\eda}{\end{align}}
\newcommand{\nn}{\nonumber}

\usepackage{bm}
\usepackage{todonotes}\topmargin -1.2cm

\usepackage[export]{adjustbox}
\usepackage{graphicx}
\usepackage{epsfig}
\usepackage{braket}
\usepackage{capt-of}
\newcommand{\om}{\omega}
\newcommand{\omi}{\omega^{-1}}
\newcommand{\ppm}{\pi^{\pm}}

\newcommand{\hp}{^{+}}
\newcommand{\hpm}{^{\pm}}
\newcommand{\hmp}{^{\mp}}
\newcommand{\tp}{\tilde{p}}
\newcommand{\uu}{_{1}}
\newcommand{\ud}{_{2}}
\newcommand{\ua}{_{a}}
\newcommand{\ub}{_{b}}
\newcommand{\us}{_{s}}
\newcommand{\2}{^{2}}
\newcommand{\cN}{\mathcal{N}}
\newcommand{\cM}{\mathcal{M}}

\newcommand{\cJ}{\mathcal{J}}
\newcommand{\cP}{\mathcal{P}}

\newcommand{\ul}{_{\lambda}}
\newcommand{\hl}{^{\lambda}}
 \newcommand{\up}{_{p}}
  \newcommand{\ut}{_{t}}
 \newcommand{\upp}{_{\pi}}
  
    \newcommand{\nk}{\slash{k}}
  \def\er{\eqref}
  \newcommand{\bel}[1]{\be\label{#1}}
  \newcommand{\bal}[1]{\ba\label{#1}}
  \newcommand{\dv}{\text{d}}
  \newcommand{\tr}{\text{tr}}
  \newcommand{\qs}{\sqrt{s}}
  \newcommand{\wik}{\bm{\hat{k}}}

  \renewcommand\slash[1]{\not \! #1}
  \newcommand{\tnp}{\slash{\tilde{p}}}
  
 \newcommand{\bp}{\slash{p}}
 \newcommand{\bk}{\slash{k}}
 
 \newcommand{\nldt}[1]{\overset{\approx}{\cN}      \ul{\phantom{\big{|}}}^{\!\!\!\!\!\!(#1)}}
  
\usepackage[normalem]{ulem}
  
\newcommand{\p}{\partial}

\newcommand{\leftsidep}[1]{\stackrel{\;\leftarrow}{\,\p}_{#1}}
\newcommand{\rightsidep}[1]{\stackrel{\;\rightarrow}{\,\p}_{#1}}

\usepackage{hyperref}
\numberwithin{equation}{section}
\renewcommand{\theequation}{\arabic{section}.\arabic{equation}}
\usepackage[noabbrev]{cleveref}    

\begin{document}


\title{Soft-photon theorem for pion-proton elastic scattering revisited}



\author{Piotr Lebiedowicz}
\email{Piotr.Lebiedowicz@ifj.edu.pl}
\affiliation{Institute of Nuclear Physics Polish Academy of Sciences, 
Radzikowskiego 152, PL-31342 Krak{\'o}w, Poland}

\author{Otto Nachtmann}
\email{O.Nachtmann@thphys.uni-heidelberg.de}
\affiliation{Institut f\"ur Theoretische Physik, Universit\"at Heidelberg,
Philosophenweg 16, D-69120 Heidelberg, Germany}

\author{Antoni Szczurek}
\email{Antoni.Szczurek@ifj.edu.pl}
\affiliation{Institute of Nuclear Physics Polish Academy of Sciences, 
Radzikowskiego 152, PL-31342 Krak{\'o}w, Poland}
\affiliation{College of Natural Sciences, 
Institute of Physics, University of Rzesz{\'o}w, 
Pigonia 1, PL-35310 Rzesz{\'o}w, Poland.}


\begin{abstract}
We discuss the reactions $\pi p \to \pi p$ and $\pi p \to \pi p \gamma$ 
from a general quantum field theory (QFT) point of view,
describing these reactions in QCD and lowest relevant
order of electromagnetism.
We consider the pion-proton elastic scattering both off shell and on shell. 
The on-shell amplitudes for $\pi^{\pm} p \to \pi^{\pm} p$ scattering
are described by two invariant amplitudes,
while the off-shell amplitudes contain eight invariant amplitudes.
We study the photon emission amplitudes
in the soft-photon limit where the c.m. photon energy 
$\omega \to 0$.
The Laurent expansion in $\omega$ 
of the $\pi^{\pm} p \to \pi^{\pm} p \gamma$ amplitudes 
is considered and the terms of the orders
$\omega^{-1}$ and $\omega^{0}$ are derived.
These terms can be expressed by the on-shell invariant amplitudes
and their partial derivatives with respect to $s$ and~$t$.
The pole term $\propto \omega^{-1}$ in the amplitudes
corresponds to Weinberg's soft-photon theorem and
is well known from the literature.
We derive the next-to-leading term $\propto \omega^{0}$
using only rigorous methods of QFT.
We give the relation of the Laurent series for
$\pi^{0} p \to \pi^{0} p \gamma$ and
Low's soft-photon theorem.
Our formulas for the amplitudes in the limit 
$\omega \to 0$ are valid 
for photon momentum $k$ satisfying $k^{2} \geqslant 0$,
$k^{0} = \omega \geqslant 0$, that is,
for both real and virtual photons.
Here we consider a limit where with $\omega \to 0$
we have also $k^{2} \to 0$.
We discuss the behavior of the corresponding cross-sections
for $\pi^{-} p \to \pi^{-} p \gamma$ with respect to $\omega$ 
for $\omega \to 0$. We consider cross sections for unpolarized
as well as polarized protons in the initial and final states.
\end{abstract}


\maketitle

\section{Introduction}
\label{sec:Introduction}

In this paper we shall study the $\pi p \to \pi p$ and $\pi p \to \pi p \gamma$ reactions. 

Let $\omega$ be the energy of the photon in the overall c.m. system. 
We are interested in the limit $\omega \to 0$, that is, 
the soft-photon limit. 
In a seminal paper \cite{Low:1958sn} F. E. Low considered 
the scattering of a charged on an uncharged scalar particle 
with photon radiation
and the scattering of a charged spin 1/2 fermion on a scalar boson
with photon radiation.
He showed that the leading terms in the amplitudes for $\omega \to 0$
come exclusively from the photon emission of the external particles. 
A soft-photon theorem, different from Low's theorem,
was given by S. Weinberg in \cite{Weinberg:1964ew,Weinberg:1965nx}.
There, photon radiation in the scattering of an arbitrary number of
charged or neutral particles was considered
and the pole term $\propto \omega^{-1}$ for $\omega \to 0$ was given.
Subsequently many authors have considered soft-photon emission
in particle collisions;
see for instance \cite{Gribov:1966hs,Burnett:1967km,Bell:1969yw,
Liou:1977qv,Liou:1978jx,
Liou:1978sz,Liou:1982fm,Liou:1987ug,
Lipatov:1988ii,
DelDuca:1990gz,
Lin:1991qk,
Korchin:1995ys,Korchin:1996up,
Timmermans:2001cm,Li:2011aq,
Gervais:2017yxv,Bern:2014vva,Lysov:2014csa,
Bonocore:2021cbv,Engel:2021ccn,Engel:2023ifn}.

In \cite{Lebiedowicz:2021byo} we reconsidered 
the problem of soft-photon production studying the reactions
\begin{eqnarray}
\pi^{-} + \pi^{0} &\to& \pi^{-} + \pi^{0}\,, 
\label{1.1} \\
\pi^{-} + \pi^{0} &\to& \pi^{-} + \pi^{0}+ \gamma\,.
\label{1.1a}
\end{eqnarray}
We worked in the framework of QCD and treated electromagnetism to lowest relevant order. As theoretical tools we used the Ward-Takahashi identity \cite{Ward:1950xp,Takahashi:1957xn} 
and the Landau conditions giving the position of singularities in amplitudes; 
see \cite{Landau:1959fi} and chapter 18 of \cite{Bjorken:1965}.
We calculated the terms of order $\omega^{-1}$ and $\omega^{0}$
for the reaction (\ref{1.1a}).
At first we were surprised to find a result for the $\omega^{0}$ term
different from the result given in \cite{Low:1958sn}.
But now we hope to have cleared up everything in the Erratum to
\cite{Lebiedowicz:2021byo} and in~\cite{Lebiedowicz:2023ell}.
We had been misled by the fact that in the literature
frequently Weinberg's version of the soft-photon theorem is used
but addressed as Low's theorem.
In \cite{Lebiedowicz:2023ell} we have shown that these two versions
of soft-photon theorems have a \textit{different} meaning
and must not be confounded.
Weinberg's version of the soft-photon theorem gives the pole term
$\propto \omega^{-1}$ in the amplitude, that is the first term
in a Laurent expansion around $\omega = 0$,
the phase-space point corresponding to no radiation.
In \cite{Lebiedowicz:2021byo} and \cite{Lebiedowicz:2023ell}
we have given the term of order $\omega^{0}$
in this Laurent expansion for (\ref{1.1a}).
Low's soft-photon theorem deals with the amplitude for (\ref{1.1a})
at a given phase-space point \textit{with} radiation and
provides an approximate expression there.
All this and the relation of Low's and Weinberg's
soft-photon theorems are discussed at length in \cite{Lebiedowicz:2023ell}.

In our present paper we continue our studies of soft-photon production in hadronic reactions. 
In \cite{Lebiedowicz:2022nnn,Lebiedowicz:2023mhe,Lebiedowicz:2023rgc} 
we treated bremsstrahlung and central-exclusive production of soft photons in proton-proton collisions at high energies. 
There we used the tensor-Pomeron model \cite{Ewerz:2013kda}
as theoretical tool. 
Now we shall study the reactions
\begin{align}
\label{1.2}
&\ppm + p \to \ppm + p\,,
\\
\label{1.3}
&\ppm + p \to \ppm + p+\gamma\,,
\end{align}
using only rigorous QFT methods. 
We are again interested in the limit $\om\to 0$, where $\om$ is the photon energy in the c.m. system. 
We shall work in the framework of QCD, the theory of hadrons, and consider electromagnetism to lowest relevant order.
Our aim is to calculate the terms of order $\omi$ and $\om^0$ 
in the Laurent expansion of the amplitudes 
for the reactions \eqref{1.3}. 
As theoretical tools we shall only use 
general relations of quantum field theory (QFT), 
the symmetry relations of QCD,
the identity due to work by Ward and Takahashi \cite{Ward:1950xp,Takahashi:1957xn},
the so called generalized Ward identity for the pion and the proton fields,
and the Landau conditions \cite{Landau:1959fi,Bjorken:1965}.
Some of our results for real photon emission
in the reactions \eqref{1.3} have already been presented in
\cite{Lebiedowicz:2023ell}.
Here we also give the detailed derivation of these results.

Our paper is organized as follows.
In Sec.~\ref{sec:2} we discuss the amplitudes 
for the $\ppm p\to \ppm p$ scattering, both off shell and on shell. 
In Sec.~\ref{sec:3} we study the reactions $\ppm p \to \ppm p\gamma$, 
first from a general point of view, 
and then in the limit of the photon energy $\omega\to 0$. 
We derive the expansion of these amplitudes to the orders $\omi$ and $\om^0$. 
Section~\ref{sec:4} deals with the cross sections 
for $\ppm p\to \ppm p\gamma$ for $\omega \to 0$.
We discuss cross sections for unpolarized protons
as well as polarized protons in the initial and final states.
Our conclusions are summarized in Sec.~\ref{sec:5}.
In Appendix~\ref{app:A} we collect kinematic relations. 
Appendix~\ref{app:B} gives 
a detailed discussion of the pion and proton propagators 
and the $\gamma \pi \pi$ and $\gamma p p$ vertex functions. 
In Appendix~\ref{app:C} we present the details of our calculations 
for the $\ppm p \to \ppm p \gamma$ amplitudes.
In Appendix~\ref{app:D} we compare our findings concerning
the soft-photon production in pion-proton scattering 
to results from the literature.
In particular, we give there the relation of Low's
approximate expression of the amplitude for the reaction
$\pi^{0} p \to \pi^{0} p \gamma$ [see Eq.~(1.8) of \cite{Low:1958sn}]
and the Laurent expansion for this amplitude as calculated
with our methods.

Throughout our paper we use the metric 
and $\gamma$-matrix conventions of \cite{Bjorken:1965}. 
The Levi-Civita symbol $\varepsilon_{\mu\nu\rho\sigma}$ 
is normalized as $\varepsilon_{0123}=1$.

Since our paper is rather long we give here a short guide 
to important equations. The off-shell and on-shell matrix elements for
$\ppm p \to \ppm p$ are given in (\ref{2.16}) and (\ref{2.23}), respectively.
The matrix element for $\ppm p \to \ppm p \gamma$ can be found in (\ref{3.5})
followed by the important discussion of energy-momentum conservation;
see Fig.~\ref{fig:2}.
In (\ref{3.16}) we define the matrix function $\mathcal{N}^{\lambda}$
for $\pi^{-} p \to \pi^{-} p \gamma$ 
which is then the main object
of our studies.
How to obtain the standard matrix elements 
from $\mathcal{N}^{\lambda}$
is shown in (\ref{3.17}).
The final results for $\mathcal{N}^{\lambda}$ 
to the orders $\omega^{-1}$ and $\omega^{0}$ 
are given in (\ref{3.49})--(\ref{3.55}).
The final results for expansions in $\omega$
for $\omega \to 0$ of the cross sections
for $\pi^{-} p \to \pi^{-} p \gamma$ 
are given in (\ref{4.16})--(\ref{4.19}) and (\ref{4.34}).

\section{The amplitudes for the on- and off-shell scattering $\ppm p \to \ppm p$}
\label{sec:2}
In this section we discuss general properties of the reactions
\begin{equation}\label{2.1}
\ppm\,(\tp\ua)+p \,(\tp\ub)\to \ppm\,(\tp\uu)+p \,(\tp\ud)
\end{equation}
off shell and on shell.
In relations which are true for both cases we shall denote
the momenta for the off- and on-shell case 
with a tilde, e.g., $\tp_{a}, \dots ,\tp_{2}$ in \eqref{2.1}. 
The corresponding on-shell momenta will be denoted without a tilde. 
In relations which are true only on shell, the momenta
will be denoted without a tilde.
Thus, the on-shell scattering reactions are denoted as
\begin{equation}\label{2.2}
\ppm\,(p\ua)+p\,(p\ub)\to \ppm\,(p\uu)+p\,(p\ud)\,.
\end{equation}
To define the amplitudes $\cM^{(0)\pm}$ for the off-shell processes \eqref{2.1} 
we consider the connected part of the following four-point functions
\begin{align}\label{2.3}
G_{4c}^{\pm}&(x\uu, x\ud , x\ua,x\ub)=
\braket{0|{\rm T}(\varphi\hpm(x\uu)\varphi\hmp(x\ua)\psi(x\ud)\overline{\psi}(x\ub))|0}_{c}.
\end{align}
Here ${\rm T}$ denotes the time-ordered product of field operators. 
We have then as defining equation for $\cM^{(0)\pm}$
\begin{align}\label{2.4}
&
i(2\pi)^{4}\delta^{(4)}(\tp\ua +\tp\ub -\tp\uu -\tp\ud)
iS_{F}(\tp\ud)
i\Delta_{F}(\tp\uu^{2})\nn\\
&\quad \times 
\cM^{(0)\pm}(\tp\uu, \tp\ud, \tp\ua, \tp \ub)
iS_{F}(\tp\ub )
i\Delta_{F}(\tp\ua^{2})\nn\\
&
=\int d^{4}x\uu\, d^{4}x\ud\, d^{4}x\ua\,d^{4}x\ub\nn\\
&\quad \times \exp[ i\tp\uu x\uu+i\tp\ud x\ud-i\tp\ua x\ua-i\tp\ub x\ub]\nn\\ 
&\quad \times G_{4c}^{\pm}(x_{1},x_{2},x_{a},x_{b})\,.
\end{align}
The propagators $S_{F}$ and $\Delta_F$ for protons and pions, respectively, are discussed in Appendix~\ref{app:B}, Sec. \ref{subsec:B1}. Given the function $G_{4c}$ the matrix-valued amplitudes $\cM^{(0)\pm}$ are well defined everywhere since the propagators $S_F$ and $\Delta_F$ are finite and nearly everywhere non zero away from the respective mass shells for protons and pions. 
The on-shell amplitudes $\cM^{(0)\pm}(p_{1},p_{2},p_{a},p_{b})$ 
are then uniquely defined as
the limit of the off-shell amplitudes 
$\cM^{(0)\pm}(\tp\uu, \tp\ud, \tp\ua, \tp\ub)$ 
as $\tilde{p}_{j}\to p_{j}$. 

Now we consider the on-shell reactions 
\begin{equation}
\label{2.5}
\ppm\,(p\ua)+ p\,(p\ub ,\lambda\ub )\to \ppm\,(p\uu)+p\,(p\ud ,\lambda\ud )\,,
\end{equation}
where $\lambda_b \,, \lambda_2\in\lbrace1/2, \, -1/2\rbrace$ are the proton's spin indices and we have 
\mbox{$p_{a}^{0}$, $p_{b}^{0}$, $p_{1}^{0}$, $p_{2}^{0}>0$}.
The standard reduction formulas for pions and protons 
(see, e.g., \cite{Bjorken:1965}) 
give for the ${\mathcal T}$-matrix element
\begin{align}\label{2.6}
&\braket{\ppm(p\uu), \, p(p\ud ,\lambda\ud)|{\mathcal T}|\ppm(p\ua), p(p\ub ,\lambda\ub)} \nn\\
&\quad =
\bar{u}(p\ud,\lambda\ud)
\cM^{(0)\pm}(p_{1},p_{2},p_{a},p_{b})
u(p\ub,\lambda\ub)\,.
\end{align}

The kinematics of the off-shell reaction \eqref{2.1} is our next topic. 
We have from \eqref{2.4} the energy-momentum conservation equation
\begin{equation}
\label{2.7}
\tp\ua +\tp\ub =\tp\uu +\tp \ud\,.
\end{equation}
We denote the invariant off-shell squared masses by
\begin{equation}
\label{2.8}
m_{a}^{2}=\tp_{a}^{2}\;,\quad
m_{b}^{2}=\tp_{b}^{2}\;,\quad
m_{1}^{2}=\tp_{1}^{2}\;,\quad
m_{2}^{2}=\tp_{2}^{2}\,.
\end{equation}
From \eqref{2.7} we see that there are three independent momenta which we choose as follows
\begin{align}
\label{2.9}
\tp_{s}=\tp\ua +\tp\ub =\tp\uu +\tp\ud\,,\nn\\
\tp_{t}=\tp\ua -\tp\uu =\tp\ud -\tp\ub\,,\nn\\
\tp_{u}=\tp\ua -\tp\ud =\tp\uu -\tp\ub\,.
\end{align}
We set
\begin{equation}
\label{2.10}
\tilde{s}=\tp_{s}^2\,,\quad
\tilde{t}=\tp_{t}^2\,,\quad
\tilde{u}=\tp_{u}^2\,,
\end{equation}
where
\begin{equation}
\label{2.11}
\tilde{s} + \tilde{t} + \tilde{u} 
= m_{a}^2 + m_{b}^2 + m_{1}^2 + m_{2}^2 \,.
\end{equation}

For the on-shell process \eqref{2.2} we have the relations \eqref{2.9} and \eqref{2.10} without tildes 
and the squared masses \eqref{2.8} are
\begin{align}
\label{2.12}
m_{a}^{2}=m_{1}^{2}=m_{\pi}^{2}\,, \quad
m_{b}^{2}=m_{2}^{2}=m_{p}^{2}\,.
\end{align}
Further kinematic relations are given in Appendix~\ref{app:A}.

The symmetries $P$ (parity), 
$C$ (charge conjugation), 
and $T$ (time reversal) hold in QCD and we recall 
the corresponding transformations of the pion and proton fields 
in Appendix~\ref{app:B}. 
Using \eqref{B2}--\eqref{B4} and \eqref{B13}--\eqref{B15} 
we find the following relations 
for $\cM^{(0)\pm}$~\eqref{2.4}.
\\
\\
$P$ invariance:
\begin{align}
\label{2.13}
&\cM^{(0)\pm}(\tp\uu ,\,\tp\ud , \,\tp\ua ,\,\tp \ub)\nn\\
&\quad =\gamma_{0}\cM^{(0)\pm}(\mathcal{P}\tp\uu ,\mathcal{P}\tp\ud , \mathcal{P}\tp\ua ,\mathcal{P}\tp \ub)\gamma_{0}\,,
\intertext{$C$ invariance:}
\label{2.14}
&\cM^{(0)\pm}(\tp\uu ,\,\tp\ud , \,\tp\ua ,\,\tp \ub)\nn\\
&\quad =S(C)\left[\cM^{(0)\mp}(\tp\uu, -\tp\ub ,\,\tp\ua , -\tp\ud)\right]^{\top}S^{-1}(C) \nn \\
&\quad =S(C)\left[\cM^{(0)\pm}(-\tp\ua, -\tp\ub ,-\tp\uu , -\tp\ud)\right]^{\top}S^{-1}(C)\,,
\intertext{$T$ invariance:}
\label{2.15}
&\cM^{(0)\pm}(\tp\uu ,\,\tp\ud ,\, \tp\ua ,\,\tp \ub)\nn \\
&\quad =S(T)\left[\cM^{(0)\pm}(\mathcal{P}\tp\ua, \mathcal{P}\tp\ub ,\mathcal{P}\tp\uu , \mathcal{P}\tp\ud)\right]^{\top}S^{-1}(T)\,.
\end{align}
Here $S(C)$, $S(T)$ and $\mathcal{P}$ are defined in \eqref{B5} and \eqref{B8}, respectively.

We can now write the following general expansions 
for $\cM^{(0)\pm}$ taking already into account 
$P$ invariance \eqref{2.13}
\begin{align}
\label{2.16}
&\cM^{(0)\pm}(\tp\uu ,\,\tp\ud ,\, \tp\ua ,\,\tp \ub)\nn\\
& \quad =
\cM_{1}^{\pm}+\tnp_{s}\cM_{2}^{\pm}+\tnp_{t}\cM_{3}^{\pm}+\tnp_{u}\cM_{4}^{\pm}\nn\\
&\qquad +i\sigma_{\mu\nu}\tp_{s}{}^{\mu}\tp_{t}{}^{\nu}\cM_{5}^{\pm}+i\sigma_{\mu\nu}\tp_{s}{}^{\mu}\tp_{u}{}^{\nu}\cM_{6}^{\pm}\nn\\
&\qquad +i\sigma_{\mu\nu}\tp_{t}{}^{\mu}\tp_{u}{}^{\nu}\cM_{7}^{\pm}\nn\\
&\qquad +i\gamma_{\mu}\gamma_{5}\varepsilon^{\mu\nu\rho\sigma}\tp_{s\nu}\tp_{t\rho}\tp_{u\sigma}\cM_{8}^{\pm}\,.
\end{align}
Here the invariant amplitudes $\cM_{j}^{\pm}$ can only depend on $\tilde{s}$, $\tilde{t}$,
and the invariant squared masses \eqref{2.8}
\begin{equation}\label{2.17}
\cM_{j}^{\pm}=\cM_{j}^{\pm}(\tilde{s},\,\tilde{t},\,m_{1}\2,\, m_{2}\2,\, m\ua\2,\,m\ub\2)\,.
\end{equation}
Both, the $C$ as well as the $T$ invariance relation, \eqref{2.14} and \eqref{2.15}, respectively, require
\begin{align}\label{2.18}
&\cM_{j}^{\pm}(\tilde{s},\,\tilde{t},\,m_{1}\2,\, m_{2}\2,\, m\ua\2,\,m\ub\2)\nn\\
&\quad =\cM_{j}^{\pm}(\tilde{s},\,\tilde{t},\,m_{a}\2,\, m\ub\2,\,m\uu\2,\,m\ud\2)\,,\nn\\
&\text{for} \; j=1,\,2,\,4,\,5,\,7,\,8\,,
\end{align}
and
\begin{align}\label{2.19}
&\cM_{j}^{\pm}(\tilde{s},\,\tilde{t},\,m_{1}\2,\, m_{2}\2,\, m\ua\2,\,m\ub\2)\nn\\
&\quad =-\cM_{j}^{\pm}(\tilde{s},\,\tilde{t},\,m_{a}\2,\, m\ub\2,\,m\uu\2,\,m\ud\2)\,,\nn\\
&\text{for} \; j=3,\,6\,.
\end{align}

Now we consider the ${\mathcal T}$-matrix element \eqref{2.6} 
for the on-shell process \eqref{2.5}. 
Inserting in \eqref{2.6} the on-shell version of $\cM^{(0)\pm}$ 
from \eqref{2.16} we get
\begin{align}\label{2.20}
&\braket{\ppm(p\uu), \, p(p\ud ,\lambda\ud)|{\mathcal T}|\ppm(p\ua), \, p(p\ub ,\lambda\ub)}\nn\\
&\quad =\bar{u}(p\ud ,\lambda\ud)\Big{[} \cM_{1}^{(\text{on})\pm}
+\slash{p}_{s}\cM_{2}^{(\text{on})\pm}
+\slash{p}_{u}\cM_{4}^{(\text{on})\pm}\nn\\
&\qquad
+i\sigma_{\mu\nu}p_{s}{}^{\mu}p_{t}{}^{\nu}\cM_{5}^{(\text{on})\pm}
+i\sigma_{\mu\nu}p_{t}{}^{\mu}p_{u}{}^{\nu}\cM_{7}^{(\text{on})\pm}\nn\\
&\qquad
+i\gamma_{\mu}\gamma_{5}\varepsilon^{\mu\nu\rho\sigma} 
p_{s\nu}p_{t\rho}p_{u\sigma}\cM_{8}^{(\text{on})\pm}\Big{ ]} u(p\ub ,\lambda\ub)\,,
\end{align}
where
\begin{align}
\label{2.21}
&\cM_{j}^{(\text{on})\pm}=\cM_{j}^{\pm}(s,t,m_{\pi}\2 ,m_{p}\2 ,m_{\pi}\2 ,m_{p}\2 )\;,\nn\\
& j=1,\dots , 8\,.
\end{align}
From \eqref{2.19} we see that on shell we have
\begin{align}\label{2.22}
\cM_{3}^{(\text{on})\pm}=0\,, \qquad 
\cM_{6}^{(\text{on})\pm}=0\,.
\end{align}
Therefore, we have omitted $\cM_{3}^{(\text{on})\pm}$
and $\cM_{6}^{(\text{on})\pm}$ in \eqref{2.20}. 
For the on-shell processes we can further simplify 
the ${\mathcal T}$-matrix element 
\eqref{2.20} by using the relations 
\eqref{A5}--\eqref{A9} from Appendix~\ref{app:A}. 
This gives
\begin{align}
\label{2.23}
&\braket{\ppm (p\uu), \,  p(p\ud ,\lambda\ud)|{\mathcal T}|\ppm(p\ua), \, p(p\ub ,\lambda\ub)} \nn\\
& \quad = \bar{u}(p\ud , \lambda\ud)\Big{[}A^{(\text{on})\pm}(s,t)\nn\\
& \qquad +\frac{1}{2}(\slash{p}_{a}+\slash{p}_{1})
B^{(\text{on})\pm}(s,t)\Big{]}u(p\ub,\lambda\ub )\,,
\end{align}
where
\begin{align}
\label{2.24}
&A^{(\text{on})\pm}(s,t)=\cM_{1}^{(\text{on})\pm}+m_{p}\cM_{2}^{(\text{on})\pm}\nn-m_{p}\cM_{4}^{(\text{on})\pm}\nn\\
&\quad +(-s+m_{p}\2 +m_{\pi}\2)\cM_{5}^{(\text{on})\pm}\nn\\
&\quad +(s+t-m_{p}\2 -m_{\pi}\2)\cM_{7}^{(\text{on})\pm}\nn \\
&\quad -m_{p}(2s+t-2m_{p}\2 -2m_{\pi}\2)\cM_{8}^{(\text{on})\pm}\,,\\
\nn\\
\label{2.25}
&B^{(\text{on})\pm}(s,t)=\cM_{2}^{(\text{on})\pm}+\cM_{4}^{(\text{on})\pm}
+2m_{p}\cM_{5}^{(\text{on})\pm} \nn\\
&\quad -2m_{p}\cM_{7}^{(\text{on})\pm}
+(4m_{p}^{2}-t)\cM_{8}^{(\text{on})\pm}\,.
\end{align}
With \eqref{2.23} we recover a standard form 
for the on-shell $\ppm p\to \ppm p$ scattering amplitudes;
see, e.g., Chap.~18.10 of \cite{Bjorken:1965}. 
We see from \eqref{2.24} and \eqref{2.25} that in the on-shell processes \eqref{2.2} only two combinations of the eight invariant amplitudes describing 
the corresponding off-shell processes \eqref{2.1} 
can be measured. 
In the next section, where we discuss 
the reactions $\ppm p\to \ppm p\gamma$, we will, however, have to deal with the off-shell amplitudes \eqref{2.16} with all eight invariant amplitudes. 
This presents one challenge of our present investigation. 

\begin{figure}[!ht]
\includegraphics[width=6cm]{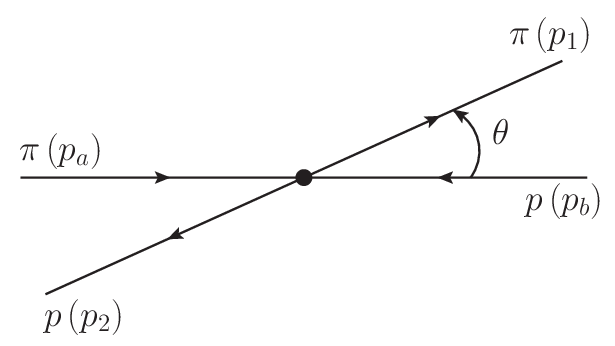}
\caption{
The scattering reaction 
$\pi(p\ua)+p(p\ub) \to \pi(p\uu)+p(p\ud)$ 
in the c.m. system.}
\label{fig:1}
\end{figure}

For completeness we recall also the standard forms 
of the $\pi p$ amplitudes obtained 
when going from the Dirac to the Pauli spinors for the protons. 
We have
\begin{align}\label{2.26}
&u(p,\lambda)=\sqrt{p^{0}+m_{p}}\left(\begin{array}{l}
\chi_{\lambda}^{\;}
\vspace{0.1cm}\\
\dfrac{\bm{\sigma}\cdot \bm{p}}{p^{0}+m_{p}}\chi_{\lambda}^{\;}
\end{array}\right),\nn\\
&
{\chi_{\lambda}^{\;}}^{\!\!\!\!\dagger}\,
\chi_{\lambda'}^{\;}
= \delta_{\lambda \lambda'}^{\;} \,, \quad
\lambda, \lambda' \in \lbrace 1/2, -1/2 \rbrace\,.
\end{align}
We consider now the reactions \eqref{2.5} 
in the c.m. system with $\theta$ the scattering angle; 
see Fig.~\ref{fig:1}.

We set here
\begin{align}\label{2.27}
&E=p\ub^{0}=p\ud^{0}=\frac{1}{2\sqrt{s}}(s+m_{p}\2-m_{\pi}\2)\,,\nn\\
&\bm{n}=\frac{\bm{p \ua}\times\bm{p \uu}}{|\bm{p \ua}\times\bm{p \uu}|}\,,
\end{align}
and obtain
\begin{align}\label{2.28}
&p\ua^{0}=\sqrt{s}-E\,,\nn\\
&|\bm{p \ua}|=|\bm{p \ub}|=|\bm{p \uu}|=|\bm{p \ud}|=\sqrt{E\2-m_{p}\2}\,,\nn\\
&\cos \theta = \frac{\bm{p \uu}\cdot\bm{p \ua}}{|\bm{p \ua}|\2}\,,\nn\\
&t=-2(E\2-m_{p}\2)(1-\cos \theta)\,,\nn\\
&|\bm{p \ua}\times\bm{p \uu}|=(E\2-m_{p}\2)\sin \theta\,.
\end{align}
Inserting all this in \eqref{2.23} we get
\begin{align}\label{2.29}
&\braket{\ppm(p\uu), \, p(p\ud ,\lambda\ud)|{\mathcal T}|\ppm(p\ua),\, p(p\ub ,\lambda\ub)}\nn\\
&\quad =8\pi\sqrt{s}\;\chi_{\lambda_{2}}^{\dagger}\Big{[}f^{\pm}(\sqrt{s},\theta)+i \bm{\sigma}\cdot \bm{n}\, g^{\pm}(\sqrt{s},\theta)\Big{]}\chi_{\lambda_{b}}\,,
\end{align}
where
\begin{align}\label{2.30}
&f^{\pm}(\sqrt{s},\theta)=\frac{1}{8\pi\sqrt{s}}
\bigg{\lbrace}(E+m_{p})\Big{[}A^{(\text{on})\pm}(s,t)\nn \\
&\quad+(\sqrt{s}-m_{p})B^{(\text{on})\pm}(s,t)\Big{]}\nn\\
&\quad +\cos \theta (E-m_{p})\Big{[}-A^{(\text{on})\pm}(s,t)\nn\\
&\quad
+(\sqrt{s}+m_{p})B^{(\text{on})\pm}(s,t)\Big{]}\bigg{\rbrace}\,,\\
\label{2.31}
&g^{\pm}(\sqrt{s},\theta)=-\sin \theta\frac{E-m_{p}}{8\pi\sqrt{s}}\nn\\
& \quad \times \Big{[}-A^{(\text{on})\pm}(s,t)+(\sqrt{s}+m_{p})B^{(\text{on})\pm}(s,t)\Big{]}\ .
\end{align}
With \eqref{2.29} we have another standard form of the $\pi p$ scattering amplitude; see for instance Chap.~16.3 of \cite{Nachtmann:1990ta}.

\section{The reaction $\pi^{-}p\to\pi^{-} p \gamma$}
\label{sec:3}

In this section we will study in detail the reaction
\begin{equation}
\pi^{-}\,(p\ua)+p\,(p\ub,\lambda\ub)\to \pi^{-}\,(p'\uu)+p\,(p'\ud,\lambda'\ud)+\gamma\,(k , \varepsilon)\,.
\label{3.1}
\end{equation}
Here $\varepsilon$ is the polarization vector of the photon and $\lambda\ub\,,\,\lambda'_{2}\in \lbrace 1/2 , -1/2\rbrace$ are the spin indices of the protons.
We are interested in the soft-photon limit and want to compare the amplitude for \eqref{3.1} to that of the process \eqref{2.5} without photon
\begin{equation}
\label{3.2}
\pi^{-}\,(p\ua)+p\,(p\ub ,\lambda\ub)\to\pi^{-}\,(p\uu)+p\,(p\ud ,\lambda\ud)\,.
\end{equation}
The energy-momentum conservation requires for \eqref{3.1}~and \eqref{3.2}
\begin{align}\label{3.3}
p\ua+p\ub&=p'\uu+p'\ud +k\,,
\intertext{and}
\label{3.4}
p\ua+p\ub&=p\uu+p\ud\,,
\end{align}
respectively.
In order not to overload the notation we consider 
here at first only $\pi^{-}p$ scattering.
But at the end of this section we give the results
for both reactions $\pi^{-} p \to \pi^{-} p \gamma$
and $\pi^{+} p \to \pi^{+} p \gamma$.

The ${\mathcal T}$-matrix element for the reaction \eqref{3.1} 
is given by
\begin{align}
\label{3.5}
&\braket{\pi^{-}(p'\uu), \, p(p'\ud,\lambda'\ud),\,\gamma(k,\varepsilon)|{\mathcal T}|\pi^{-}(p\ua),\,p (p\ub,\lambda\ub)}\nn\\
&\quad =\varepsilon{}^{*}_{\lambda}\,\bar{u}(p'\ud ,\lambda'\ud) \cM^{\lambda}(p'\uu , p'\ud ,k, p\ua, p\ub)\, u(p\ub , \lambda\ub)\ .
\end{align}
For real photon emission we have $k^2=0$. 
But we shall consider $\cM^{\lambda}$ 
for $k$ corresponding to real 
or virtual photon emission 
$k^2 \geqslant 0$, $k^0 \geqslant 0$.

As mentioned in the introduction we treat 
the reaction \eqref{3.1} in full QCD 
but only to leading order in the electromagnetic interaction. 
The precise definition of $\bar{u} \cM^{\lambda} u$ 
in \eqref{3.5} using the framework of QFT 
is given in Appendix~\ref{app:C};
see Eqs.~\eqref{C2} and \eqref{3.17} below.

We want to study the amplitude \eqref{3.5} 
in the limit \mbox{$k \to 0$}. 
We see immediately from \eqref{3.3} 
that with changing $k$ also $p'\uu$ and $p'\ud$ must change. 
Setting in \eqref{3.5} $p'\uu=p\uu$ and $p'\ud=p\ud$ 
we would for $k\neq 0$ violate energy-momentum conservation. 
Therefore, we proceed as in 
Sec.~III of \cite{Lebiedowicz:2021byo}
and Secs.~II--IV of \cite{Lebiedowicz:2023ell} where
detailed discussions of this issue 
for the reaction $\pi \pi \to \pi \pi \gamma$ are given.

We work in the overall c.m. system of the reactions
\eqref{3.1} and \eqref{3.2}.
Given initial conditions for the momenta of 
the reaction \eqref{3.2},
that is, given the c.m. energy $\sqrt{s}$ and the unit vector
$\bm{\hat{p}_{a}} = \bm{p_{a}}/|\bm{p_{a}}|$,
we can vary from the momenta in \eqref{3.2} only the direction
of the outgoing pion, i.e. the unit vector
$\bm{\hat{p}_{1}} = \bm{p_{1}}/|\bm{p_{1}}|$.
The phase space of \eqref{3.2} is the unit sphere.
Given the same initial conditions for \eqref{3.1}
we have there as free momentum variables in the final state
the four-vector $k$ and 
$\bm{\hat{p}'_{1}} = \bm{p'_{1}}/|\bm{p'_{1}}|$.
For~small~$k$, say $|k^{\mu}| \ll |\bm{p_{a}}|$
($\mu = 0, \ldots, 3$) $\bm{\hat{p}'_{1}}$
can vary over the whole unit sphere.
The phase space of \eqref{3.1} is,
in our case, given by $(k, \bm{\hat{p}'_{1}})$
and is six dimensional.

\begin{figure}[!ht]
\includegraphics[height=3cm]{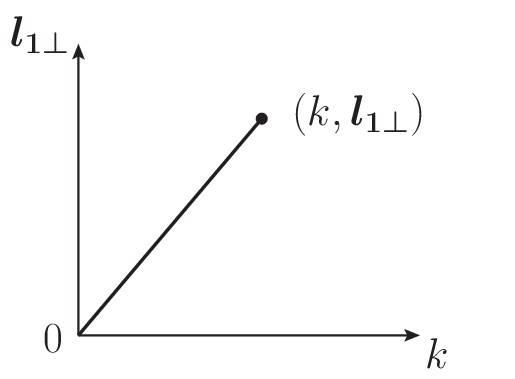}
\caption{Sketch of the six-dimensional neighborhood $(k, \bm{l_{1 \perp}})$
of the phase-space point $(k = 0, \bm{\hat{p}'_{1}} = \bm{\hat{p}_{1}})$;
see (\ref{3.5b}).}
\label{fig:4}
\end{figure}

In the following we shall give the expansion of the amplitude
$\cM_{\lambda}$ (\ref{3.5}) around the phase-space point
\mbox{$(k = 0, \bm{\hat{p}'_{1}} = \bm{\hat{p}_{1}})$}.
In a small neighborhood of this point we set
\begin{eqnarray}
\bm{\hat{p}\uu'} = \bm{\hat{p}\uu} - 
\frac{\bm{l_{1 \perp}}}{|\bm{p_{1}}|}\,,
\label{3.5b}
\end{eqnarray}
where
$|\bm{l_{1 \perp}}| = {\mathcal O}(\omega)$,
$\bm{l_{1 \perp}} \cdot \bm{\hat{p}\uu} = 
0 + {\mathcal O}(\omega^{2})$.
This neighborhood is then parametrized by 
$(k, \bm{l_{1 \perp}})$.
A sketch of this neighborhood is shown in Fig.~\ref{fig:4}.

\begin{figure}[!ht]
\includegraphics[height=3cm]{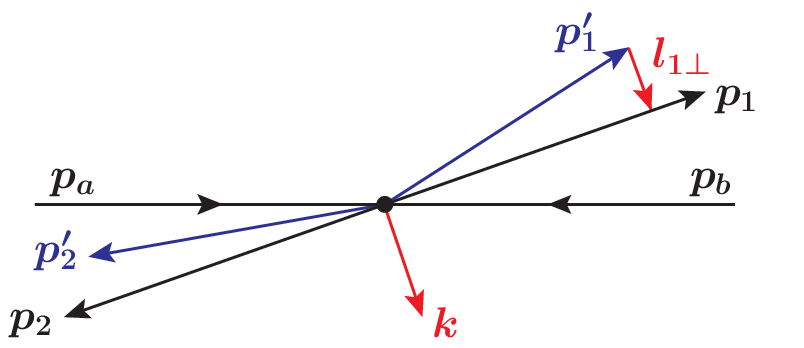}
\caption{Sketch of the momentum configurations for the reactions \eqref{3.1} and \eqref{3.2} in the c.m. system.}
\label{fig:2}
\end{figure}

In order to calculate the four-momenta corresponding to 
$(k, \bm{l_{1 \perp}})$
we set
\begin{equation}
p'\uu=p\uu-l\uu\,,\quad p'\ud=p\ud-l\ud\,,
\label{3.6}
\end{equation}
and we get for $l_{1,2}$ the conditions
\begin{align}\label{3.7}
l\uu+l\ud&=k\,,\nn\\
(p\uu-l\uu)\2&=m_{\pi}\2\,,\nn\\
(p\ud-l\ud)\2&=m_{p}\2\,,
\end{align}
which we can also write as
\begin{align}\label{3.8}
l\uu+l\ud&=k\,,\nn\\
2\,(p\uu\cdot l\uu)&=l_{1}\2\,,\nn\\
2\,(p\ud\cdot l\ud)&=l_{2}\2\,.
\end{align}
These are 6 equations for the 8 unknown components of 
$l\uu$ and $l\ud$. 
We expect that the solution of \eqref{3.6} will have 2 free parameters and this is indeed the case; see~\cite{Lebiedowicz:2021byo}. 
As~discussed there we shall also here consider the situation that the momenta $p'\uu$ and $p'\ud$ in \eqref{3.1} are close to the momenta $p\uu$ and $p\ud$, respectively, in \eqref{3.2}. 
That is, we shall assume that all components of 
$k$, $l\uu$, and $l\ud$, are of order $\omega$ and then study 
the limit $\omega \to 0$. 
We want to extract the terms of order $\omega^{-1}$ 
and $\omega^{0}$ from the amplitude \eqref{3.5}.
For this we need $l\uu$ and $l\ud$ as solutions of \eqref{3.8} only to order $\omega$. 
Neglecting, therefore, $l\uu\2$ and $l\ud\2$ in \eqref{3.8} we get a system of linear equations for $l\uu$ and $l\ud$
\begin{align}
\label{3.9}
l\uu+l\ud&=k\,,\nn\\
p\uu\cdot l\uu&=0\,,\nn\\
p\ud\cdot l\ud&=0\,.
\end{align}
To obtain the solution of \eqref{3.9} we go,
as mentioned above,
to the c.m. system of the reactions \eqref{3.1} and \eqref{3.2} 
where we have
\begin{align}
\label{3.9a}
&\bm{p \ua}+\bm{p \ub}=\bm{p \uu}+\bm{p \ud}=0\,;\\
\label{3.9b}
&(p\uu{}^{\mu})=\left(\begin{array}{l}
p\uu^{0}
\vspace{0.1cm}\\
|\bm{p \uu}|\,\bm{\hat{p}\uu}
\end{array}\right),\quad |\, \bm{\hat{p}\uu}|=1\;,
\nn\\
&(p\ud{}^{\mu})=\left(\begin{array}{l}
p\ud^{0}
\vspace{0.1cm}\\
-|\bm{p \ud}|\,\bm{\hat{p}\uu}
\end{array}\right);
\\
\label{3.10}
& p_{a}^{0} = p_{1}^{0} = \frac{1}{2 \sqrt{s}}(s + m_{\pi}^{2} - m_{p}^{2})\,, \nn \\
& p_{b}^{0} = p_{2}^{0} = \frac{1}{2 \sqrt{s}}(s - m_{\pi}^{2} + m_{p}^{2})\,, \nn \\
& |\bm{p_{a}}| = |\bm{p_{b}}| = |\bm{p_{1}}| = |\bm{p_{2}}| \nn \\
& \quad = \sqrt{(p_{a}^{0})^{2} - m_{\pi}^{2}}
= \sqrt{(p_{b}^{0})^{2} - m_{p}^{2}}
\,.\\
\intertext{We set}
\label{3.11}
&(k^{\mu})=
\left( \begin{array}{l}
k^{0}
\vspace{0.1cm}\\
k_{\parallel}\,\bm{\hat{p}\uu}+\bm{k_{\perp}}
\end{array}\right) , \nn\\
&k^{0}=\omega\,,\qquad \bm{k_{\perp}}\cdot\bm{\hat{p}\uu}=0\,.
\end{align}
The solution of \eqref{3.9} reads then
\begin{align}
(l\uu{}^{\mu})&=\left( \begin{array}{l}
\dfrac{p\ud\cdot k}{\sqrt{s}}
\vspace{0.1cm}\\
\dfrac{p_{1}^{0}}{|\bm{p\uu}| \sqrt{s}} (p\ud\cdot k) \bm{\hat{p}\uu} +\bm{l_{1 \perp}}
\end{array}\right),\nn \\
(l\ud{}^{\mu})&=\left( \begin{array}{l}
\dfrac{p\uu\cdot k}{\sqrt{s}}
\vspace{0.1cm}\\
\bm{k}-\dfrac{p_{1}^{0}}{|\bm{p\uu}| \sqrt{s}} (p\ud\cdot k) \bm{\hat{p}\uu} -\bm{l_{1 \perp}}
\end{array}\right),
\label{3.13}
\end{align}
where
\begin{equation}
\label{3.14}
\bm{l_{1\perp}}\cdot \bm{\hat{p}_{1}}=0\,.
\end{equation}
The vector $\bm{l_{1\perp}}$ remains undetermined and corresponds 
to the 2 free parameters of the solution of \eqref{3.9}. 
In the following we require $\bm{l_{1\perp}}$ to be of order $\omega$. 
With \eqref{3.13}, \eqref{3.14} 
we have given the solution of \eqref{3.8} 
correct to order~$\omega$.
We illustrate the situation 
in the c.m. system in Fig.~\ref{fig:2}.

We can also easily check from (\ref{3.6}) and (\ref{3.13}) 
that we have
\begin{eqnarray}
\bm{\hat{p}\uu'} = \bm{\hat{p}\uu} - 
\frac{\bm{l_{1 \perp}}}{|\bm{p_{1}}|}
+ {\mathcal O}(\omega^{2})\,.
\label{3.16a}
\end{eqnarray}
Therefore, (\ref{3.6}) plus (\ref{3.13}) give
the momenta $p'_{1}$ and $p'_{2}$
of the final pion and proton, respectively, 
corresponding to the phase-space point $(k, \bm{l_{1 \perp}})$
in the neighborhood of 
\mbox{$(k = 0, \bm{\hat{p}'_{1}} = \bm{\hat{p}_{1}})$}; see (\ref{3.5b}).

Our simple kinematic discussion clearly shows that it makes no sense trying to expand the amplitude \eqref{3.5} with respect to the photon momentum $k$ keeping $p'\uu=p\uu$ and $p'\ud= p\ud$ fixed to their values in \eqref{3.2}. The momentum configuration $p\uu$, $p\ud$, $k$ leads \textit{outside} of the physical region for $k\neq 0$, since then we would have from \eqref{3.4}
\begin{equation}
\label{3.15}
p\ua+p\ub\neq p\uu +p\ud +k\,.
\end{equation}

Now we come to another problem in the comparison of the amplitudes for \eqref{3.1} and \eqref{3.2}. 
In \eqref{3.5} we have the spinor $\bar{u}(p'\ud, \lambda'\ud)$ which must change with changing~$p'\ud$. We~could fix, by some arbitrary convention, 
these spinors for each $p'\ud$ in order to have a uniquely defined function for the ${\mathcal T}$-matrix element in \eqref{3.5}. 
We avoid this arbitrariness by considering 
instead the following matrix function
\begin{align}
\label{3.16}
&\mathcal{N}^{\lambda}(p'\uu, p'\ud, k, p\ua ,p\ub)\nn\\
&\quad =\sum_{\lambda'\ud , \lambda\ub} u(p'\ud ,\lambda'\ud)\bar{u}(p'\ud ,\lambda'\ud) \nn\\
&\qquad \times \mathcal{M}^{\lambda}(p'\uu, p'\ud, k, p\ua ,p\ub) u(p\ub ,\lambda\ub)\bar{u}(p\ub ,\lambda\ub)\nn\\
&\quad =(\slash{p}'\ud +m_{p})
\mathcal{M}^{\lambda}(p'\uu, p'\ud, k, p\ua ,p\ub)(\slash{p}\ub +m_{p})\,.
\end{align}
$\mathcal{N}^{\lambda}$ is unambiguously defined and contains all information on the matrix elements \eqref{3.5} for all spin configurations, since we have
\begin{align}
\label{3.17}
&\bar{u}(p'\ud ,\lambda'\ud) \mathcal{M}^{\lambda}u(p\ub ,\lambda\ub )=\frac{1}{(2m_{p})\2}\bar{u}(p'\ud ,\lambda'\ud)\mathcal{N}^{\lambda}u(p\ub ,\lambda\ub)\,.
\end{align}

\begin{figure}[!ht]
(a)\quad \includegraphics[width=.3\textwidth]{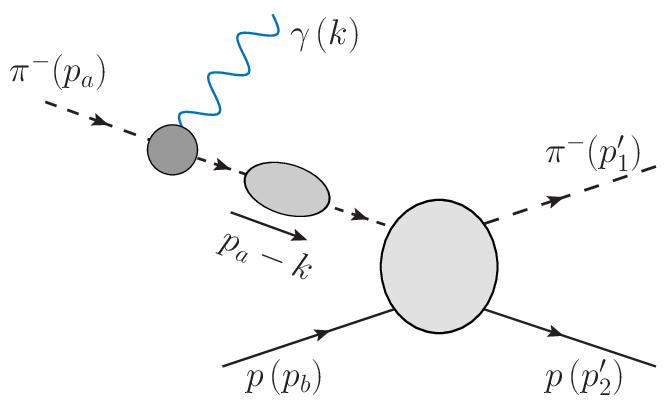}\\
(b)\quad \includegraphics[width=.3\textwidth]{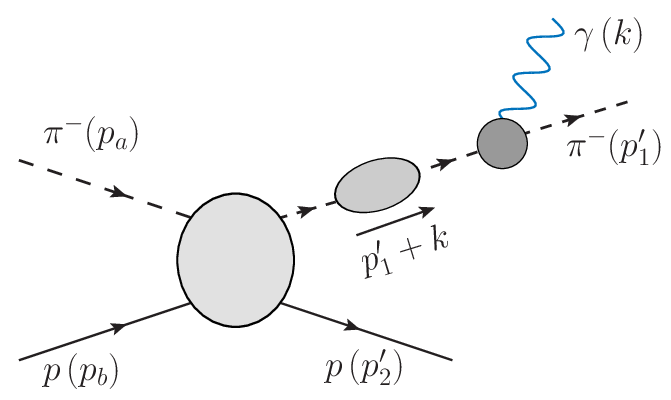}\\
(c)\quad \includegraphics[width=.3\textwidth]{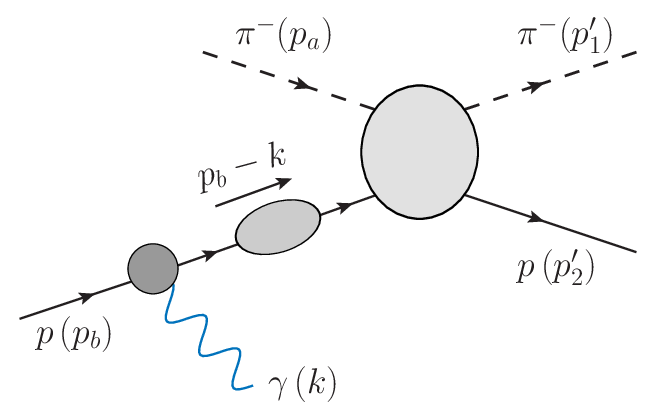}\\
(d)\quad \includegraphics[width=.3\textwidth]{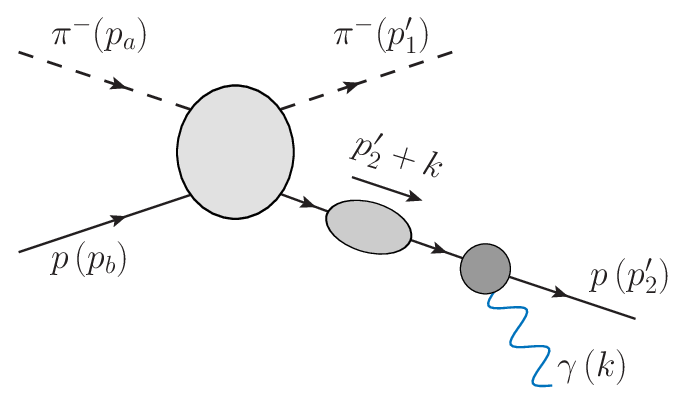}\\
(e)\quad \qquad \;\includegraphics[width=.26\textwidth]{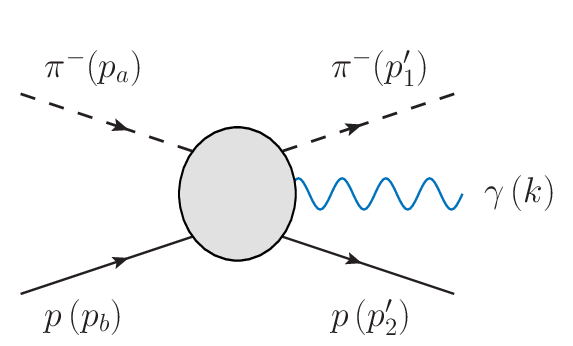}
\caption{
Diagrams for single photon emission
in the reaction $\pi^{-} p \to \pi^{-} p \gamma$.
The shaded blobs in (a)--(d) correspond to
the $\pi$ and $p$ full propagators,
the full $\gamma \pi \pi$ and $\gamma pp$ vertices,
and the off-shell $\pi p \to \pi p$ amplitudes, respectively.
The diagrams in (a)--(d) give contributions to
the amplitude which are singular for $k \to 0$.
The diagram of (e) represents the remaining part of
the amplitude which is non singular for $k \to 0$.}
\label{fig:3}
\end{figure}

In the following we shall, therefore, mainly work with $\mathcal{N}^{\lambda}$ instead of $\mathcal{M}^{\lambda}$.
The definition of $\mathcal{N}^{\lambda}$
using the reduction formulas of QFT is given in \eqref{C2}.
We have five types of diagrams for~$\mathcal{N}^{\lambda}$; 
see Fig.~\ref{fig:3}.
The diagrams Figs.~\ref{fig:3}(a)
and~\ref{fig:3}(b) correspond to the emission of the photon from the external $\pi^{-}$ lines. 
In Figs.~\ref{fig:3}(c) and \ref{fig:3}(d) we have the diagrams for photon emission by the external proton lines. 
Fig.~\ref{fig:3}(e) corresponds to the structure term, that is, all the remaining photon emissions.
We have
\begin{equation}
\label{3.18}
\mathcal{N}\ul=\cN\ul^{(a)}+\cN\ul^{(b)}+\cN\ul^{(c)}+\cN\ul^{(d)}+\cN\ul^{(e)}\,,
\end{equation}
and gauge invariance requires
\begin{equation}
\label{3.19}
k^{\lambda}\mathcal{N}_{\lambda}=0\ .
\end{equation}
Note that in the diagrams of Figs.~\ref{fig:3}(a)--(d) we have the complete $\pi$ and $p$ propagators, the complete $\gamma\pi\pi$ and $\gamma pp$ vertex functions, and the off-shell $\pi p\to \pi p$ scattering amplitudes.
We have
\begin{align}
\label{3.20}
\cN\ul^{(a)}&=-e (\slash{p}'\ud+m_{p})\mathcal{M}^{(0,a)}(\slash{p}\ub+m_{p})\nn\\
&\quad \times \Delta_{F}\big{[}(p\ua-k)\2\big{]}\widehat{\Gamma}\ul^{(\gamma\pi\pi)}(p\ua-k, \, p\ua)\,,\\
\nn \allowdisplaybreaks\\
\label{3.21}
\cN\ul^{(b)}&=-e\widehat{\Gamma}\ul^{(\gamma\pi\pi)}(p'\uu,\,p'\uu+k)\Delta_{F}\big{[}(p'\uu+k)\2\big{]}\nn\\
&\quad \times 
(\slash{p}'\ud+m_{p})\mathcal{M}^{(0,b)}(\slash{p}\ub+m_{p})\,,\\
\nn \allowdisplaybreaks\\
\label{3.22} 
\cN\ul^{(c)}&=e (\slash{p}'\ud+m_{p})\mathcal{M}^{(0,c)}S_{F}(p\ub -k)\nn\\
&\quad \times \widehat{\Gamma}\ul^{(\gamma pp)}(p\ub-k,\,  p\ub)(\slash{p}\ub+m\up)\,,\\ \allowdisplaybreaks
\nn \\
\label{3.23}
\cN\ul^{(d)}&= e (\slash{p}'\ud+m_{p})\widehat{\Gamma}\ul^{(\gamma pp)}(p'\ud,\,p'\ud+k)\nn\\
&\quad \times S_{F}(p'\ud+k)\mathcal{M}^{(0,d)}(\slash{p}\ub+m\up)\,.\allowdisplaybreaks
\end{align}
Here the propagators $\Delta_{F}$, $S_{F}$,
and the vertex functions $\widehat{\Gamma}\ul^{(\gamma\pi\pi)}$, $\widehat{\Gamma}\ul^{(\gamma pp)}$ are defined and discussed in Appendix~\ref{app:B} and we have set for the off-shell~$\pi^{-} p\to \pi^{-} p$ amplitudes occurring 
\begin{align}
\label{3.24}
\mathcal{M}^{(0,a)}&=\mathcal{M}^{(0)}(p'\uu,\, p'\ud,\, p\ua-k ,\, p\ub )\,,\\
\label{3.25}
\mathcal{M}^{(0,b)}&=\mathcal{M}^{(0)}(p'\uu+k,\,p'\ud,\, p\ua ,\, p\ub )
\,,\\
\label{3.26}
\mathcal{M}^{(0,c)}&=\mathcal{M}^{(0)}(p'\uu,\, p'\ud,\, p\ua ,\, p\ub -k)\,,\\
\label{3.27}
\mathcal{M}^{(0,d)}&=\mathcal{M}^{(0)}(p'\uu,\, p'\ud+k,\, p\ua ,\, p\ub )\,.
\end{align}
From the generalized Ward identities \eqref{B33} and \eqref{B55} we get
\begin{align}
\label{3.28}
k^{\lambda}\cN\ul^{(a)}&=\phantom{-}e\cN^{(0,a)}=\phantom{-}e(\slash{p}'\ud+m\up)\cM^{(0,a)}(\slash{p}\ub+m\up)\,,\\
\label{3.29}
k^{\lambda}\cN\ul^{(b)}&=-e\cN^{(0,b)}=-e(\slash{p}'\ud+m\up)\cM^{(0,b)}(\slash{p}\ub+m\up)\,,\\
\label{3.30}
k^{\lambda}\cN\ul^{(c)}&=-e\cN^{(0,c)}=-e(\slash{p}'\ud+m\up)\cM^{(0,c)}(\slash{p}\ub+m\up)\,,\\
\label{3.31}
k^{\lambda}\cN\ul^{(d)}&=\phantom{-}e\cN^{(0,d)}=\phantom{-}e(\slash{p}'\ud+m\up)\cM^{(0,d)}(\slash{p}\ub+m\up)\,.
\end{align}
The structure term $\cN\ul^{(e)}$ will be determined to the order $\omega^{0}$ from the gauge-invariance relation \eqref{3.19} which, together with \eqref{3.18} and \eqref{3.28}--\eqref{3.31}, gives
\begin{align}
\label{3.32}
k^{\lambda}\cN\ul^{(e)}
&=
-k^{\lambda}\big{[}\cN\ul^{(a)}+\cN\ul^{(b)}+\cN\ul^{(c)}+\cN\ul^{(d)}\big{]}\nn\\
&=
e\big{[}-\cN^{(0,a)}+\cN^{(0,b)}+\cN^{(0,c)}-\cN^{(0,d)}\big{]}\,.
\end{align}

One of the main purposes of our present paper is the calculation of
the expansion of $\mathcal{N}_{\lambda}$ (\ref{3.16}) for $\omega \to 0$
up to the order $\omega^{0}$.
To be precise, we shall expand $\mathcal{N}_{\lambda}$ around
the phase-space point $(k = 0, \bm{\hat{p}_{1}})$ setting
\begin{align}
\label{3.34a}
& k =
\omega \left( \begin{array}{l}
1
\vspace{0.1cm}\\
\bm{\tilde{k}}
\end{array}\right), \quad 
\omega \geqslant 0, \quad
\bm{\tilde{k}}^{2} \leqslant 1\,,\nn\\
& \bm{l_{1 \perp}} = \omega \;\bm{\tilde{l}_{1 \perp}}\,,
\quad |\bm{\tilde{l}_{1 \perp}}| = {\mathcal O}(1)\,.
\end{align}
We keep $\bm{\tilde{k}}$ and $\bm{\tilde{l}_{1 \perp}}$ fixed 
and vary only $\omega$.
In this way we obtain a Laurent expansion 
for $\mathcal{N}_{\lambda}$
where we shall give the pole term $\propto \omega^{-1}$
and the next-to-leading term $\propto \omega^{0}$.
Note that with (\ref{3.34a}) we have
\begin{align}
\label{3.34b}
k^{2}=\omega^{2}(1 - \bm{\tilde{k}}^{2}) \geqslant 0\,.
\end{align}
That is, $k^{2}$ has to be counted as 
${\mathcal O}(\omega^{2})$.
But, as we shall see, we must keep $k^{2}$ in the appropriate places,
since we want to give the above expansion for real ($k^{2} = 0$)
and virtual ($k^{2} > 0$) photon emission.
The latter should be relevant for the reaction
\begin{align}
\pi^{-}\,(p\ua)+p\,(p\ub,\lambda\ub) \to \,& 
\pi^{-}\,(p'\uu)+p\,(p'\ud,\lambda'\ud) \nn \\
& + [\gamma^{*}\,(k) \to e^{+}(k_{1}) + e^{-}(k_{2})]\,.
\label{3.34c}
\end{align}
Of course, in (\ref{3.34c}) we have always $k^{2} \geqslant 4 m_{e}^{2}$.
Thus, the limit $k \to 0$ according to (\ref{3.34b}) cannot be reached.
But the electron mass is very small on a hadronic scale and, therefore, 
the limit (\ref{3.34a}) and (\ref{3.34b}) 
should be of relevance for (\ref{3.34c}).

The calculation of $\cN\ul^{(a)},\dots ,\cN\ul^{(d)}$ up to the order 
$\omega^{0}$ is, in principle, straightforward but very lengthy and tedious. 
We shall here only sketch the calculation of $\cN\ul^{(a)}$. 
The complete calculations for $\cN\ul^{(a)},\dots ,\cN\ul^{(d)}$ 
are presented in Appendix~\ref{app:C}.

In $\cN\ul^{(a)}$ \eqref{3.20} the momentum $p\ua$ is on shell, 
$p\ua\2=m_{\pi}\2$. 
Therefore, we can use the result from \eqref{B41} which implies
\begin{align}
\label{3.33}
&\Delta_{F}\big{[}(p\ua-k)\2\big{]}\,\widehat{\Gamma}\ul^{(\gamma\pi\pi)}(p\ua-k ,\,p\ua)\nn\\
&\quad =\frac{(2p\ua -k)\ul}{-2p\ua\cdot k +k\2+i\varepsilon}+{\mathcal O}(\omega)\,.
\end{align}

Note that here we have to keep $k^{2}$ (\ref{3.34b})
in order to get the correct result to the orders
$\omega^{-1}$ and $\omega^{0}$ upon expansion:
\begin{align}
\label{3.35a}
&\frac{(2p\ua - k)\ul}{-2p\ua\cdot k + k\2}
= 
\frac{(2p\ua - k)\ul}{-2p\ua\cdot k} 
\left( 1 + \frac{k^{2}}{2p\ua\cdot k} \right)
+ {\mathcal O}(\omega) \nn \\
&\qquad = 
- \frac{p_{a \lambda}}{p\ua \cdot k} + \frac{1}{2(p\ua \cdot k)^{2}}
\left( k_{\lambda} p\ua \cdot k - p_{a \lambda} k^{2} \right)
+ {\mathcal O}(\omega)\,.
\end{align}

This determines the terms of order $\omega^{-1}$ and $\omega^{0}$ in $\Delta_{F}\,\widehat{\Gamma}\ul^{(\gamma\pi\pi)}$. 
From \eqref{3.20}, \eqref{3.24}, and \eqref{3.28} 
we see that we have to calculate
\begin{align}
\label{3.34}
\cN^{(0,a)}&=(\slash{p}'\ud
+m\up)\cM^{(0,a)}(\slash{p}\ub+m\up)
\intertext{to the orders $\omega^{0}$ and $\omega^{1}$ 
in order to obtain $\cN\ul^{(a)}$ 
to the orders $\omega^{-1}$ and $\omega^{0}$. 
For the off-shell amplitude \eqref{3.24}}
\label{3.35}
\cM^{(0,a)}&=\cM^{(0)}(p'\uu ,\,p'\ud,\,p\ua-k,\,p\ub)
\end{align}
we have to use \eqref{2.16} where the momenta $\tilde{p}_{s}$, $\tilde{p}_{t}$, $\tilde{p}_{u}$ are given by
\begin{align}
\label{3.36}
\tilde{p}_{s}&=p\ua+p\ub-k=p_{s}-k\,,\nn\\
\tilde{p}_{t}&=p\ud-p\ub-l\ud=p_{t}-l\ud\,,\nn\\
\tilde{p}_{u}&=p\uu-p\ub-l\uu=p_{u}-l\uu\,.
\end{align}
\begingroup
\allowdisplaybreaks
Here and in the following $p_{s}$, $p_{t}$, $p_{u}$ 
are the momenta corresponding to 
the on-shell $\pi p \to \pi p$ reaction \eqref{3.2}:
\begin{align}
\label{3.37}
&p_{s}=p\ua+p\ub\,,\nn\\
&p_{t}=p\ud-p\ub\,,\nn\\
&p_{u}=p\uu-p\ub\,, \nn\\
&s=p_{s}\2\,,\quad t=p_{t}\2\,,\quad u=p_{u}\2\,.
\end{align}
We have from \eqref{3.35}, \eqref{3.36}
\begin{align}
&\tilde{s}=\tilde{p}_{s}\2=(p_{s}-k)\2=s-2p_{s}\cdot k+{\mathcal O}(\omega\2)\,,\nn\\
&\tilde{t}=\tilde{p}_{t}\2=(p_{t}-l_{2})\2=t-2p_{t}\cdot l_{2}+{\mathcal O}(\omega\2)\,,\nn\\
&m\ua\2=(p\ua-k)\2=m_{\pi}\2-2p_{a}\cdot k+{\mathcal O}(\omega\2)\ ,\nn\\
&m\uu\2=m_{\pi}\2\,,\nn\\
&m\ub\2=m\ud\2=m\up\2\ ,
\end{align}
and, therefore, from \eqref{2.16} and \eqref{2.17}
\begin{align}
\label{3.39}
\cN^{(0,a)}&=(\slash{p}'\ud+m\up)\Big{\lbrace}\cM\uu^{(a)}+(\slash{p}_{s}-\slash{k})\cM\ud^{(a)}\nn\\
&\quad +(\slash{p}_{t}-\!\not{l}\ud)\cM_{3}^{(a)}+(\slash{p}_{u}-\!\not{l}\uu)\cM_{4}^{(a)}\nn\\
&\quad +i\sigma_{\mu\nu}(p_{s}-k)^{\mu}(p_{t}-l\ud)^{\nu}\cM_{5}^{(a)}\nn\\
&\quad +i\sigma_{\mu\nu}(p_{s}-k)^{\mu}(p_{u}-l\uu)^{\nu}\cM_{6}^{(a)}\nn\\
&\quad +i\sigma_{\mu\nu}(p_{t}-l\ud)^{\mu}(p_{u}-l\uu)^{\nu}\cM_{7}^{(a)}\nn\\
&\quad +i\gamma_{\mu}\gamma_{5} 
\varepsilon^{\mu\nu\rho\sigma}(p_{s}-k)_{\nu}(p_{t}-l\ud)_{\rho}(p_{u}-l\uu)_{\sigma}\cM_{8}^{(a)}\Big{\rbrace}\nn\\
&\quad \times (\slash{p}\ub+m\up)
+{\mathcal O}(\omega^{2})\,,
\end{align}
where
\begin{align}\label{3.40}
\cM_{j}^{(a)}&=\cM_{j}(\tilde{s},\,\tilde{t},\,m_{\pi}\2,\,m\up\2,\,m\ua\2,\,m\up\2)\nn\\
&=\cM_{j}(s-2p_{s}\cdot k,\, t-2p_{t}\cdot l\ud,\,m_{\pi}\2,
m\up\2,
\nn\\
&\qquad \quad \; m_{\pi}\2-2p\ua\cdot k,\,m_{p}\2)+{\mathcal O}(\omega\2)\nn\\
&=\cM_{j}^{(\text{on})}(s,t,\,m_{\pi}\2,\,m\up\2,\,m_{\pi}\2,\,m\up\2)\nn\\
&\quad -2(p_{s}\cdot k)\, \cM_{j},_{s}(s,t,\,m_{\pi}\2,\,m\up\2,\,m_{\pi}\2,\,m\up\2)\nn\\
&\quad -2(p_{t}\cdot l\ud)\, \cM_{j},_{t}(s,t,\,m_{\pi}\2,\,m\up\2,\,m_{\pi}\2,\,m\up\2)\nn\\
&\quad -2(p_{a}\cdot k)\, \cM_{j},_{m_{a}\2}(s,t,\,m_{\pi}\2,\,m\up\2,\,m_{a}\2,\,m\up\2)\Big{|}_{m_{a}\2 = m_{\pi}\2} \nn\\
&\quad+{\mathcal O}(\omega\2)\,,  \\
(j&=1,\dots , 8)\,. \nn
\end{align}
Here  $\cM_{j},_{s}$,  $\cM_{j},_{t}$, $\cM_{j},_{m_{a}\2}$ denote the partial derivatives of $\cM_{j}$ with respect to $s$, $t$, and $m\ua\2$.

Now we insert \eqref{3.40} into \eqref{3.39} and put together the terms of order $\omega^{0}$ and $\omega^{1}$. This gives a lengthy expression for $\cN^{(0,a)}$ 
which is given in \eqref{C11} of Appendix~\ref{app:C}.
Multiplying with \eqref{3.33} we get for $\cN\ul^{(a)}$ 
(\ref{3.20}) the terms of order 
$\omega^{-1}$ and $\omega^{0}$
as given in \eqref{C22}. 
In an analogous way we calculate in Appendix~\ref{app:C} 
the terms of order $\omega^{-1}$ and $\omega^{0}$ 
for $\cN\ul^{(b)}$, $\cN\ul^{(c)}$, and $\cN\ul^{(d)}$, 
Eqs.~\eqref{3.21}--\eqref{3.23}. 
Finally we determine the structure term $\cN\ul^{(e)}$.
This term is non-singular for $\omega\to 0$ and to the order $\omega^{0}$ it can be uniquely determined from \eqref{3.32} using the results for $\cN^{(0,a)},\dots ,\cN^{(0,d)}$ 
to the orders $\omega^{0}$ and $\omega^{1}$; 
see Appendix~\ref{app:C}.

We give now the final result for $\cN\ul$ \eqref{3.16} to the orders $\omega^{-1}$ and $\omega^{0}$. 
Let $A^{(\text{on})\mp}(s,t)$ and $B^{(\text{on})\mp}(s,t)$ be the invariant amplitudes for $\pi^{\mp}p$ elastic scattering as defined in \eqref{2.23}--\eqref{2.25}.
Here and in the following we indicate again by the superscript 
$-\,(+)$ if we are dealing with $\pi^{-}p \;(\pi^{+}p)$ scattering. 
We set:
\begin{align}
\label{3.41}
A,_{s}^{\!\!(\text{on})\mp}(s,t)&=\frac{\partial}{\partial s}A^ {(\text{on})\mp}(s,t)\,,\nn\\
A,_{t}^{\!\!(\text{on})\mp}(s,t)&=\frac{\partial}{\partial t}A^ {(\text{on})\mp}(s,t)\,,\nn\\
B,_{s}^{\!\!(\text{on})\mp}(s,t)&=\frac{\partial}{\partial s}B^ {(\text{on})\mp}(s,t)\,,\nn\\
B,_{t}^{\!\!(\text{on})\mp}(s,t)&=\frac{\partial}{\partial t}B^ {(\text{on})\mp}(s,t)\,.
\end{align}
We get then
\begin{equation}
\label{3.42}
\cN\ul^{(\pi^{-}p\to\pi^{-} p\gamma)}=\cN\ul^{(a+b+e1)-}+\cN\ul^{(c+d+e2)-}\,,
\end{equation}
where
\begin{widetext}
\begin{align}
\cN\ul^{(a+b+e1)-}=&-e(\slash{p}'_{2}+m\up)\bigg{[}A^{(\text{on})-}
+\frac{1}{2}(\slash{p}\ua+\slash{p}'\uu)
B^{(\text{on})-}\bigg{]}(\slash{p}_{b}+m\up)
\bigg{[}\frac{(2p\ua-k)\ul}{-2p\ua\cdot k+k\2}+\frac{(2p'\uu+k)\ul}{2p'\uu\cdot k+k\2}\bigg{]}\nn\\
&-e(\slash{p}\ud+m\up)
\bigg{\lbrace}
\bigg{[}A,_{s}^{\!\!(\text{on})-}+\frac{1}{2}(\slash{p}\ua+\slash{p}\uu)
B,_{s}^{\!\!(\text{on})-}\bigg{]}\bigg{[}2(p_{s}\cdot k)\frac{p_{a\lambda}}{p\ua\cdot k}-2p_{s\lambda}\bigg{]}\nn\\
&+\bigg{[}
A,_{t}^{\!\!(\text{on})-}+\frac{1}{2}(\slash{p}\ua+\slash{p}\uu)
B,_{t}^{\!\!(\text{on})-}\bigg{]}(-2p_{t}\cdot l\ud)
\bigg{[}\frac{p_{a\lambda}}{(-p\ua\cdot k)}+\frac{p_{1\lambda}}{p\uu\cdot k}\bigg{]}\nn\\
&+B^{(\text{on})-}\bigg{[}\frac{1}{2}\slash{k}\Big{(}\frac{p_{a\lambda}}{p\ua\cdot k}+\frac{p_{1\lambda}}{p\uu\cdot k}\Big{)}-\gamma\ul\bigg{]}
\bigg{\rbrace}
(\slash{p}\ub+m\up)
+\mathcal{O}(\omega)\,,
\label{3.43}
\end{align}
\begin{align}
\cN\ul^{(c+d+e2)-}=&\;
e(\slash{p}'_{2}+m\up)\bigg{[}A^{(\text{on})-}+\frac{1}{2}(\slash{p}\ua+\slash{p}'\uu)B^{(\text{on})-}\bigg{]}(\slash{p}_{b}+m\up)
\bigg{[}\frac{(2p\ub-k)\ul}{-2p\ub\cdot k+k\2}
+
\frac{(2p'\ud+k)\ul}{2p'\ud\cdot k+k\2} \bigg{]}
\nn\\
&+e(\slash{p}\ud+m\up)\bigg{[}A,_{s}^{\!\!(\text{on})-}+\frac{1}{2}(\slash{p}\ua+\slash{p}\uu)B,_{s}^{\!\!(\text{on})-}\bigg{]}\bigg{[}2(p_{s}\cdot k)\frac{p_{b\lambda}}{p\ub\cdot k}-2p_{s\lambda}\bigg{]}(\slash{p}\ub+m\up)\nn\\
&+e(\slash{p}\ud+m\up)\bigg{[}A,_{t}^{\!\!(\text{on})-}+\frac{1}{2}(\slash{p}\ua+\slash{p}\uu)B,_{t}^{\!\!(\text{on})-}\bigg{]}
\bigg{[}-2(p_{t}\cdot l\uu)\frac{p_{b\lambda}}{p\ub\cdot k}+2(p_{t}\cdot l\uu)\frac{p_{2\lambda}}{p\ud\cdot k}\bigg{]}(\slash{p}\ub+m\up)\nn\\
&
+e(\slash{p}_{2}+m\up)\bigg{[}A^{(\text{on})-}+\frac{1}{2}(\slash{p}\ua+\slash{p}\uu)B^{(\text{on})-}\bigg{]}(k\ul-\slash{k}\gamma\ul)(\slash{p}\ub+m_{p})\frac{1}{(-2p\ub\cdot  k)}\nn\\
&-e\frac{1}{(2p\ud\cdot  k)}(\slash{p}\ud+m\up) (k\ul-\gamma\ul\slash{k})\bigg{[}A^{(\text{on})-}+\frac{1}{2}(\slash{p}\ua+\slash{p}\uu)B^{(\text{on})-}\bigg{]}(\slash{p}_{b}+m\up)\nn\\
&+e(\slash{p}\ud+m\up)\bigg{[}A^{(\text{on})-}+\frac{1}{2}(\slash{p}\ua+\slash{p}\uu)B^{(\text{on})-}\bigg{]}\bigg{[}m\up(k\ul-\nk\gamma\ul)+(p_{b\lambda}\nk-(p\ub\cdot k)\gamma\ul)\bigg{]}(\slash{p}_{b}+m\up) \nn\\
&\quad \times 
\frac{F\ud(0)}{m\up}\frac{1}{(-2p\ub\cdot k)}
-e\frac{F\ud(0)}{m\up}\frac{1}{(2p\ud\cdot k)}(\slash{p}_{2}+m\up)\nn\\
&\quad \times \bigg{[}m\up(k\ul-\gamma\ul\nk)+(p_{2\lambda}\nk-(p\ud\cdot k)\gamma\ul)\bigg{]}
\bigg{[}A^{(\text{on})-}+\frac{1}{2}(\slash{p}\ua+\slash{p}\uu)B^{(\text{on})-}\bigg{]}(\slash{p}\ub+m\up)+\mathcal{O}(\omega)\,.
\label{3.44}
\end{align}
\end{widetext}
Here the momenta 
$p_{a}$, $p_{b}$, $p_{1}$, $p_{2}$, $p'_{1}$, $p'_{2}$, $k$, $l_{1}$, $l_{2}$
are as defined in \eqref{3.3}, \eqref{3.4}, \eqref{3.6}, and \eqref{3.13}.
The definition of $p_{s}$, $p_{t}$, $p_{u}$, $s$, $t$, $u$ 
is given in \eqref{3.37}.

To get the corresponding expressions for the reaction $\pi^{+}p \to p^{+}p\gamma$ we make the following replacements in \eqref{3.43} and \eqref{3.44}:
\begin{align}
\label{3.45}
&\mathcal{N}\ul^{(a+b+e1)+}=-\cN\ul^{(a+b+e1)-}\Big{|}_{\mathop{^{A^{(\text{on})-}\to A^{(\text{on})+}}                  
_{B^{(\text{on})-}\to B^{(\text{on})+}}}}\,,\\
\label{3.46}
&\mathcal{N}\ul^{(c+d+e2)+}=\cN\ul^{(c+d+e2)-}\Big{|}
_{\mathop{^{A^{(\text{on})-}\to A^{(\text{on})+}}                  
_{B^{(\text{on})-}\to B^{(\text{on})+}}}}\,.
\end{align}
Of course, these replacements also apply to the derivatives 
$A,_{s}^{\!\!(\text{on})},\dots, B,_{t}^{\!\!(\text{on})}$.
Note the minus sign in \eqref{3.45} which 
is due to the opposite charge of $\pi^{-}$ and $\pi^{+}$ 
which gives \eqref{B25}.
We get then
\begin{equation}
\label{3.47}
\cN^{(\pi^{+}p\to\pi^{+}p\gamma)}\ul=\mathcal{N}\ul^{(a+b+e1)+}+\mathcal{N}\ul^{(c+d+e2)+}\,.
\end{equation}
The main results of our paper are the above expressions for $\cN^{(\pi^{-}p\to\pi^{-}p\gamma)}\ul$ and $\cN^{(\pi^{+}p\to\pi^{+}p\gamma)}\ul$, Eqs.~\eqref{3.42}--\eqref{3.47}, which give these matrix amplitudes to the orders $\omega^{-1}$ and $\omega^{0}$ in the expansion for $\omega\to 0$. We have written \eqref{3.43} and \eqref{3.44} in a (hopefully) transparent form where we have not strictly grouped together the $\omega^{-1}$ and the $\omega^{0}$ terms.
Of course, this could easily be done by substituting $p'\ud=p\ud-l\ud\,$, $\,p'\uu=p\uu-l\uu$ and making suitable expansions,
(\ref{3.35a}), and
\begin{widetext}
\begin{align} 
&\frac{(2p'\uu+k)\ul}{2p'\uu\cdot k +k\2}=
\frac{p_{1\lambda}}{p\uu\cdot k}
-\frac{p_{1\lambda}}{2(p\uu\cdot k)\2}
\big{(}-2(l\uu\cdot k)+k\2\big{)} 
+\frac{1}{2(p\uu\cdot k)}(-2l\uu+k)\ul
+{\mathcal O}(\omega)\,,
\nn \\
&\frac{(2p\ub - k)\ul}{-2p\ub\cdot k + k\2}
= 
- \frac{p_{b \lambda}}{p\ub \cdot k} + \frac{1}{2(p\ub \cdot k)^{2}}
\left( k_{\lambda} p\ub \cdot k - p_{b \lambda} k^{2} \right)
+ {\mathcal O}(\omega)\,,
\nn \\
&\frac{(2p'\ud+k)\ul}{2p'\ud\cdot k +k\2}=
\frac{p_{2\lambda}}{p\ud\cdot k}
-\frac{p_{2\lambda}}{2(p\ud\cdot k)\2}
\big{(}-2(l\ud\cdot k)+k\2\big{)} 
+\frac{1}{2(p\ud\cdot k)}(-2l\ud+k)\ul
+{\mathcal O}(\omega)\,.
\label{3.48}
\end{align}
Note that $k$, $l\uu$ and $l\ud$ are of order $\omega$. 
But we think that substituting such expansions in \eqref{3.43} and \eqref{3.44} would not increase their legibility. 

We give the explicit expansions of $\cN\ul$ 
in powers of $\omega$ in the following 
only for the case of real photon emission, 
that is, for $k^{2} = 0$. 
We write
\begin{align}
\label{3.49}
&\cN\ul^{(\ppm p\to \ppm p\gamma)} \equiv \cN\ul\hpm 
=\frac{1}{\omega}\widehat{\cN}\ul^{(0)\pm}+\widehat{\cN}\ul^{(1)\pm}+{\mathcal O}(\omega)\,,\\
\nn \\
\label{3.50}
&\widehat{\cN}\ul^{(0)\pm}=\widehat{\cN}\ul^{(a+b+e1)(0)\pm}+\widehat{\cN}\ul^{(c+d+e2)(0)\pm}\,,\\
\nn\\
\label{3.51}
&\widehat{\cN}\ul^{(1)\pm}=\widehat{\cN}\ul^{(a+b+e1)(1)\pm}+\widehat{\cN}\ul^{(c+d+e2)(1)\pm}\,.
\end{align}

Neglecting some gauge terms we obtain from (\ref{3.35a}), 
\er{3.42}--\er{3.48}
\bal{3.52}
\widehat{\cN}\ul^{(a+b+e1)(0)\pm}=&\pm e(\bp\ud+m\up)\Big{[}A^{(\text{on})\pm}+\frac{1}{2}(\bp\ua +\bp\uu )B^{(\text{on})\pm}\Big {]}(\bp\ub +m\up)\;
\omega\Big{[}-\frac{p_{a\lambda}}{p\ua\cdot k}+\frac{p_{1\lambda}}{p\uu\cdot k}\Big{]}\,,
\\
\nn \\ 
\label{3.53}
\widehat{\cN}\ul^{(a+b+e1)(1)\pm}=
&\pm e(\bp\ud+m\up)\Big{[}A^{(\text{on})\pm}+\frac{1}{2}(\bp\ua +\bp\uu )B^{(\text{on})\pm}\Big {]}(\bp\ub +m\up)\frac{1}{(p\uu\cdot k)\2}\Big{[}p_{1\lambda}(l\uu\cdot k)-l_{1\lambda}(p\uu\cdot k)\Big{]}\nn\\
&\pm e(-\!\not{l}\ud)\Big{[}A^{(\text{on})\pm}+\frac{1}{2}(\bp\ua +\bp\uu )B^{(\text{on})\pm}\Big {]}(\bp\ub +m\up)
\Big{[}-\frac{p_{a\lambda}}{p\ua \cdot k}+\frac{p_{1\lambda}}{p\uu \cdot k}\Big{]}\nn\\
&\pm e(\bp\ud+m\up) \frac{1}{2}(-\!\not{l}\uu)B^{(\text{on})\pm}(\bp\ub +m\up)
\Big{[}-\frac{p_{a\lambda}}{p\ua \cdot k}+\frac{p_{1\lambda}}{p\uu \cdot k}\Big{]}\nn\\
&\pm e (\bp\ud +m\up)\bigg{\lbrace}\Big{[}A,_{s}^{\!\!(\text{on})\pm}
+\frac{1}{2}(\bp\ua+\bp\uu)
B,_{s}^{\!\!(\text{on})\pm}\Big{]}
\Big{[} 2(p\us \cdot k) \frac{p_{a\lambda}}{p\ua\cdot k}-2p_{s\lambda}\Big{]}\nn\\
&+\Big{[}A,_{t}^{\!\!(\text{on})\pm}+\frac{1}{2}(\bp\ua+\bp\uu)B,_{t}^{\!\!(\text{on})\pm}\Big{]}
2(p\ut \cdot l\ud)
\Big{[}
\frac{p_{a\lambda}}{p\ua \cdot k}-\frac{p_{1\lambda}}{p\uu \cdot k}\Big{]}\nn\\
&+B^{(\text{on})\pm}\Big{[}\frac{1}{2}\bk \Big{(}\frac{p_{a\lambda}}{p\ua \cdot k}+\frac{p_{1\lambda}}{p\uu \cdot k}\Big{)}-\gamma_{\lambda}\Big{]}\bigg{\rbrace}(\bp\ub+m\up )
\,,
\\
\nn\\
\label{3.54}
\widehat{\cN}\ul^{(c+d+e2)(0)\pm}=&\;
e(\bp\ud+m\up)
\Big{[}A^{(\text{on})\pm}+\frac{1}{2}(\bp\ua +\bp\uu)
       B^{(\text{on})\pm}\Big {]}(\bp\ub +m\up)\;
\omega\Big{[}
-\frac{p_{b\lambda}}{p\ub \cdot k}+\frac{p_{2\lambda}}{p\ud \cdot k}\Big{]}\,, 
\\
\nn\\
\label{3.55}
\widehat{\cN}\ul^{(c+d+e2)(1)\pm}=&\;
e(\bp\ud+m\up)
\Big{[}A^{(\text{on})\pm}+\frac{1}{2}(\bp\ua +\bp\uu )B^{(\text{on})\pm}\Big {]}(\bp\ub +m\up)
\frac{1}{(p\ud\cdot k)\2}\Big{[}p_{2\lambda} (l\ud \cdot k )-l_{2\lambda}(p\ud \cdot k)\Big{]}\nn\\
&+e(-\!\not{l}\ud)
\Big{[}A^{(\text{on})\pm}+\frac{1}{2}(\bp\ua +\bp\uu )B^{(\text{on})\pm}\Big {]}(\bp\ub +m\up)
\Big{[}
-\frac{p_{b\lambda}}{p\ub \cdot k}+\frac{p_{2\lambda}}{p\ud \cdot k}\Big{]}
\nn\\
&+e(\bp\ud +m\up)
\frac{1}{2}(-\!\not{l}\uu) B^{(\text{on})\pm}(\bp\ub +m\up)\Big{[}
-\frac{p_{b\lambda}}{p\ub \cdot k}+\frac{p_{2\lambda}}{p\ud \cdot k}\Big{]}
\nn\\
&+e(\bp\ud +m\up)
\Big{[}A,_{s}^{\!\!(\text{on})\pm}+\frac{1}{2}(\bp\ua+\bp\uu)B,_{s}^{\!\!(\text{on})\pm}\Big{]}(\bp\ub +m\up)
\Big{[}2(p\us\cdot k)\frac{p_{b\lambda}}{p\ub \cdot k}-2p_{s \lambda}\Big{]}
\nn\\
&+e(\bp\ud +m\up)\Big{[}A,_{t}^{\!\!(\text{on})\pm}+\frac{1}{2}(\bp\ua+\bp\uu)B,_{t}^{\!\!(\text{on})\pm}\Big{]}(\bp\ub +m\up)
2(p\ut \cdot l\uu)\Big{[}-\frac{p_{b\lambda}}{p\ub \cdot k}+\frac{p_{2\lambda}}{p\ud \cdot k}\Big{]}\nn \allowdisplaybreaks\\
&+e(\bp\ud +m\up)
\Big{[}A^ {(\text{on})\pm}+\frac{1}{2}(\bp\ua+\bp\uu)B^ {(\text{on})\pm}\Big{]}
(k\ul -\bk \gamma\ul)(\bp\ub +m\up)\frac{1}{(-2p\ub\cdot k)}\nn \\
&-e\frac{1}{(2p\ud\cdot k)}(\bp\ud +m\up)(k\ul -\gamma\ul \bk)
\Big{[}A^ {(\text{on})\pm}+\frac{1}{2}(\bp\ua+\bp\uu)B^ {(\text{on})\pm}\Big{]}(\bp\ub +m\up)\nn\\
&+e(\bp\ud + m\up)\Big{[}A^ {(\text{on})\pm}+\frac{1}{2}(\bp\ua+\bp\uu)B^ {(\text{on})\pm}\Big{]}
\Big{[}
m\up(k\ul-\bk\gamma\ul)+\big{(}p_{b\lambda}\bk -(p\ub\cdot k )\gamma\ul\big{)}\Big{]}(\bp\ub+m\up) \nn\\
&\quad \times 
\frac{F\ud(0)}{m\up}\frac{1}{(-2p\ub\cdot k)}
-e\frac{F\ud(0)}{m\up}\frac{1}{(2p\ud\cdot k)}(\bp\ud + m\up)\Big{[} m\up (k\ul -\gamma\ul \bk)+\big{(}p_{2\lambda}\bk -(
p\ud \cdot k)\gamma\ul \big{)}\Big{]}\nn\\
&\quad \times \Big{[}A^ {(\text{on})\pm}+\frac{1}{2}(\bp\ua+\bp\uu)B^ {(\text{on})\pm}\Big{]}(\bp\ub +m\up)\,.
\end{align}

\end{widetext}

With \eqref{3.49}--\eqref{3.55} we have given
the terms of order $\omega^{-1}$ and $\omega^{0}$
of the Laurent expansions of the amplitudes $\cN\ul\hpm$.
We remind the reader that we set $k$ and $\bm{l_{1 \perp}}$
according to (\ref{3.34a}), here with $\bm{\tilde{k}}^{2} = 1$,
and that $l_{1}$ and $l_{2}$ (\ref{3.13})
are then proportional to $\omega$.
The pole term $\widehat{\cN}\ul^{(0)\pm} / \omega$ in \eqref{3.49}
corresponds exactly to Weinberg's soft-photon theorem;
see Sec.~II~1, Eq.~(2.3) of \cite{Weinberg:1965nx}.

In Appendix~\ref{app:D} we compare our results to the results for
soft-photon production in the scattering of a charged spin 1/2 particle
on an uncharged scalar particle given in \cite{Low:1958sn}.
As already discussed in \cite{Lebiedowicz:2023ell} 
Low's results \cite{Low:1958sn}
give approximate expressions valid at a given phase-space point
$(p'_{1}, p'_{2}, k)$, not expansions of the amplitudes around
a phase-space point.
We also discuss in Appendix~\ref{app:D} the relation of our
work to the results from \cite{Liou:1978sz}.

\endgroup

\section{Cross sections for $\ppm p \to \ppm p \gamma $ for $\omega\to 0$}
\label{sec:4}

In Sec.~\ref{sec:3} we have discussed the matrix amplitudes
\begin{equation}
\cN\ul^{(\ppm p\to \ppm p \gamma)}
\equiv
\cN\ul\hpm(p'\uu , p'\ud  , k ,  p\ua ,  p\ub)\,;
\label{4.1}
\end{equation}
see \eqref{3.16}, \eqref{3.42}, and \eqref{3.47}.
We consider now the reactions [cf. \eqref{3.1}]
\bel{4.1a}
\ppm\,(p\ua)+p\,(p\ub , \lambda\ub )\to \ppm\,(p'\uu)+p\,(p'\ud , \lambda'\ud) +\gamma\,(k,\varepsilon)
\ee
with real photon emission. 
We have then from \er{3.5} and \er{3.17} with $k\2 =0$
\bal{4.2}
&\braket{\ppm (p'\uu), \, p(p'\ud , \lambda'\ud) ,\, \gamma (k,\varepsilon)|{\mathcal T}|\ppm (p\ua),\, p(p\ub , \lambda\ub )}\nn\\
&\quad =(\varepsilon\hl)^{*}\frac{1}{(2m\up)\2}
\bar{u}(p'\ud , \lambda'\ud)
\cN\hpm\ul(p'\uu , p'\ud  , k , p\ua ,  p\ub) u(p\ub , \lambda\ub)\,.
\end{align}
We shall first consider differential cross sections 
for unpolarized protons in the initial state 
and no observation of the proton and photon polarizations 
in the final state. We have then
\bal{4.3}
&\dv \sigma (\ppm p\to \ppm p \gamma) \nn\\
&\quad =\frac{1}{2w(s,m\up\2 , m\upp\2)}
\frac{\dv^{3}k}{(2\pi)^{3}2k^{0}}\,
\frac{\dv^{3}p'\uu}{(2\pi)^{3}2p\uu^{\prime 0}}\,
\frac{\dv^{3}p'\ud}{(2\pi)^{3}2p\ud^{\prime 0}}\nn\\
&\qquad \times (2\pi)^{4}\delta^{(4)}(p'\uu+p'\ud+k-p\ua-p\ub)\nn\\
&\qquad \times \frac{1}{(2m\up)\2}\frac{1}{2}(-g^{\lambda\mu})\;
\tr \big{[}\cN\ul\hpm\overline{\cN_{\mu}\hpm}\, 
\big{]}\,,
\end{align}
where
\bel{4.4}
w(x,y,z)=[x\2+y\2+z\2-2xy-2yz-2xz]^{1/2}\,,
\ee
and $\overline{\cN_{\mu}\hpm}$
denotes, as usual, 
the Dirac adjoint matrix to $\cN_{\mu}\hpm$.

Now we go to the overall c.m. system where we have
\bal{4.5}
p\ua^{0}+p\ub^{0}
&=p\uu^{\prime 0}+p\ud^{\prime 0}+\omega=\sqrt{s}\,,\nn\\
\bm{p \ua}+\bm{p \ub}
&=\bm{p \uu'} 
 +\bm{p \ud'}
 +\bm{k}=0\,,\nn\\
 p\uu^{ \prime 0}
&=\sqrt{|\bm{p \uu'} |\2 + m\upp\2} \,, \nn\\
 p\ud^{ \prime 0}
&=\sqrt{|\bm{p \ud'} |\2 + m\up\2} \,.
 \end{align}
Here and in the following we set
\bal{4.6}
&\bm{\hat{p} \ua}=\bm{p \ua} / |\bm{p \ua}|\,,
\quad \bm{\hat{p} \uu'} =\bm{p \uu'} / |\bm{p \uu'}|\,,
\quad \bm{\hat{p} \ud'} =\bm{p \ud'} / |\bm{p \ud'}|\,, \nn \\
&\bm{\hat{k}}=\bm{k}/|\bm{k}|\,, 
\quad \omega = k^{0}=|\bm{k}|\,,
\end{align}
and we consider always small $\omega$,
\bal{4.7a}
\omega \ll |\bm{p \ua}| = \frac{w(s,m_{p}^{2},m_{\pi}^{2})}{2 \sqrt{s}}\,.
\end{align}

As we have discussed in Sec.~\ref{sec:3} we have for
given initial conditions in (\ref{4.2}) as independent
momentum variables in the final state $k$ and $\bm{\hat p\uu'}$,
that is, $\omega$, $\bm{\hat{k}}$, and $\bm{\hat p\uu'}$.
Of course, we can instead also choose $\omega$, $\bm{\hat{k}}$,
and $\bm{\hat p\ud'}$ as independent variables.

\begin{figure}[!ht]
\includegraphics[width=.38\textwidth]{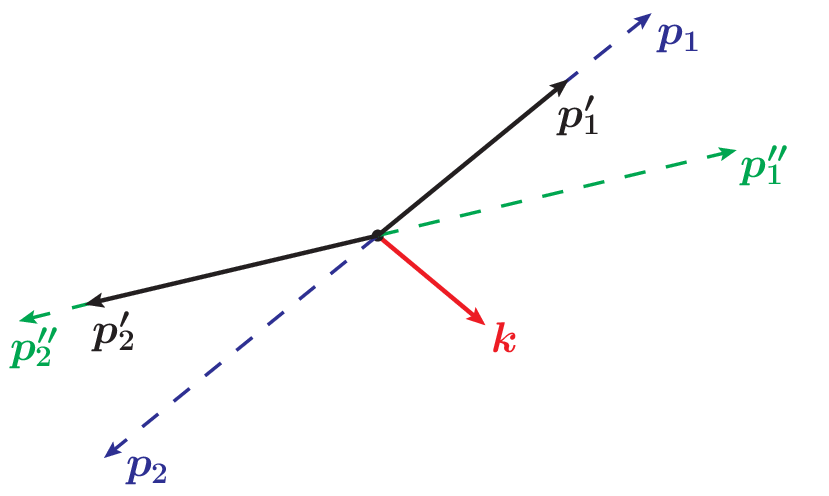}
\caption{The phase-space configuration $(p_{1}', p_{2}', k)$
and the different starting points for
the expansions for $\sigma_{1}$ and $\sigma_{2}$ of (\ref{4.7b}),
$(p_{1}, p_{2}, k=0)$ and $(p_{1}'', p_{2}'', k=0)$, respectively.}
\label{fig:5}
\end{figure}

In the following we shall, therefore, discuss the cross sections
\bal{4.7b}
\sigma_{1} = \frac{\omega \,\dv\sigma}{\dv \omega \, 
\dv \Omega_{\hat{k}} \, \dv \Omega_{\hat{p}_{1}'}}\,, \quad 
\sigma_{2} = \frac{\omega \,\dv\sigma}{\dv \omega \, 
\dv \Omega_{\hat{k}} \, \dv \Omega_{\hat{p}_{2}'}}\,, \quad 
\sigma_{3} = \frac{\omega \,\dv\sigma}{\dv \omega}\,,
\end{align}
where $\dv \Omega_{\hat{k}}$, $\dv \Omega_{\hat{p}_{1}'}$,
and $\dv \Omega_{\hat{p}_{2}'}$ are the solid-angle elements
to $\bm{\hat{k}}$, $\bm{\hat{p}_{1}'}$, and $\bm{\hat{p}_{2}'}$ 
in the overall c.m. system, respectively.
The analogous cross sections for $\pi^{-} \pi^{0} \to \pi^{-} \pi^{0} \gamma$
were discussed in Sec.~VI of \cite{Lebiedowicz:2023ell}.
In order not to overload the notation we shall restrict ourselves
in the following to the discussion of the reaction
$\pi^{-} p \to \pi^{-} p \gamma$ and set from (\ref{4.1})
\begin{equation}
\cN\ul^{-}(p'\uu, p'\ud, k, p\ua, p\ub) \equiv 
\cN\ul(p'\uu, p'\ud, k, p\ua, p\ub) \,.
\label{4.7c}
\end{equation}
Our aim is to give the expansions of the cross sections (\ref{4.7b})
for $\omega \to 0$ to the orders $\omega^{0}$ and $\omega$.
For this it is important to choose the expansion of $\cN\ul$ (\ref{4.7c})
appropriately. For $\sigma_{1}$ in (\ref{4.7b}) we shall use
the expansion in the phase space parametrized by $(k, \bm{\hat{p}_{1}'})$
starting from $(k = 0, \bm{\hat{p}_{1}'} = \bm{\hat{p}_{1}})$
and, keeping $\bm{\hat{p}_{1}'}$ constant;
see Fig.~\ref{fig:5}.
For calculating $\sigma_{2}$ it is convenient to keep $\bm{\hat{p}_{2}'}$
fixed in the corresponding expansion for which we choose a different starting
point in phase space; see Fig.~\ref{fig:5}.
We set
\begin{align}
\label{4.7d}
\bm{\hat{p}_{1}''} = - \bm{\hat{p}_{2}'}
\end{align}
and get from (\ref{3.6}) and (\ref{3.13}),
\begin{align}
\label{4.7e}
\bm{\hat{p}_{1}'} = \bm{\hat{p}_{1}''}
- \frac{\omega}{|\bm{p_{1}}|}
\left[ 
\bm{\hat{k}} (\bm{\hat{p}_{1}''})^{2} - 
(\bm{\hat{p}_{1}''} \cdot \bm{\hat{k}})
\bm{\hat{p}_{1}''}
\right] 
+ {\cal O}(\omega^{2})\,.
\end{align}
As the starting point of the calculation for $\sigma_{2}$
in (\ref{4.7b}) we choose $(p_{1}'', p_{2}'', k = 0)$,
where
\begin{eqnarray}
\bm{p_{1}''} &=& - \bm{p_{2}''} = \bm{\hat{p}_{1}''} \,|\bm{p_{1}}|
= \bm{\hat{p}_{1}''} \frac{1}{2 \sqrt{s}}\, w(s,m_{p}^{2},m_{\pi}^{2})\,,\nn\\
p_{1}''^{0} &=& \frac{1}{2 \sqrt{s}} (s - m_{p}^{2} + m_{\pi}^{2})\,, \nn\\
p_{2}''^{0} &=& \frac{1}{2 \sqrt{s}} (s + m_{p}^{2} - m_{\pi}^{2})\,, \nn\\
p_{a} + p_{b} &=& p_{1}'' + p_{2}''\,, \nn\\
t'' &=& (p_{a} - p_{1}'')^{2} = (p_{b} - p_{2}'')^{2}\,.
\label{4.7f}
\end{eqnarray}
Of course, we must get the same result for 
$\cN\ul(p'\uu, p'\ud, k, p\ua, p\ub)$ irrespective of the starting points
of the expansions, either $(p_{1}, p_{2}, k=0)$ 
with $\bm{\hat{p}_{1}} = \bm{\hat{p}'_{1}}$, or
$(p_{1}'', p_{2}'', k=0)$ with
$\bm{\hat{p}_{1}''} = - \bm{\hat{p}_{2}'}$; see Fig.~\ref{fig:5}.
We get from (\ref{3.49}) and (\ref{4.1}) for $\pi^{-} p \to \pi^{-} p \gamma$
\begin{align}
\label{4.7g}
\omega \cN\ul(p'\uu, p'\ud, k, p\ua, p\ub) 
& = \widehat{\cN}\ul^{(0,1)}+\omega\widehat{\cN}\ul^{(1,1)}+{\mathcal O}(\omega^{2}) \nn\\
& = \widehat{\cN}\ul^{(0,2)}+\omega\widehat{\cN}\ul^{(1,2)}+{\mathcal O}(\omega^{2}) \,.
\end{align}
Here
\begin{align}
\widehat{\cN}\ul^{(0,1)} &\equiv
\widehat{\cN}\ul^{(0,1)}(s, \bm{\hat{p}_{a}}, \bm{\hat{p}_{1}'}, \bm{\hat{k}}) \,,
\label{4.7h}\\
\widehat{\cN}\ul^{(1,1)} &\equiv
\widehat{\cN}\ul^{(1,1)}(s, \bm{\hat{p}_{a}}, \bm{\hat{p}_{1}'}, \bm{\hat{k}}; \bm{\hat{p}_{1}'}) \,,
\label{4.7h1}
\end{align}
are given by $\widehat{\cN}\ul^{(j)-}$ from (\ref{3.50}), (\ref{3.51})
setting for \mbox{$i = 1,2$} $l_{i} = l_{i}'$, where $l_{i}'$ is given by
(\ref{3.13}) with $\bm{l_{1\perp}}=0$, corresponding to
$\bm{\hat{p}_{1}'} = \bm{\hat{p}_{1}}$.
For
\begin{align}
\widehat{\cN}\ul^{(0,2)} &\equiv
\widehat{\cN}\ul^{(0,2)}(s, \bm{\hat{p}_{a}}, \bm{\hat{p}_{1}''}, \bm{\hat{k}}) \,,
\label{4.7i}\\
\widehat{\cN}\ul^{(1,2)} &\equiv
\widehat{\cN}\ul^{(1,2)}(s, \bm{\hat{p}_{a}}, \bm{\hat{p}_{1}'}, \bm{\hat{k}}; \bm{\hat{p}_{1}''}) \,,
\label{4.7i1}
\end{align}
we have to insert again $\widehat{\cN}\ul^{(j)-}$ from (\ref{3.50}), (\ref{3.51}),
but with the replacements there and in (\ref{3.13}):
\begin{eqnarray}
&&p_{1} \to p_{1}'', \quad p_{2} \to p_{2}''\,, \nn \\
&&\bm{l_{1 \perp}} \to \bm{l_{1 \perp}''} = \bm{k} (\bm{\hat{p}_{1}''})^{2}
- (\bm{k} \cdot \bm{\hat{p}_{1}''}) \bm{\hat{p}_{1}''} \,, \nn \\
&&l_{i} \to l_{i}'' \;(i = 1, 2)\,.
\label{4.7j}
\end{eqnarray}
In (\ref{4.7h1}) and (\ref{4.7i1}) we indicate as last variables on the r.h.s.
the starting points in phase space for the expansions,
$(k = 0, \bm{\hat{p}_{1}'})$ and $(k = 0, \bm{\hat{p}_{1}''})$,
respectively; see Fig.~\ref{fig:5}.

Now we come to the calculation of $\sigma_{1}$ (\ref{4.7b}).
For fixed values of $\sqrt{s}$, $\bm{\hat{p} \ua}$, 
$\bm{\hat p\uu'}$, $\omega$, 
and $\bm{\hat{k}}$, 
which we choose as independent variables here, 
we can determine from \er{4.5} 
$|\bm{p \uu'}|$, $\bm{p \ud'}$, 
$p\uu^{\prime 0}$, and $p\ud^{\prime 0}$. 
From (\ref{4.5}) we get
\bal{4.7k}
p\ud^{\prime 0}
=\sqrt{(\bm{p \uu'} + \bm{k})\2 + m\up\2}\,,
\end{align}
from which we find
\bal{4.7l}
p\ud^{\prime 0}
=\Big{[}|\bm{p \uu'} |\2 +2|\bm{p \uu'}|\omega (\bm{\hat p\uu'} \cdot \bm{\hat{k}})+\omega\2+m\up\2\Big{]}^{1/2}\,.
\end{align}
Using now the first Eq. of (\ref{4.5}) we have
\bal{4.7m}
\sqrt{|\bm{p \uu'}|\2 + m_{\pi}^{2}}
+ p\ud^{\prime 0} = \sqrt{s} - \omega
\end{align}
which gives, together with (\ref{4.7l}),
the result for $|\bm{p \uu'}|$ as
\bal{4.7}
|\bm{p \uu'}|=&\frac{1}{2\big{[}(\qs-\omega)\2-\omega\2(\bm{\hat{p}\uu'} \cdot \wik )\2\big{]}}\nn\\
& \times \bigg{\lbrace} \Big{[}\Big{(}(\qs-\omega)\2-(\omega\2+m\up\2-m\upp\2)\Big{)}\2\omega\2(\bm{\hat p\uu'}\cdot \wik)\2
\nn\\
&\quad +\Big{(}(\qs -\omega)\2-\omega\2(\bm{\hat p\uu'}\cdot\wik)\2\Big{)}
\nn\\
&\quad \times \Big{(}(\qs -\omega)^{4}
-2(\qs-\omega)\2(\omega\2+m\up\2+m\upp\2)\nn\\
&\quad +(\omega\2+m\up\2-m\upp\2)\2\Big{)}
 \Big{]}^{1/2}\nn\\
&\quad -\Big{[}(\qs-\omega)\2-(\omega\2+m\up\2-m\upp\2)\Big{]}\omega(\bm{\hat p\uu'}\cdot\wik)
 \bigg{\rbrace}\,.
\end{align}
From \er{4.3} we get the following differential cross section with respect to $\omega$ and the solid angles of $\wik$ and $\bm{\hat p\uu'}$:
\bal{4.8}
&\dv\sigma(\pi^{-} p \to \pi^{-} p \gamma)=
\frac{1}{2^{4}(2\pi)^{5}w(s,m\up\2 ,m\upp\2)}\nn\\
& \quad \times 
\omega \, 
\dv\omega \, 
\dv \Omega_{\hat{k}} \, 
\dv \Omega_{\hat{p} \uu'} \, 
J^{(\pi)}(s,\omega, \bm{\hat p\uu'} \cdot \wik)\,
\mathcal{K}(p'\uu, p'\ud, k, p\ua, p\ub)
\,.
\end{align}
Here
\begin{align}
\label{4.9}
\mathcal{K}(p'\uu, p'\ud, k, p\ua, p\ub)
=
\frac{1}{(2m\up)\2}\frac{1}{2}(-g^{\lambda\mu})
\;\tr(\cN\ul \overline{\cN_{\mu}})\,,
\end{align}
with $\cN\ul$ from (\ref{4.7c}) and
$J^{(\pi)}$ a phase-space function which we calculate
with the methods explained 
in Appendix~B of \cite{Lebiedowicz:2021byo}
with the result
\begin{align}
\label{4.10} 
&J^{(\pi)}(s,\omega, \bm{\hat p\uu'} \cdot \wik) \nn \\
&\quad = |\bm{p \uu'}|\2\Big{[}p^{\prime 0}_{2}|\bm{p \uu'}|+
p^{\prime 0}_{1}\big{(}|\bm{p \uu'}|+\omega(\bm{\hat p\uu'}\cdot \wik)\big{)}\Big{]}^{-1}\,.
\end{align}

We are interested in the limit $\omega \to 0$.
The expansion of $J^{(\pi)}$ for $\omega \to 0$
is easily calculated as
\bal{4.11a}
&J^{(\pi)}(s, \omega, \bm{\hat{p}\uu'} \cdot \bm{\hat{k}}) =
J^{(\pi,1)}(s, \omega, \bm{\hat{p}\uu'} \cdot \bm{\hat{k}})
+ {\cal O}(\omega^{2})\,, \nn\\
&J^{(\pi,1)}(s, \omega, \bm{\hat{p}\uu'} \cdot \bm{\hat{k}}) =
\frac{w(s, m\up\2, m\upp\2)}{2s} - \frac{\omega}{\sqrt{s}} \nn\\
&\quad \times
\bigg{[}
\frac{s(m_{p}^{2}+m_{\pi}^{2}) - 
(m_{p}^{2}-m_{\pi}^{2})^{2}}{s\, w(s, m\up\2, m\upp\2)} 
+ \frac{s-m_{p}^{2}+m_{\pi}^{2}}{s}
\bm{\hat{p}\uu'} \cdot \bm{\hat{k}}
\bigg{]} \,.
\end{align}

Now we consider $\mathcal{K}$ (\ref{4.9}) for $\omega\to 0$.
From (\ref{4.7g}) we find
%
\bal{4.11a1} 
&\mathcal{K}(p'\uu, p'\ud, k, p\ua, p\ub)
=
\frac{1}{\omega\2}\Big{[}\widehat{\mathcal{K}}^{(0,j)}
+ \omega \widehat{\mathcal{K}}^{(1,j)}
+ {\mathcal O}(\omega\2)\Big{]}\,, \nn\\
&(j = 1,2)\,,
\end{align} 
where with (\ref{4.7h})--(\ref{4.7i1}) we have
\begin{eqnarray}
&& \widehat{\mathcal{K}}^{(0,j)} =
\frac{1}{(2m\up)\2}\frac{1}{2}(-g^{\lambda\mu})\;
\tr\Big{[}\widehat{\cN}\ul^{(0,j)}
\overline{\widehat{\cN}_{\mu}^{(0,j)}}\,\Big{]}\,,\nn \\
&& \widehat{\mathcal{K}}^{(1,j)} =
\frac{1}{(2m\up)\2}\frac{1}{2}(-g^{\lambda\mu})\nn\\
&& \qquad \qquad \times \tr \Big{[}\widehat{\cN}\ul^{(0,j)}
 \overline{\widehat{\cN}_{\mu}^{(1,j)}}+\widehat{\cN}\ul^{(1,j)}
 \overline{\widehat{\cN}_{\mu}^{(0,j)}}\,\Big{]}\,, \nn\\
&& (j = 1,2)\,.
\label{4.11b} 
\end{eqnarray}

Inserting (\ref{4.11a1}) and \er{4.11b} with $j = 1$ in (\ref{4.8})
we obtain our final result for the expansion
of the cross section $\sigma_{1}$ of \er{4.7b}:
\bal{4.16}
\sigma_{1} & = 
\frac{\omega\, \dv\sigma(\pi^{-} p \to \pi^{-} p \gamma)}
{\dv \omega \, 
 \dv \Omega_{\hat{k}} \, 
 \dv \Omega_{\hat{p}\uu'} }
=\frac{1}{2^{4}(2\pi)^{5}w(s,m\up\2 ,m\upp\2)}\nn\\
& \quad \times
\Big{[}J^{(\pi,1)}(s,\omega, \bm{\hat p\uu'} \cdot \wik) + {\mathcal O}(\omega\2)\Big{]}
\nn\\
& \quad \times
\Big{[}\widehat{\mathcal{K}}^{(0,1)}(s,\bm{{\hat p}_{a}}, \bm{\hat p\uu'}, \wik)
+ \omega \widehat{\mathcal{K}}^{(1,1)}(s,\bm{{\hat p}_{a}}, \bm{\hat p\uu'}, \wik;\bm{\hat p\uu'})\nn\\
&  \quad 
+ {\mathcal O}(\omega\2)\Big{]} \nn\\
& = \frac{1}{2^{4}(2\pi)^{5}w(s,m\up\2 ,m\upp\2)} \nn \\
&\quad \times \bigg\lbrace 
\frac{w(s, m\up\2, m\upp\2)}{2s} 
\widehat{\mathcal{K}}^{(0,1)}(s,\bm{{\hat p}_{a}}, \bm{\hat p\uu'}, \wik) \nn\\
&\quad 
- \frac{\omega}{\sqrt{s}} 
\bigg{[}
\frac{s(m_{p}^{2}+m_{\pi}^{2}) - 
(m_{p}^{2}-m_{\pi}^{2})^{2}}{s\, w(s, m\up\2, m\upp\2)} \nn \\
&\quad 
+ \frac{s-m_{p}^{2}+m_{\pi}^{2}}{s}
\bm{\hat{p}\uu'} \cdot \bm{\hat{k}}
\bigg{]}
\widehat{\mathcal{K}}^{(0,1)}(s,\bm{{\hat p}_{a}}, \bm{\hat p\uu'}, \wik) \nn \\
&\quad 
+ \omega 
\frac{w(s, m\up\2, m\upp\2)}{2s} 
\widehat{\mathcal{K}}^{(1,1)}(s,\bm{{\hat p}_{a}}, \bm{\hat p\uu'}, \wik;\bm{\hat p\uu'})
+ {\mathcal O}(\omega\2)
\bigg\rbrace \,. \nn \\
\end{align}
Here we indicate in $\widehat{\mathcal{K}}^{(i,j)}$
again the functional dependencies as obtained from 
\er{4.7h}, \er{4.7h1}, \er{4.11a1}, and \er{4.11b}.

The calculation of the cross section $\sigma_{2}$ from \er{4.7b}
is completely analogous to the one for $\sigma_{1}$.
The result for the expansion in $\omega$ reads as follows
\bal{4.17}
\sigma_{2} = & \; \frac{\omega\, \dv\sigma(\pi^{-} p \to \pi^{-} p \gamma)}
{\dv \omega \, 
 \dv \Omega_{\hat{k}} \, 
 \dv \Omega_{\hat{p}\ud'} }
=\frac{1}{2^{4}(2\pi)^{5}w(s,m\up\2 ,m\upp\2)}\nn\\
& \times
\Big{[}J^{(p,1)}(s,\omega, \bm{\hat p\ud'} \cdot \wik) + {\mathcal O}(\omega\2)\Big{]}
\nn\\
& \times
\Big{[}\widehat{\mathcal{K}}^{(0,2)}(s,\bm{{\hat p}_{a}}, \bm{\hat p\uu''}, \wik)
+ \omega \widehat{\mathcal{K}}^{(1,2)}(s,\bm{{\hat p}_{a}}, \bm{\hat p\uu'}, \wik;\bm{\hat p\uu''})\nn\\
&
+ {\mathcal O}(\omega\2)\Big{]}
\,.
\end{align}
Here $J^{(p,1)}$ is a phase-space function which we discuss
in Appendix~\ref{app:A}, see (\ref{A12})--(\ref{A16}),
and $\widehat{\mathcal{K}}^{(i,2)}$ $(i=0,1)$
are obtained from (\ref{4.11b}) setting there $j = 2$.

Our next topic is the discussion of the cross section $\sigma_{3}$
from \er{4.7b}.
We can calculate $\sigma_{3}$ by integration over either 
$\sigma_{1}$ or $\sigma_{2}$.
We show now that the result for the expansion of $\sigma_{3}$
is the same to the orders of 
$\omega^{0}$ and $\omega^{1}$ which we consider.
We have from (\ref{4.16})
\bal{4.18}
\sigma_{3} = & \; \omega \frac{\dv\sigma(\pi^{-} p \to \pi^{-} p \gamma)}
{\dv \omega}
=\frac{1}{2^{4}(2\pi)^{5}w(s,m\up\2 ,m\upp\2)}\nn\\
& \times
\int \dv \Omega_{\hat{k}}
\int \dv \Omega_{\hat{p}\uu'}
\Big{[}J^{(\pi,1)}(s,\omega, \bm{\hat p\uu'} \cdot \wik) + {\mathcal O}(\omega\2)\Big{]}
\nn\\
& \times
\Big{[}\widehat{\mathcal{K}}^{(0,1)}(s,\bm{{\hat p}_{a}}, \bm{\hat p\uu'}, \wik)
+ \omega \widehat{\mathcal{K}}^{(1,1)}(s,\bm{{\hat p}_{a}}, \bm{\hat p\uu'}, \wik;\bm{\hat p\uu'})\nn\\
&
+ {\mathcal O}(\omega\2)\Big{]}
\,.
\end{align}
On the other hand we get from (\ref{4.17})
\bal{4.19}
\sigma_{3} = & \; \omega \frac{\dv\sigma(\pi^{-} p \to \pi^{-} p \gamma)}
{\dv \omega}
=\frac{1}{2^{4}(2\pi)^{5}w(s,m\up\2 ,m\upp\2)}\nn\\
& \times
\int \dv \Omega_{\hat{k}}
\int \dv \Omega_{\hat{p}\ud'}
\Big{[}J^{(p,1)}(s,\omega, \bm{\hat p\ud'} \cdot \wik) + {\mathcal O}(\omega\2)\Big{]}
\nn\\
& \times
\Big{[}\widehat{\mathcal{K}}^{(0,2)}(s,\bm{{\hat p}_{a}}, \bm{\hat p\uu''}, \wik)
+ \omega \widehat{\mathcal{K}}^{(1,2)}(s,\bm{{\hat p}_{a}}, \bm{\hat p\uu'}, \wik;\bm{\hat p\uu''})\nn\\
&
+ {\mathcal O}(\omega\2)\Big{]}
\,.
\end{align}
Now we shall prove that Eqs.~(\ref{4.18}) and (\ref{4.19}) 
give the same expansion
for $\sigma_{3}$ up to the order $\omega$.
In Appendix~\ref{app:A}, Eq.~(\ref{A18}), 
we show that for fixed $\wik$
we get from (\ref{4.7e}) the following transformation
of the measure
\bal{4.20}
&\dv \Omega_{\hat{p}\uu'} 
J^{(\pi,1)}(s,\omega, \bm{\hat p\uu'} \cdot \wik)  \nn \\
&\quad =
\dv \Omega_{\hat{p}\ud'} 
\Big{[}J^{(p,1)}(s,\omega, \bm{\hat p\ud'} \cdot \wik) 
+ {\mathcal O}(\omega\2)\Big{]}\,.
\end{align}
Inserting this in (\ref{4.18}) and using (\ref{4.11a1})
and (\ref{4.11b}) we get (\ref{4.19}), as it must be.

\begingroup
\allowdisplaybreaks

Now we shall discuss cross sections for polarized initial proton
and observation of the polarization of the final-state proton.
We want to show that our formalism 
is well suited to deal with this case.

The covariant density matrix for an initial polarized proton
of (\ref{4.1a}) is
\bal{4.21}
\rho_b = \frac{1}{2} 
(1 + \gamma_{5} \! \slash{s}_{b}) 
(\slash{p}_{b} + m_{p})\,,
\end{align}
where $s_{b}$ is a four-vector satisfying
\bal{4.22}
s_{b} \cdot p_{b} = 0\,, \qquad 
0 \leqslant -s_{b} \cdot s_{b} \leqslant 1\,.
\end{align}
The normalization of $\rho_{b}$ is
\bal{4.23}
\tr \rho_{b} = 2 m_{p}\,.
\end{align}
Working in the c.m. system of the reaction (\ref{4.1a}) we have
\begin{align}
&p_{b}=\left(\begin{array}{l}
p_{b}^{0}
\vspace{0.1cm}\\
|\bm{p_{b}}|\,\bm{\hat{p}_{b}}
\end{array}\right),
\quad 
s_{b}=\left(\begin{array}{l}
r_{b \parallel} \dfrac{|\bm{p_{b}}|}{m_{p}}
\vspace{0.1cm}\\
r_{b \parallel} \dfrac{p_{b}^{0}}{m_{p}} \bm{\hat{p}_{b}} + \bm{r_{b \perp}}
\end{array}\right),
\label{4.24}\\
\intertext{with}
&\bm{r_{b \perp}} \cdot \bm{\hat{p}_{b}} = 0\,,
\quad 
\bm{r_{b}} = r_{b \parallel} \bm{\hat{p}_{b}} + \bm{r_{b \perp}}\,,
\quad
- s_{b} \cdot s_{b} = \bm{r_{b}}^{2}\,.
\label{4.25}
\end{align}
After a rotation-free transformation 
to the rest system $\text{R}$
of proton $p\,(p_{b})$ we get
\begin{align}
s_{b, \text{R}}=\left(\begin{array}{l}
0
\vspace{0.1cm}\\ 
\bm{r_{b}}
\end{array}\right),
\quad 
\rho_{b, \text{R}}=
2m_{p}
\left(\begin{array}{c c}
\frac{1}{2}( 1 + \bm{r_{b}} \cdot \bm{\sigma})  & 0
\vspace{0.1cm}\\
0  & 0
\end{array}\right).
\label{4.26}
\end{align}
For an unpolarized proton we have $\bm{r_{b}} = 0$,
a pure spin state has $|\bm{r_{b}}| = 1$.
A state of helicity $\lambda_{b} = \pm 1/2$
in the c.m. system corresponds to 
$r_{b \parallel} = \pm 1$.
For $0 \leqslant |\bm{r_{b}}| < 1$
we have a mixed spin state.

Now we consider the final proton $p\,(p_{2}')$
in (\ref{4.1a}) and assume that its polarization is measured.
The covariant density matrix of the final proton is
\bal{4.27}
\rho_2 = \frac{1}{2} 
(1 + \gamma_{5} \! \slash{s}_{2}\!\!') 
(\slash{p}_{2}\!\!' + m_{p})\,,
\end{align}
where
\begin{align}
&s_{2}'=\left(\begin{array}{l}
r_{2 \parallel} \dfrac{|\bm{p_{2}'}|}{m_{p}}
\vspace{0.1cm}\\
r_{2 \parallel} \dfrac{p_{2}^{\prime 0}}{m_{p}} \bm{\hat{p}_{2}'} + \bm{r_{2 \perp}}
\end{array}\right),
\label{4.28} \\
&\bm{r_{2 \perp}} \cdot \bm{\hat{p}_{2}'} = 0\,,
\quad 
\bm{r_{2}} = r_{2 \parallel} \bm{\hat{p}_{2}} + \bm{r_{2 \perp}}\,,
\nn \\
&0 \leqslant - s_{2}' \cdot s_{2}' = \bm{r_{2}}^{2} \leqslant 1\,.
\label{4.29}
\end{align}
Similarly to (\ref{4.26}) the vector $\bm{r_{2}}$ determines
the polarization of the proton $p\,(p_{2}')$
after rotation-free transformation to the rest system.

We shall now assume that in the reaction (\ref{4.1a})
we have an initial proton with polarization described by
$\bm{r_{b}}$, 
\mbox{$0 \leqslant \bm{r_{b}}^{2} \leqslant 1$}.
For the final proton we assume observation of a pure spin state
corresponding to $\bm{r_{2}}$
with $\bm{r_{2}}^{2} = 1$.
We assume summation over the photon polarizations.
The corresponding cross section reads
\begin{align}
&\frac{\omega\, \dv\sigma_{\rm pol}(\pi^{-} p \to \pi^{-} p \gamma)}
{\dv \omega \, 
 \dv \Omega_{\hat{k}} \, 
 \dv \Omega_{\hat{p}\ud'} }
=\frac{1}{2^{4}(2\pi)^{5}w(s,m\up\2 ,m\upp\2)}\nn\\
\nn \\
& \quad \times
J^{(p)}(s, \omega, \bm{\hat p\ud'} \cdot \wik) \, \omega\2
\mathcal{K}_{\rm pol}(p_{1}',p_{2}',s_{2}',k,p_{a},p_{b},s_{b})
\,.
\label{4.30}
\end{align}
Here $J^{(p)}$ is given in (\ref{A12})
and with $\cN_{\lambda}$ from (\ref{4.7c}) we get
\begin{align}
&\mathcal{K}_{\rm pol}(p_{1}',p_{2}',s_{2}',k,p_{a},p_{b},s_{b})
= 
\frac{1}{(2m\up)\2} \frac{1}{4}(-g^{\lambda\mu}) \nn \\
&\quad \times 
\tr \Big{[}
(1 + \gamma_{5} \! \slash{s}_{2}\!\!')  
\cN\ul(p_{1}',p_{2}',k,p_{a},p_{b}) \nn \\
&\quad \times 
(1 + \gamma_{5} \! \slash{s}_{b}) 
\overline{\cN_{\mu}}(p_{1}',p_{2}',k,p_{a},p_{b})
\Big{]}\,.
\label{4.31}
\end{align}

Now we want to calculate the expansion in $\omega$
for $\omega \to 0$ of the cross section (\ref{4.30}).
The expansion of $J^{(p)}$ is given in (\ref{A16}).
For $\cN_{\lambda}$ we shall use the second line of (\ref{4.7g}).
We also have to expand $s_{2}'$ (\ref{4.28})
keeping $r_{2 \parallel}$ and $\bm{r_{2 \perp}}$ fixed.
Using (\ref{3.6}) and (\ref{3.13})
we get with (\ref{4.7d})--(\ref{4.7f}) and (\ref{4.28})
\bal{4.32}
s_{2}' = s_{2}'' - 
\omega s_{2}^{(1)} + {\mathcal O}(\omega\2)\,,
\end{align}
where
\begin{align}
s_{2}''=
\left(\begin{array}{l}
r_{2 \parallel} \dfrac{|\bm{p_{2}''}|}{m_{p}}
\vspace{0.1cm}\\
r_{2 \parallel} \dfrac{p_{2}^{\prime \prime 0}}{m_{p}} \bm{\hat{p}_{2}''} + \bm{r_{2 \perp}}
\end{array}\right), 
\quad 
s_{2}^{(1)} = \frac{r_{2 \parallel}}{m_{p}} 
\frac{p_{1}'' \cdot k}{\omega |\bm{p_{2}''}| \sqrt{s}}
p_{2}''\,.
\label{4.33}
\end{align}
Inserting all this in (\ref{4.30}) and (\ref{4.31})
and collecting the terms of orders $\omega^{0}$ and $\omega$
we find as final result for the expansion of our
polarized cross section
\begin{widetext}
\bal{4.34}
\frac{\omega\, \dv\sigma_{\rm pol}(\pi^{-} p \to \pi^{-} p \gamma)}
{\dv \omega \, 
 \dv \Omega_{\hat{k}} \, 
 \dv \Omega_{\hat{p}\ud'} }
= \, & \frac{1}{2^{4}(2\pi)^{5}w(s,m\up\2 ,m\upp\2)}
\frac{1}{(2m\up)\2} \frac{1}{4}(-g^{\lambda\mu})\nn\\
&\times
\bigg{\lbrace}\bigg{[}
\frac{w(s, m\up\2, m\upp\2)}{2s} - \frac{\omega}{\sqrt{s}} 
\bigg{(}
\frac{s(m_{p}^{2}+m_{\pi}^{2}) - 
(m_{p}^{2}-m_{\pi}^{2})^{2}}{s\, w(s, m\up\2, m\upp\2)} 
+ \frac{s+m_{p}^{2}-m_{\pi}^{2}}{s}
\bm{\hat{p}\ud'} \cdot \bm{\hat{k}}
\bigg{)}\bigg{]}
 \nn \\
&\times 
\tr \Big{[}
(1 + \gamma_{5} \! \slash{s}_{2}\!\!'')  
\widehat{\cN}\ul^{(0,2)} 
(1 + \gamma_{5} \! \slash{s}_{b}) 
\overline{\widehat{\cN}_{\mu}^{(0,2)}}
\Big{]}
+ \omega \frac{w(s, m\up\2, m\upp\2)}{2s}
 \nn \\
& \times 
\tr \Big{[}
(1 + \gamma_{5} \! \slash{s}_{2}\!\!'')  
\widehat{\cN}\ul^{(1,2)} 
(1 + \gamma_{5} \! \slash{s}_{b}) 
\overline{\widehat{\cN}_{\mu}^{(0,2)}}
+ 
(1 + \gamma_{5} \! \slash{s}_{2}\!\!'')  
\widehat{\cN}\ul^{(0,2)} 
(1 + \gamma_{5} \! \slash{s}_{b}) 
\overline{\widehat{\cN}_{\mu}^{(1,2)}}
 \nn \\
& 
- \gamma_{5} \! \slash{s}_{2}^{\,(1)} 
\widehat{\cN}\ul^{(0,2)} 
(1 + \gamma_{5} \! \slash{s}_{b}) 
\overline{\widehat{\cN}_{\mu}^{(0,2)}}
\Big{]}
\bigg{\rbrace}
+ {\mathcal O}(\omega\2)
\,.
\end{align}

\end{widetext}

\endgroup


\section{Conclusions}
\label{sec:5}

In this article we have considered soft-photon production in 
$\ppm$-proton elastic scattering. That is, we studied the reactions 
$\ppm p\to \ppm p$ and $\ppm p\to \ppm p\gamma$ 
for the photon energy $\omega \to 0$. 
We considered the Laurent series in $\omega$ for
the amplitudes $\ppm p\to \ppm p\gamma$.
The terms of order $\omega^{-1}$
are well known \cite{Weinberg:1965nx}
and we reproduced them in our calculations.
The evaluation of the terms of order $\omega^{0}$ is the central topic of our paper.

The main issues which we had to consider and the main results which we obtained were the following.
\begin{itemize}
\item[(i)] 
In going from
\bal{5.1}
\quad \ppm\,(p\ua)+p\,(p\ub) &\to \ppm\,(p\uu)+p\,(p\ud)
\intertext{to}
\label{5.2}
\quad \ppm\,(p\ua)+p\,(p\ub) &\to \ppm\,(p'\uu)+p\,(p'\ud)+\gamma(k)
\end{align}
for $k\neq 0$ a corresponding change of the momenta $(p\uu , p\ud)\to (p'\uu , p'\ud)$ has to be taken into account.
Keeping $p'\uu=p\uu$ and $p'\ud =p\ud$ 
for $k\neq 0$ violates energy-momentum conservation; 
see \mbox{\er{3.6}--\er{3.14}} and Fig.~\ref{fig:2}.
We discussed the phase space for (\ref{5.1}) and (\ref{5.2}).
We studied the expansion in $\omega$ of the amplitude (\ref{5.2})
around a phase-space point of no radiation
\mbox{$(p_{1}, p_{2}, k = 0)$}, where
\mbox{$(p_{1}, p_{2})$} from (\ref{5.1})
is close to \mbox{$(p_{1}', p_{2}')$}
but otherwise arbitrary.
\item[(ii)]
In the diagrams for \er{5.2} we have to deal with the \textit{off-shell} amplitudes for $\ppm p \to \ppm p$; 
see Fig.~\ref{fig:3}. 
Whereas the on-shell amplitude 
for $\pi^{-}p\to \pi^{-}p$ scattering is described 
by two invariant amplitudes, 
$A^{(\text{on})-}(s,t)$ and $B^{(\text{on})-}(s,t)$, 
see \er{2.23}--\er{2.25}, 
the off-shell amplitude for $\pi^{-}p \to \pi^{-}p$ 
contains eight invariant amplitudes; 
see \er{2.16}--\er{2.19}.
The same is, of course, true for $\pi^{+}p\to \pi^{+}p$.
In the calculation of $\ppm p \to \ppm p \gamma$ 
all these eight off-shell invariant amplitudes have to be taken into account. 
The final result 
for the terms of order $\omega^{-1}$ and $\omega^{0}$
in the $\ppm p \to \ppm p \gamma$ amplitudes 
can be completely expressed 
in terms of the on-shell amplitudes 
$A^{(\text{on})\pm}(s,t)$, $B^{(\text{on})\pm}(s,t)$, 
and their partial derivatives with respect to $s$ and $t$;
see \er{3.41}--\er{3.47}.

\item[(iii)]
The formulas (\ref{3.42})--(\ref{3.47})
which we give for 
the $\ppm p \to \ppm p \gamma$ amplitudes 
in the limit \mbox{$\omega \to 0$} 
are valid both for real photons ($k\2 = 0$) 
and for virtual photons ($k\2 > 0$).
Here we consider this limit \mbox{$\omega \to 0$} 
as specified in (\ref{3.34a}), (\ref{3.34b}).

\item[(iv)]
We have discussed in detail the expansions 
of the $\ppm p \to \ppm p \gamma$ amplitudes
in $\omega$ for the case of real photon emission;
see \eqref{3.49}--\eqref{3.55}.
The pole term $\propto \omega^{-1}$ 
in (\ref{3.49}) corresponds to Weinberg's
soft-photon theorem and is the first term 
in the Laurent expansion around $\omega = 0$.
We have given the next-to-leading term
$\propto \omega^{0}$
in this Laurent expansion.

\item[(v)] 
Our results for the amplitudes of $\ppm p \to \ppm p \gamma$ 
with real photon emission,
see \er{3.49}--\er{3.55},
were then used to discuss the expansions of different cross sections in $\omega$ for $\omega \to 0$.
This was the subject of Sec.~\ref{sec:4}.
We calculated the cross-sections
$\sigma_{1} = \omega \dv\sigma /(\dv \omega \, 
\dv \Omega_{\hat{k}} \, \dv \Omega_{\hat{p}_{1}'})$
and
$\sigma_{2} = \omega \dv\sigma /(\dv \omega \, 
\dv \Omega_{\hat{k}} \, \dv \Omega_{\hat{p}_{2}'})$
for $\pi^{-} p \to \pi^{-} p \gamma$.
The cross section 
$\sigma_{3} = \omega \dv\sigma / \dv \omega$
can be obtained by integration over $\sigma_{1}$ or $\sigma_{2}$.
We have shown that these two ways 
of calculation of $\sigma_{3}$
give the same result up to the order $\omega$
which we consider.
This represents a consistency check of our calculations
of the expansions in $\omega$.
We have then considered the cross section for a polarized
proton in the initial state and observation of a polarized
proton in the final state.
We have shown that our methods are very well suited
for deriving the expansion in $\omega$ for $\omega \to 0$
also for this case; see (\ref{4.34}).
\item[(vi)] 
We emphasize that our results represent a strict theorem in the framework of QCD which applies at all c.m. energies~$\sqrt{s}$. 
Thus, the theorem can be tested experimentally both at low values of $\sqrt{s}$, available for instance 
in the HADES experiment
\cite{Rathod:2022pof} 
at GSI, Darmstadt, and at high energy pion-proton scattering. 
\end{itemize}

We end with an outlook. Using the methods which we have developed 
for $\pi p$ scattering with soft-photon production one could also tackle proton-proton scattering with soft-photon production. We see no problem of principle there, except that the calculations will be very long and complex. Will there be at the end a theorem analogous to the $\pi p$ scattering case? To answer this question will need much more work. 

\begin{acknowledgments}
The authors thank C. Ewerz for reading the main part of the manuscript 
and for useful suggestions.
This work was partially supported by
the Polish National Science Centre under Grant
No. 2018/31/B/ST2/03537.
\end{acknowledgments}


\appendix

\renewcommand\theequation{\thesection\arabic{equation}} 


\section{Kinematic relations}
\label{app:A}

In this appendix we collect useful formulas which we need in the main chapters.

For the momenta $\tilde{p}\ua$, $\tilde{p}\ub$, $\tilde{p}\uu$, $\tilde{p}\ud$ of the general off-shell reaction \er{2.1} we note in addition
to \er{2.7}--\er{2.11} the following relations
\begin{align}
\label{A1}
\tilde{p}\ua &=\frac{1}{2}(\tilde{p}_{s}+\tilde{p}_{t}+\tilde{p}_{u})\,,\nn\\
\tilde{p}\ub &=\frac{1}{2}(\tilde{p}_{s}-\tilde{p}_{t}-\tilde{p}_{u})\,,\nn\\
\tilde{p}\uu &=\frac{1}{2}(\tilde{p}_{s}-\tilde{p}_{t}+\tilde{p}_{u})\,,\nn\\
\tilde{p}\ud &=\frac{1}{2}(\tilde{p}_{s}+\tilde{p}_{t}-\tilde{p}_{u})\,;\\
\label{A2}\nn\\
\tilde{p}_{s}+\tilde{p}_{u} &=\tilde{p}\ua +\tilde{p}\uu \,, \nn\\
\tilde{p}_{s}-\tilde{p}_{u} &=\tilde{p}\ub +\tilde{p}\ud \,;\\
\nn\\
\label{A3}
\tilde{p}\ua \cdot\tilde{p}\ub &=\frac{1}{2}(\tilde{s}-m\ua\2-m\ub\2)\ ,\nn\\
\tilde{p}\uu \cdot\tilde{p}\ud &=\frac{1}{2}(\tilde{s}-m\uu\2-m\ud\2)\ ,\nn\\
\tilde{p}\ua \cdot\tilde{p}\uu &=\frac{1}{2}(-\tilde{t}+m\ua\2+m\uu\2)\ ,\nn\\
\tilde{p}\ub \cdot\tilde{p}\ud &=\frac{1}{2}(-\tilde{t}+m\ub\2+m\ud\2)\ ,\nn\\
\tilde{p}\ua \cdot\tilde{p}\ud &=\frac{1}{2}(-\tilde{u}+m\ua\2+m\ud\2)\ ,\nn\\
&=\frac{1}{2}(\tilde{s}+\tilde{t}-m\ub\2-m\uu\2)\,,\nn\\
\tilde{p}\uu \cdot\tilde{p}\ub &=\frac{1}{2}(-\tilde{u}+m\ub\2+m\uu\2)\ ,\nn\\
&=\frac{1}{2}(\tilde{s}+\tilde{t}-m\ua\2-m\ud\2)\,.
\end{align}
For three four-vectors $a$, $b$, $c$ we have
\begin{align}
\label{A4}
i\gamma_{\mu}\gamma_{5}\varepsilon^{\mu\nu\rho\sigma}a_{\nu}b_{\rho}c_{\sigma}&=
\frac{1}{6}(
 \slash{a}\slash{b}\slash{c}\,
+\slash{c}\slash{a}\slash{b}\,
+\slash{b}\slash{c}\slash{a}\nn\\
&\quad -\slash{b}\slash{a}\slash{c}\,
-\slash{a}\slash{c}\slash{b}\,
-\slash{c}\slash{b}\slash{a})\,.
\end{align}

For on-shell momenta $p\ua$, $p\ub$, $p\uu$, $p\ud$ 
in the reactions \eqref{2.2} 
we have for the masses \eqref{2.12} 
and we get the following relations with Dirac spinors 
$u(p\ub)$ and $\bar{u}(p\ud)$:
\begin{align}
\label{A5}
&\bar{u}(p\ud) \!\! \slash{p}_{s}u(p\ub)
=\bar{u}(p\ud)\Big{[}m_{p}+\frac{1}{2}(\slash{p}\ua+\slash{p}\uu)\Big{]}u(p\ub)\,,\\
\label{A6}
&\bar{u}(p\ud) \!\! \slash{p}_{u}u(p\ub)
=\bar{u}(p\ud)\Big{[}-m_{p}+\frac{1}{2}(\slash{p}\ua+\slash{p}\uu)\Big{]}u(p\ub)\,,\\
\label{A7}
&\bar{u}(p\ud) i\sigma_{\mu\nu}p_{s}{}^{\mu}p_{t}{}^{\nu}u(p\ub)\nn\\
&\quad
=\bar{u}(p\ud)\Big{[}-s+m_{p}\2+m_{\pi}\2+m_{p}(\slash{p}\ua+\slash{p}\uu)\Big{]}u(p\ub)\,,\\
\label{A8}
&\bar{u}(p\ud) i\sigma_{\mu\nu}p_{t}{}^{\mu}p_{u}{}^{\nu}u(p\ub)\nn\\
&\quad
=\bar{u}(p\ud)\Big{[}s+t-m_{p}\2-m_{\pi}\2-m_{p}
(\slash{p}\ua+\slash{p}\uu)\Big{]}u(p\ub)\,,\\
\label{A9}
&\bar{u}(p\ud)i\gamma_{\mu}\gamma_{5}\varepsilon^{\mu\nu\rho\sigma}p_{s\nu}p_{t\rho}p_{u\sigma}u(p\ub)\nn\\
&\quad
=\bar{u}(p\ud)\Big{[}-m_{p}(2s+t-2m_{p}\2-2m_{\pi}\2)\nn\\
&\quad \quad +\frac{1}{2}(\slash{p}\ua+\slash{p}\uu)(4m_{p}\2-t)
\Big{]}u(p\ub)\,.
\end{align}
Useful relations for the momenta of the on-shell reaction
\eqref{2.2}, $\pi p \to \pi p$, are as follows:
\begin{align}
\label{A10}
&
-\frac{1}{2}(p_{a} + p_{1}, p_{b} + p_{2})
-(p_{b} \cdot p_{2})+m_{p}^{2} = -s + m_{p}^{2} + m_{\pi}^{2}\,,\nn\\
&
\frac{1}{2}(p_{a} + p_{1}, p_{b} + p_{2})
-(p_{b} \cdot p_{2})+m_{p}^{2} = s + t - m_{p}^{2} - m_{\pi}^{2}\,,\nn\\
&
(p_{a} + p_{1}, p_{b} + p_{2}) = 2s + t - 2m_{p}^{2} - 2m_{\pi}^{2}\,,\nn\\
&
m_{p}^{2} + (p_{b} \cdot p_{2}) 
= \frac{1}{2} (4 m_{p}^{2} - t)\,.
\end{align}

Next we discuss relations which we need for the calculations of cross sections in Sec.~\ref{sec:4}.
The general formula for the cross section
$\sigma_{2}$ of (\ref{4.7b}) reads
\bal{A11}
\sigma_{2} = & \; \frac{\omega\, \dv\sigma(\pi^{-} p \to \pi^{-} p \gamma)}
{\dv \omega \, 
 \dv \Omega_{\hat{k}} \, 
 \dv \Omega_{\hat{p}\ud'} }
=\frac{1}{2^{4}(2\pi)^{5}w(s,m\up\2 ,m\upp\2)}\nn\\
& \times
J^{(p)}(s,\omega, \bm{\hat p\ud'} \cdot \wik)\,
\mathcal{K}(p_{1}',p_{2}',k,p_{a},p_{b})\,.
\end{align}
Here $\mathcal{K}$ is given in (\ref{4.9}),
(\ref{4.11a1}) and
$J^{(p)}$ is the phase-space function
\begin{align}
\label{A12} 
&J^{(p)}(s,\omega, \bm{\hat p\ud'} \cdot \wik) \nn \\
&\quad = \int_{0}^{\infty} \dv |\bm{p \ud'}|
\frac{|\bm{p \ud'}|}{p^{\prime 0}_{2} p^{\prime 0}_{1}}
\delta(p^{\prime 0}_{1}+p^{\prime 0}_{2}+\omega-\sqrt{s})
\nn \\
&\quad = |\bm{p \ud'}|\2 
\Big{[} \big{(} p^{\prime 0}_{1} +
p^{\prime 0}_{2} \big{)} |\bm{p \ud'}| +
p^{\prime 0}_{2}
\omega(\bm{\hat p\ud'}\cdot \wik)\Big{]}^{-1}\,,
\end{align}
where from (\ref{4.5})
\bal{A13}
\bm{p \uu'} &=- \bm{p \ud'} -\bm{k} \,,\nn\\
p\uu^{ \prime 0}
&=\sqrt{|\bm{p \uu'} |\2 + m\upp\2}  \nn\\
&=\Big{[}|\bm{p \ud'} |\2 +2|\bm{p \ud'}|\omega (\bm{\hat p\ud'} \cdot \bm{\hat{k}}) 
+ \omega\2 + m\upp\2 \Big{]}^{1/2}\,,\nn\\
p\ud^{ \prime 0} 
&=\sqrt{|\bm{p \ud'} |\2 + m\up\2} \,.
 \end{align}
From
\bal{A14}
p\uu^{\prime 0}+p\ud^{\prime 0}=\sqrt{s}-\omega
\end{align}
we get
\bal{A15}
|\bm{p \ud'}|=&\frac{1}{2\big{[}(\qs-\omega)\2-\omega\2(\bm{\hat{p}\ud'} \cdot \wik )\2\big{]}}\nn\\
& \times \bigg{\lbrace} \Big{[}\Big{(}(\qs-\omega)\2-(\omega\2-m\up\2+m\upp\2)\Big{)}\2\omega\2(\bm{\hat p\uu'}\cdot \wik)\2
\nn\\
&\quad +\Big{(}(\qs -\omega)\2-\omega\2(\bm{\hat p\ud'}\cdot\wik)\2\Big{)}
\nn\\
&\quad \times \Big{(}(\qs -\omega)^{4}
-2(\qs-\omega)\2(\omega\2+m\up\2+m\upp\2)\nn\\
&\quad +(\omega\2-m\up\2+m\upp\2)\2\Big{)}
 \Big{]}^{1/2}\nn\\
&\quad -\Big{[}(\qs-\omega)\2-(\omega\2-m\up\2+m\upp\2)\Big{]}\omega(\bm{\hat p\ud'}\cdot\wik)
 \bigg{\rbrace}\,.
\end{align}
Inserting $p\uu^{\prime 0}$, $p\ud^{\prime 0}$,
and $|\bm{p \ud'}|$ from
(\ref{A13})--(\ref{A15}) in (\ref{A12}) gives
the general expression for $J^{(p)}$ in (\ref{A12}).
The expansion of $J^{(p)}$ in $\omega$ reads
\begin{align}
\label{A16} 
&J^{(p)}(s,\omega, \bm{\hat p\ud'} \cdot \wik)
= J^{(p,1)}(s,\omega, \bm{\hat p\ud'} \cdot \wik)
+ {\mathcal O}(\omega\2)\,,\nn \\
&J^{(p,1)}(s, \omega, \bm{\hat{p}\ud'} \cdot \bm{\hat{k}}) =
\frac{w(s, m\up\2, m\upp\2)}{2s} - \frac{\omega}{\sqrt{s}} \nn\\
&\quad \times
\bigg{[}
\frac{s(m_{p}^{2}+m_{\pi}^{2}) - 
(m_{p}^{2}-m_{\pi}^{2})^{2}}{s\, w(s, m\up\2, m\upp\2)} 
+ \frac{s+m_{p}^{2}-m_{\pi}^{2}}{s}
\bm{\hat{p}\ud'} \cdot \bm{\hat{k}}
\bigg{]} \,.
\end{align}
Inserting (\ref{A16}) in (\ref{A11}) and using (\ref{4.11a1})
with $j = 2$ for $\mathcal{K}$ we arrive
at the expression for $\sigma_{2}$ from (\ref{4.17})
which is proved in this way.

Now we come to the transformation formula for
the measure going from $\bm{\hat{p}\uu'}$ to $\bm{\hat{p}\uu''}$ according to (\ref{4.7e}).
For fixed $\omega$ and $\bm{\hat{k}}$
we get from (B4) of \cite{Lebiedowicz:2023ell}
\begin{eqnarray}
\dv \Omega_{\hat{p}\uu'} &=&
\dv \Omega_{\hat{p}\uu''}\,
{\rm det} \left( \frac{\partial \hat{p}_{1i}'}{\partial \hat{p}_{1e}''} \right) \nn\\
&=& 
\dv \Omega_{\hat{p}\uu''}
\Big{[}
1 + \frac{2 \omega}{|\bm{{p}_{1}}|} 
\bm{\hat{p}_{1}''} \cdot \bm{\hat{k}} +
{\cal O}(\omega^{2}) \Big{]} \nn \\
&=&
\dv \Omega_{\hat{p}\uu''}
\Big{[}
1 + 4 \frac{\omega \sqrt{s}}{w(s,m_{p}^{2},m_{\pi}^{2})} 
\bm{\hat{p}_{1}''} \cdot \bm{\hat{k}} +
{\cal O}(\omega^{2}) \Big{]}. \quad \qquad 
\label{A17} 
\end{eqnarray}
With $J^{(\pi,1)}$ from (\ref{4.11a})
and $J^{(p,1)}$ from (\ref{A16}) we get
\bal{A18}
&\dv \Omega_{\hat{p}\uu'} 
J^{(\pi,1)}(s,\omega, \bm{\hat p\uu'} \cdot \wik) \nn \\
&\quad =
\dv \Omega_{\hat{p}\uu''} 
\Big{[}J^{(p,1)}(s,\omega, -\bm{\hat p\uu''} \cdot \wik) 
+ {\mathcal O}(\omega\2)\Big{]}\,.
\end{align}
Using (\ref{4.7d}) proves (\ref{4.20}).

\section{Propagators and vertices}
\label{app:B}

In this appendix we recall the symmetry relations for fields, propagators and vertices which are valid in QCD. This is standard material from QFT; 
see, e.g., \cite{Bjorken:1965} 
and \cite{Nachtmann:1990ta,Weinberg:1995_I}. 
Then we use these relations and the generalized Ward identities to derive the general forms of propagators, photon-vertex functions and of their products as we need them for our calculations.

\subsection{$P$, $C$, and $T$ relations for fields and the general forms of the pion and proton propagators}
\label{subsec:B1}

We start by considering the renormalized field operators for the proton and for the pions:
\begin{equation}
\label{B1}
\psi(\bm{x},t)\;,\quad \varphi^{\pm}(\bm{x},t)\;,\quad \varphi^{0}(\bm{x},t)\ .
\end{equation}
In QCD these fields are not the fundamental ones which are the quark and gluon fields. But so called interpolating local fields \eqref{B1} with the correct quantum numbers exist and can easily be constructed in the QCD framework. These constructions are not unique but this is no problem for us. For the pions interpolating fields are obtained from the divergence of the axial currents with suitable normalization factors. For the proton such fields have, for instance, be given in \cite{Chung:1981wm,Ioffe:1981kw}.

From the fields in (\ref{B1})
$\psi$ annihilates protons and creates antiprotons, 
$\varphi^{\pm}$ annihilate $\pi^{\pm}$ and create $\pi^{\mp}$, $\varphi^{0}$ annihilates and creates $\pi^{0}$. 
We work in QCD where parity ($P$), charge conjugation ($C$),
and time reversal ($T$) are good symmetries. 
We also consider $\Theta=CPT$ which always is a good symmetry.
With the corresponding transformations: $U(P)$, $U(C)$ unitary, $V(T), V(\Theta)$ antiunitary, we have the following relations
\begin{align}
\label{B2}
&U(P)\psi(\bm{x},t)U^{-1}(P)=\gamma_{0}\psi(-\bm{x},t)\,,\nn\\
&U(P)\varphi^{\pm}(\bm{x},t)U^{-1}(P)=-\varphi^{\pm}(-\bm{x},t)\,,\nn\\
&U(P)\varphi^{0}(\bm{x},t)U^{-1}(P)=-\varphi^{0}(-\bm{x},t)\,, \\
\nn\\
\label{B3}
&U(C)\psi(\bm{x},t)U^{-1}(C)=S(C)\overline{\psi}^{\top}(\bm{x},t)\,,\nn\\
&U(C)\varphi^{\pm}(\bm{x},t)U^{-1}(C)=\varphi^{\mp}(\bm{x},t)\,,\nn\\
&U(C)\varphi^{0}(\bm{x},t)U^{-1}(C)=\varphi^{0}(\bm{x},t)\,,
\\
\nn\\
\label{B4}
&\big{(}V(T)\psi(\bm{x},t)V^{-1}(T)\big{)}^{\dagger}=S(T)\overline{\psi}^{\top}(\bm{x},-t)\,,\nn\\
&\big{(}V(T)\varphi^{\pm}(\bm{x},t)V^{-1}(T)\big{)}^{\dagger}=-\varphi^{\mp}(\bm{x},-t)\,,\nn\\
&\big{(}V(T)\varphi^{0}(\bm{x},t)V^{-1}(T)\big{)}^{\dagger}=-\varphi^{0}(\bm{x},-t)\,,
\\
\nn\\
\label{B4a}
&\big{(}V(\Theta)\psi(x)V^{-1}(\Theta)\big{)}^{\dagger}=-\gamma_{5}\psi(-x)\,,\nn\\
&\big{(}V(\Theta)\varphi^{\pm}(x)V^{-1}(\Theta)\big{)}^{\dagger}=\varphi^{\pm}(-x)\,,\nn\\
&\big{(}V(\Theta)\varphi^{0}(x)V^{-1}(\Theta)\big{)}^{\dagger}=\varphi^{0}(-x)\,.
\end{align}
Here $S(C)$ and $S(T)$ are given by
\begin{align}
\label{B5}
S(C)=i\gamma\2\gamma^{0}\,, \qquad
S(T)=i\gamma\2\gamma_{5}\,.
\end{align}
All these are standard relations of QFT; 
see for instance \cite{Nachtmann:1990ta,Weinberg:1995_I}. 
For the electromagnetic current operator 
$\mathcal{J}^{\lambda}(x)$ we have
in QCD with $q(x)$ the quark fields
and $Q_{q}$ their charges in units of the proton charge 
$e=\sqrt{4\pi \alpha_{\rm em}}$
\bel{B6}
\cJ^{\lambda}(x)=e\, \sum_{q} Q_{q}\,\bar{q}(x)\,\gamma^{\lambda}\,q(x) 
\,.
\ee
The current is conserved
\bel{B6a}
\frac{\partial}{\partial x^{\lambda}}\cJ^{\lambda}(x)=0\,,
\ee
and the transformation properties of this current operator are
\bal{B7}
&U(P)\cJ\hl (\bm{x},t)U^{-1}(P) = \mathcal{P}\hl{}_{\lambda'} \cJ^{\lambda'}(-\bm{x},t)\,,\nn\\
&U(C)\cJ\hl (\bm{x},t)U^{-1}(C) = - \cJ^{\lambda}(\bm{x},t)\,,\nn\\
&\big{(}V(T)\cJ\hl (\bm{x},t)V^{-1}(T)\big{)}^{\dagger} = \mathcal{P}\hl{}_{\lambda'} \cJ^{\lambda'}(\bm{x},-t)\,,\nn\\
&\big{(}V(\Theta)\cJ\hl (x)V^{-1}(\Theta)\big{)}^{\dagger} = - \cJ^{\lambda}(-x)\,,
\end{align}
where
\bel{B8}
(\mathcal{P}\hl{}_{\lambda'}) = \text{diag}(1,\, -1,\, -1, \, -1) \,.
\ee

Our next step is to discuss the general structure of the pion and proton propagators following from the above 
$P$, $C$, $T$, and $\Theta$ relations. 
From Poincar{\'e} and $CPT$ invariance one finds that the propagators for $\pi^{-}$ and $\pi^{+}$ 
are the same and we have
\bal{B9}
i\int\frac{\dv^{4}p}{(2\pi)^{4}} e^{-ip(x-y)}\Delta_{F}(p\2 )
&
= \braket{0|{\rm T}( \varphi^{-} (x)\,\varphi\hp (y) )|0}\nn\\
&
= \braket{0|{\rm T}( \varphi\hp (x)\,\varphi^{-} (y) )|0}
\,.
\end{align}
Here ${\rm T}$ denotes the time-ordered product of the field operators for the pions.
The $P$, $C$, and $T$ relations following for the propagator from \er{B2}--\er{B4} are automatically fulfilled by \er{B9}. 
The normalization conditions 
for the renormalized pion propagator are
\bal{B10}
&\Delta_{F}^{-1}(p\2 )\big{|}_{p\2 = m_{\pi}\2} =0\,,\nn\\
\frac{\partial}{\partial p\2}
&\Delta_{F}^{-1}(p\2 )\big{|}_{p\2 = m_{\pi}\2} =1\,.
\end{align}
In QCD we find from the Landau conditions
\cite{Landau:1959fi,Bjorken:1965}
that $\Delta_{F}^{-1}(p\2 )$ is analytic at 
$p\2 = m_{\pi}\2$ with the nearest singularity at $p\2=(3m\upp)\2$ where the cut on the real $p\2$~axis starts. Therefore, we can write the general form for  $\Delta_{F}^{-1}(p\2 )$ as
\bal{B11}
 \Delta_{F}^{-1}(p\2 )&=p\2 -m\upp\2
 +(p\2 -m\upp\2)\2 \, C(p\2 -m\upp\2)
 \end{align}
with $C(p\2 - m\upp\2)$ an analytic function
for $|p\2 - m\upp\2|<8m\upp\2$,
\bel{B11a}
C(p\2 -m\upp\2)=c_{0}+c_{1}(p\2 -m\upp\2)+\dots \ .
\ee
The self-energy function of the pion is
\be
-(p\2 -m\upp\2)\2 \, C(p\2 -m\upp\2)\,.\nn
\ee

Now we turn to the proton propagator 
\bal{B12}
i\int\frac{\dv^{4}p}{(2\pi)^{4}} &e^{-ipx}S_{F}(p)
=\braket{0|{\rm T}( \psi(x)\,\overline{\psi}(0) )|0}\,.
\end{align}
From the $P$-invariance relation for the proton field \er{B2} we get
\bal{B13}
\braket{0|{\rm T}( \psi (\bm{x},t)\,\overline{\psi}(0) )|0}
&=
\gamma_{0}
\braket{0|{\rm T}( \psi (-\bm{x},t)\,\overline{\psi}(0) )|0}\gamma_{0}\,,
\nn\\
S_{F}(\bm{p},p^{0})& =\gamma_{0}S_{F}(-\bm{p},p^{0})\gamma_{0}\,.
\end{align}
The $C$ and $T$ relations, 
\er{B3} and \er{B4}, give
\bel{B14}
S_{F}(p)=S(C)S^{\top}_{F}(-p)S^{-1}(C)\,,
\ee
and
\bel{B15}
S_{F}(\bm{p},p^{0})=S(T)S^{\top}_{F}(-\bm{p},p^{0})S^{-1}(T)\,,
\ee
respectively. The most general ansatz for $S_{F}^{-1}(p)$ 
compatible with \er{B13}--\er{B15} is
\bel{B16}
S_{F}^{-1}(p)=C_{0}(p\2)+C_{1}(p\2)\!\bp \,.
\ee
The functions $C_{0}(p\2)$ and $C_{1}(p\2)$ are analytic 
at $p\2 = m_{p}\2$ with the nearest singularity at 
\bel{B17}
p\2=(m\up +m\upp)\2 \,.
\ee
To discuss the normalization conditions 
for the proton propagator as it is usual for fermions 
(see, e.g., \cite{Weinberg:1995_I}) 
we write in \er{B16} 
\mbox{$p^2 = \slash{p}^2$}
and consider $S_{F}^{-1}$ as a function~of~$\bp$
\bel{B18}
S_{F}^{-1}(\bp)=C_{0}(\bp\2)+C_{1}(\bp\2) \!\bp \,.
\ee
The proton mass is given by the zero point of the matrix valued function $S_{F}^{-1}(\bp)$
\bel{B19}
S_{F}^{-1}(\bp)\big{|}_{\slash{\;p} = m\up}=0\,,
\ee
and the normalization condition is
\bel{B20}
\frac{\partial}{\partial\!\!\bp}
S_{F}^{-1}(\bp)\big{|}_{\slash{\;p} = m\up}=1\,.
\ee
From \er{B17}--\er{B20} 
we find that \mbox{$S_{F}^{-1}(\bp)$} 
must have the general form
\bal{B20a}
S_{F}^{-1}(\bp)&=\bp -m\up -\Sigma(\bp) \nn\\
&=
(\bp -m\up)\Big{[} 1+
(\bp -m\up)(a\up \!\bp +b\up m\up)\Big{]}\,,
\end{align}
where $\Sigma(\bp)$ is the proton self-energy function 
and $a\up$, $b\up$ are functions of $p\2-m\up\2$, 
analytic at $p\2 -m\up\2=0$. That is, we have
\bal{B20b}
a\up \equiv a(p\2-m\up\2)=& \,a_0+(p\2-m\up\2)a_{1}\nn\\
&+(p\2-m\up\2)\2 a\ud +\dots ,\nn\\
\nn\\
b\up \equiv b(p\2-m\up\2)=&\, b_0+(p\2-m\up\2)b_{1}\nn\\
&+(p\2-m\up\2)\2 b\ud +\dots ,
\end{align}
where the expansions are valid for
\bel{B20c}
|p\2 - m\up\2 |< (m\up + m\upp)\2 -m\up\2 = 2m\up m\upp + m\upp\2 \,;
\ee
see \er{B17}.

The last task is to calculate $S_{F}(p)$ from \er{B20a} and this gives
\bal{B21}
S_{F}(p)=&\frac{1}{\bp-m\up +i\varepsilon}\nn\\
&+\Big{\lbrace}-(\bp +m\up)\Big{[} a\up + (p\2-m\up\2)a\up\2\nn\\
&\qquad +m\up\2(a\up\2-b\up\2 )\Big{]} +m\up(a\up-b\up)
\Big{\rbrace}\nn\\
&\times \Big{\lbrace}1+2(a\up p\2 -b\up m\up\2 )+ (a\up p\2-b\up m\up\2 )\2\nn\\
&\qquad -p\2m\up\2 (a\up-b\up)\2\Big{\rbrace}^{-1}\,.
\end{align}

\subsection{Vertex functions and generalized Ward identity for pions}
\label{sec:B2}

\begingroup
\allowdisplaybreaks

We consider first the $\gamma \pi\pi$ vertex functions 
which are defined by
\begin{align}\label{B22}
&\braket{0|{\rm T}(\varphi\hpm(y)\cJ\ul(x)\varphi\hmp(z))|0}\nn\\
&\quad =\int\frac{\dv^{4}p'}{(2\pi)^{4}}\, \frac{\dv^{4}p}{(2\pi)^{4}}e^{-ip'(y-x)}e^{-ip(x-z)}\nn\\
&\qquad \times 
i\Delta_{F}(p^{\prime 2})
\left[ -\Gamma\ul^{(\gamma\ppm\ppm)}(p', p) \right]
i\Delta_{F}(p^{2})\,.
\end{align}
Using the relations \er{B3} and \er{B4a} we get from the definition \er{B22} and from $\Theta$ and $C$ invariance
\bal{B23}
\Gamma\ul^{(\gamma\pi\hp\pi\hp)}(p,p')&=\phantom{-}\Gamma\ul^{(\gamma\pi^{-}\pi^{-})}(-p',-p)\,,\\
\label{B24}
\Gamma\ul^{(\gamma\pi^{-}\pi^{-})}(p',p)&=-\Gamma\ul^{(\gamma\pi\hp\pi\hp)}(p,p')\,,\\
\label{B25}
\Gamma\ul^{(\gamma\pi\hp\pi\hp)}(p',p)&=-\Gamma\ul^{(\gamma\pi^{-}\pi^{-})}(p',p)\,,
\end{align}
respectively.
Therefore, we can set
\begin{align}
\label{B26}
\Gamma\ul^{(\gamma\pi^{-}\pi^{-})}(p',p)
&=-\Gamma\ul^{(\gamma\pi\hp\pi\hp)}(p',p) \nn \\
&=e\, \widehat{\Gamma}\ul^{(\gamma\pi\pi)}(p',p)\,,
\end{align}
where $\widehat{\Gamma}\ul$ satisfies
\bel{B27}
\widehat{\Gamma}\ul^{(\gamma\pi\pi)}(p',p)=\widehat{\Gamma}\ul^{(\gamma\pi\pi)}(p,p')=-\widehat{\Gamma}\ul^{(\gamma\pi\pi)}(-p',-p)\,.
\ee
The most general ansatz for $\widehat{\Gamma}\ul$ is, therefore, 
\bal{B28}
&\widehat{\Gamma}\ul^{(\gamma\pi\pi)}(p',p)\nn\\
&\quad
=(p'+p) \ul A\Big{[}p^{\prime\, 2}-m\upp\2 ,\,p\2-m\upp\2 ,\, (p'-p)\2\Big{]}\nn\\
&\qquad +(p'-p)\ul B\Big{[}p^{\prime\, 2}-m\upp\2 ,\,p\2-m\upp\2 ,\, (p'-p)\2\Big{]}\,.
\end{align}
Here $A$ and $B$ are analytic functions for
\bal{B29}
|p^{\prime\, 2}-m\upp\2 |&<8m\upp\2\,,\nn\\
|p\2-m\upp\2|&<8m\upp\2\,,\nn\\
|(p' -p)\2|&<4m\upp\2 \,,
\end{align}
which satisfy:
\bal{B30}
&A\Big{[}p^{\prime\,  2}-m\upp\2 ,\,p\2-m\upp\2 ,\, (p'-p)\2\Big{]}\nn\\
&\quad =A\Big{[}p^{  2}-m\upp\2 ,\,p^{\prime\,  2}-m\upp\2 ,\, (p'-p)\2\Big{]}\,, \nn\\
&B\Big{[}p^{\prime\,  2}-m\upp\2 ,\,p\2-m\upp\2 ,\, (p'-p)\2\Big{]}\nn\\
&\quad =-B\Big{[}p^{  2}-m\upp\2 ,\,p^{\prime\,  2}-m\upp\2 ,\, (p'-p)\2\Big{]}\,.
\end{align}
Due to the analyticity properties of $A$ and $B$ \er{B29} we can write
\bal{B31}
&B\Big{[}p^{\prime\,  2}-m\upp\2 ,\,p\2-m\upp\2 ,\, (p'-p)\2\Big{]}\nn\\
&\quad =(p^{\prime\,  2}-p\2)\widetilde{B} \Big{[}p^{\prime\,  2}-m\upp\2 ,\,p^{  2}-m\upp\2 ,\, (p'-p)\2\Big{]}\,,
\end{align}
where $\widetilde{B}$ is symmetric under the exchange $p'\leftrightarrow p$. \\
The most general ansatz for 
$\widehat{\Gamma}\ul^{(\gamma\pi\pi)}$ \er{B28} 
reads then
\bal{B32}
&\widehat{\Gamma}\ul^{(\gamma\pi\pi)}(p',p)\nn\\
&=(p'+p)\ul A\Big{[}p^{\prime\,  2}-m\upp\2 ,\,p\2-m\upp\2 ,\, (p'-p)\2\Big{]}\nn\\
&\quad +(p'-p)\ul (p^{\prime\, 2}-p\2)\widetilde{B} \Big{[}p^{\prime\,  2}-m\upp\2 ,\,p^{  2}-m\upp\2 ,\, (p'-p)\2\Big{]}
\end{align}
with $A$ and $\widetilde{B}$ symmetric 
under $p'\leftrightarrow p$.

Now we study the consequences of the generalized Ward identity: 
\bel{B33}
(p'-p)^{\lambda}\, \widehat{\Gamma}\ul^{(\gamma\pi\pi)}(p',p)=
\Delta_{F}^{-1}(p^{\prime\, 2})-\Delta_{F}^{-1}(p\2)\,.
\ee
We can ask if \er{B33} holds for our local interpolating pion fields. 
The answer is yes. The proof of \er{B33} just relies on current conservation \er{B6a}, the locality of the pion fields, and their charge assignment. From these one finds the following commutation relations
\bal{B33a}
&\Big{[}\cJ^{0}(x),\, \varphi\hpm (y)\Big{]}\delta (x^{0}-y^{0})=0
\qquad \text{for }x\neq y\ ,\nn\\
&\Big{[}Q,\, \varphi\hpm (y)\Big{]}=\mp\,  e \, \varphi\hpm(y) \,,
\end{align}
where $Q$ is the charge operator
\bel{B33b}
Q=\int \dv^{3} x\,  \cJ^{0}(\bm{x}, x^{0})\,.
\ee
And with these relations one can prove the generalized Ward identity; see \cite{Takahashi:1957xn} 
and chapter~10.4 of \cite{Weinberg:1995_I}.
Inserting now into \er{B33} Eqs.~\er{B11} and \er{B32} 
we get
\begin{widetext}
\vspace{-0.5cm}
\bal{B34}
&\Big{[}(p^{\prime\,  2}-m\upp\2 )-(p\2-m\upp\2)\Big{]}
A\Big{[}p^{\prime\,  2}-m\upp\2 ,\,p\2-m\upp\2 ,\, (p'-p)\2\Big{]}\nn\\
&\quad +(p'-p)\2\Big{[}(p^{\prime\,  2}-m\upp\2 )-(p\2-m\upp\2)\Big{]}
\widetilde{B} \Big{[}p^{\prime\,  2}-m\upp\2 ,\,p^{  2}-m\upp\2 ,\, (p'-p)\2\Big{]}\nn\\
& =(p^{\prime\,  2}-m\upp\2 )+(p^{\prime\,  2}-m\upp\2 )\2 \, C(p^{\prime\,  2}-m\upp\2 )
-(p\2-m\upp\2 )-(p\2-m\upp\2 )\2\,   
C(p\2-m\upp\2 )\,.
\end{align}
\end{widetext}
Considering $p^{\prime\,  2}=p\2 \rightarrow m\upp\2$ and then $p' =p$ we get
\bel{B35}
A(0,0,0)=1
\ee
which gives from \er{B32}
\bel{B36}
\widehat{\Gamma}\ul^{(\gamma \pi\pi)}(p',p)\big{|}
_{\mathop{^{p'=p}                  
_{p^{\prime 2} = p\2 = m\upp\2}}}
=2p\ul\,.
\ee
This is exactly the normalization condition for this vertex function.

Now we consider the case
\bel{B37}
p\2=m\upp\2\;,\quad p'=p-k\,,
\ee
which gives
\bel{B37a}
p'-p=-k\;,\quad p^{\prime\, 2}-m\upp\2 = -2p \cdot k+k\2\,.
\ee
We are interested in the quantity
\bel{B38}
\Delta_{F}\big{[}(p-k)\2\big{]}\, \widehat{\Gamma}\ul^{(\gamma\pi\pi)}(p-k,\, p)\,.
\ee
From \er{B11} and \er{B34} we get then
\bal{B39}
&A\Big{[}p^{\prime\,  2}-m\upp\2 ,\, 0 ,\, k\2\Big{]}+k\2\widetilde{B} \Big{[}p^{\prime\,  2}-m\upp\2 ,\,0,\, k\2\Big{]}\nn\\
&\quad =1+(p^{\prime\,  2}-m\upp\2 )\;  C (p^{\prime\,  2}-m\upp\2 )\nn\\
&\quad =\Delta_{F}^{-1}(p^{\prime\,  2})
\,(p^{\prime\,  2}-m\upp\2 +i\varepsilon)^{-1}\,,
\\
\nn\\
\label{B40}
&A\Big{[}p^{\prime\,  2}-m\upp\2 ,\, 0 ,\, k\2\Big{]}
=\Delta_{F}^{-1}(p^{\prime\,  2})(p^{\prime\,  2}-m\upp\2 +i\varepsilon)^{-1}\nn\\
&\qquad \qquad \qquad \qquad \quad \;-k\2\widetilde{B} \Big{[}p^{\prime\,  2}-m\upp\2 ,\,0,\, k\2\Big{]}\,.
\end{align}
Inserting this in \er{B32} and using \er{B37} we find finally
\bal{B41}
&\Delta_{F}\big{[}(p-k)\2\big{]}\, 
\widehat{\Gamma}\ul^{(\gamma \pi\pi)}(p-k,\, p)\big{|}_{p\2=m\upp\2}\nn\\
&\quad =\frac{(2p-k)\ul}{-2p\cdot k + k\2 +i\varepsilon}
-\big{[}(2p-k)\ul\,  k\2 -k\ul (2p\cdot k -k\2)\big{]}\nn\\
&\qquad \times \big{[}-2p\cdot k +k\2+i\varepsilon \big{]}^{-1}\nn\\
&\qquad \times \big{[} 1+(-2p\cdot k +k\2)\; C(-2p\cdot k + k\2)\big{]}^{-1}\nn\\
&\qquad \times
\widetilde{B}\big{[}-2p\cdot k +k\2 ,\, 0,\, k\2\big{]}\,.
\end{align}
We also consider the case
\bel{B42}
p^{\prime \, 2}=m\upp\2 \, ,\quad p=p'+k
\ee
which gives
\bel{B43}
p'-p=-k\, ,\quad p\2-m\upp\2=2p'\cdot k +k\2 \,.
\ee
Here we study
\bel{B44}
\widehat{\Gamma}\ul^{(\gamma \pi\pi)}(p',\, p'+k)
\;
\Delta_{F}\big{[}(p'+k)\2\big{]}\,.
\ee
We find, in a completely analogous way as shown above for the derivation of \er{B41}, the following result:
\bal{B45}
&\widehat{\Gamma}\ul^{(\gamma \pi\pi)}(p', \, p'+k)
\;\Delta_{F}\big{[}(p'+k)\2\big{]}\nn\\
&\quad =\frac{(2p'+k)\ul}{2p'\cdot k + k\2 +i\varepsilon}
-\big{[}(2p'+k)\ul\,  k\2 -k\ul (2p'\cdot k +k\2)\big{]}\nn\\
&\qquad \times \big{[}2p'\cdot k +k\2+i\varepsilon \big{]}^{-1}\nn\\
&\qquad \times \big{[} 1+(2p'\cdot k +k\2)\; C(2p'\cdot k + k\2)\big{]}^{-1}\nn\\
&\qquad \times
\widetilde{B}(0,\, 2p'\cdot k +k\2 ,\, k\2)\ .
\end{align}

The results \er{B41} and \er{B45} are used 
in Sec.~\ref{sec:3} and Appendix~\ref{app:C}.

\endgroup

\subsection{Vertex function and generalized Ward identity for the proton}
\label{subsec:B3}

The defining equation for the $\gamma pp$ vertex function is
\bal{B46}
&\braket{0|{\rm T}(\psi (y)\,\cJ_{\mu}(x)\, \overline{\psi}(z))|0}\nn\\
&\quad =\int\frac{\dv^{4}p'}{(2\pi)^{4}}\, \frac{\dv^{4}p}{(2\pi)^{4}}e^{-ip'(y-x)}e^{-ip(x-z)}\nn\\
&\qquad \times 
iS_{F}(p')
\left[-\Gamma_{\mu}^{(\gamma pp)}(p', p) \right]
iS_{F}(p)\,,
\end{align}
where $\cJ_{\mu}$ is given in \er{B6}. We set
\bel{B47}
\Gamma_{\mu}^{(\gamma pp)}(p', p)=-e\, \widehat{\Gamma}_{\mu}^{(\gamma pp)}(p', p)\,.
\ee
The $P$, $C$, and $T$ relations, see \er{B2}--\er{B8}, 
lead to
\bal{B48}
\widehat{\Gamma}^{(\gamma pp)\mu}(p', p)&
=\cP^{\mu}{}_{\nu}\gamma_{0}\widehat{\Gamma}^{(\gamma pp)\nu}(\cP p',\cP  p)\gamma_{0}\,,\\
\label{B49}
\widehat{\Gamma}^{(\gamma pp)\mu}(p', p)&=-S(C)\Big{(}\widehat{\Gamma}^{(\gamma pp)\mu}(-p,-p')\Big{)}^{\top}S^{-1}(C)\,,\\
\label{B50}
\widehat{\Gamma}^{(\gamma pp)\mu}(p', p)&=\cP^{\mu}{}_{\nu} S(T)\Big{(}\widehat{\Gamma}^{(\gamma pp)\nu}(\cP p , \cP p')\Big{)}^{\top}S^{-1}(T)\,,
\end{align}
respectively. 
The most general ansatz for 
$\widehat{\Gamma}_{\mu}^{(\gamma pp)}$ 
compatible with $P$, $C$, and $T$ invariance, which hold in QCD, is then obtained as follows. $P$ invariance \er{B48} requires for $\widehat{\Gamma}_{\mu}^{(\gamma pp)}$ the structure
\bal{B51}
&\widehat{\Gamma}^{(\gamma pp)\mu}(p', p)=(p'+p)^{\mu} A_{1}+
(p'-p)^{\mu} A_{2}\nn\\
&\quad +\Big{[} g^{\mu\rho} A_{3}+(p'+p)^{\mu}(p'+p)^{\rho} A_{4}\nn\\
&\qquad +(p'+p)^{\mu}(p'-p)^{\rho} A_{5}\nn\\
&\qquad +(p'-p)^{\mu}(p'+p)^{\rho} A_{6}\nn\\
&\qquad +(p'-p)^{\mu}(p'-p)^{\rho} A_{7}\Big{]}\gamma_{\rho} \nn\\
&\quad +\Big{[}g^{\mu\rho}(p'+p)^{\sigma} A_{8}+g^{\mu\rho}(p'-p)^{\sigma} A_{9}\nn\\
&\qquad+(p'+p)^{\mu}(p'+p)^{\rho}(p'-p)^{\sigma} A_{10}\nn\\
&\qquad+(p'-p)^{\mu}(p'+p)^{\rho}(p'-p)^{\sigma} A_{11}\Big{]} i\sigma_{\rho\sigma}\,,
\end{align}
where
\bel{B52}
A_{j}
= A_{j}\big{[}p^{\prime\, 2}-m\up\2 ,\; p\2 - m\up\2 , \; (p' - p)\2 \big{]}
\ee
are analytic functions for 
\bal{B53}
|p^{\prime\, 2}-m\up\2 |&< 2m\up m\upp + m\upp\2\ ,\nn\\
|p\2 - m\up\2 |&< 2m\up m\upp + m\upp\2\ ,\nn\\
|(p' - p)\2 |&< 4m\upp\2 \,.
\end{align}
From $C$ and $T$ invariance \er{B49}, \er{B50}, we find that $A_{j}$ must be symmetric under 
the exchange \mbox{$p'\leftrightarrow p$}
for
\mbox{$j=1,\, 3 ,\,4,\, 7,\, 9,\, 10$} 
and antisymmetric for 
\mbox{$j = 2,\, 5,\, 6,\, 8,\, 11$}. 
Therefore we can set
\bal{B54}
&A_{j}\Big{[}p^{\prime\, 2}-m\up\2 ,\; p\2-m\up\2 , \; (p' -p)\2\Big{]}\nn\\
&\quad =( p^{\prime\, 2}-p\2)\widetilde{ A}_{j}\Big{[}p^{\prime\, 2}-m\up\2 ,\; p\2-m\up\2 , \; (p' -p)\2\Big{]}\,,\nn\\
& \text{for } j=2,\, 5,\, 6,\, 8,\, 11\,,
\end{align}
where the $\widetilde{A}_{j}$ 
are symmetric under $p' \leftrightarrow p$.

Now we explore the consequences of the generalized Ward identity
\bel{B55}
(p'-p)^{\mu}\,\widehat{\Gamma}_{\mu}^{(\gamma pp)} (p', p) = S_{F}^{-1}(p')-S_{F}^{-1}(p)\,.
\ee
The proof of the generalized Ward identity \er{B55} for the local interpolating proton field is analogous to the proof for the pion field; see \er{B33}--\er{B33b}. Of course, we have to take into account that the proton fields are Fermi fields.
Inserting in \er{B55} $\widehat{\Gamma}_{\mu}^{(\gamma pp)}$ from \er{B51}, \er{B54}, 
and $S_{F}^{-1}(p)$ from \er{B20a} we find the following relations, setting
\bel{B56}
q = p'-p \,,
\ee
\bal{B57}
&(p^{\prime 2}-p\2) A_{1} +q\2 (p^{\prime 2}-p\2)\widetilde{ A}\ud\nn\\
&\quad =-2m\up (p^{\prime 2}a_{p'}-p\2 a\up)\nn\\
&\qquad +m\up\Big{[} (p^{\prime 2}+m\up\2)b_{p'}-(p\2+m\up\2)b\up\Big{]}\,,\\
\nn\\
\label{B58}
& A_{3}+ (p^{\prime 2}-p\2)\2 \widetilde{ A}_{5}+q\2  A_{7}\nn\\
&\quad =1+ \frac{1}{2}(p^{\prime 2}+m\up\2)a_{p'}-m\up\2 b_{p'}\nn\\
&\qquad + \frac{1}{2}(p^{2}+m\up\2)a_{p}-m\up\2 b_{p}
\,,\\
\nn\\
\label{B59}
&(p^{\prime 2}-p\2)\Big {[} A_{4}+q\2\widetilde{ A}_{6}\Big{]}\nn\\
&\quad =\frac{1}{2}(p^{\prime 2}+m\up\2)a_{p'}-m\up\2 b_{p'}- \frac{1}{2}(p^{2}+m\up\2)a_{p}+m\up\2 b_{p}
\,,\\
\label{B60}
&\widetilde{ A}_{8}- A_{10}-q\2 \widetilde{ A}_{11}=0\,.
\end{align}
We can use \er{B57}--\er{B60} 
for expressing $A_{1}$, $A_{3}$, $A_{4}$, 
and $\widetilde{A}_{8}$ in terms of the other functions, 
$\widetilde{A}_{2}$, $\widetilde{A}_{5}$, 
$\widetilde{A}_{6}$, $A_{7}$, 
$A_{9}$, $A_{10}$, $\widetilde{A}_{11}$,
and $a\up$, $b\up$ \er{B20b}:
\bal{B61}
& A_{1}(p^{\prime 2}-m\up\2 , \, p\2-m\up\2 ,\, q\2)\nn\\
&\quad =-q\2 \widetilde{ A}_{2}(p^{\prime 2}-m\up\2 , \, p\2-m\up\2 ,\, q\2)\nn\\
&\qquad -m\up(a_{p'}+a\up)+\frac{1}{2} m\up(b_{p'}+b\up)\nn\\
&\qquad -m\up(p^{\prime 2} +p\2)\frac{a_{p'}-a\up}{p^{\prime 2} -p\2} \nn\\
&\qquad +\frac{m\up}{2}(p^{\prime 2} +p\2+2m\up\2)\,\frac{b_{p'}-b\up}{p^{\prime 2} -p\2} \,,\\  \nn \\
\label{B62}
& A_{3}(p^{\prime 2}-m\up\2 , \, p\2-m\up\2 ,\, q\2)\nn\\
&\quad =1+\frac{1}{2}(p^{\prime 2}+m\up\2)a_{p'}-m\up\2 b_{p'}\nn\\
&\qquad + \frac{1}{2}(p^{2}+m\up\2)a_{p}-m\up\2 b_{p}
\nn\\
&\qquad -(p^{\prime 2}-p\2)\2\widetilde{ A}_{5}(p^{\prime 2}-m\up\2 , \, p\2-m\up\2 ,\, q\2)\nn\\
&\qquad -q\2 { A}_{7}(p^{\prime 2}-m\up\2 , \, p\2-m\up\2 ,\, q\2)\,,\\
\nn 
 \\
\label{B63}
& A_{4}(p^{\prime 2}-m\up\2 , \, p\2-m\up\2 ,\, q\2)\nn \\
&\quad =
\frac{1}{4}(a_{p'}+a\up)
+\frac{1}{4}(p^{\prime 2}+p\2+2m\up\2)\, \frac{a_{p'}-a\up}{p^{\prime 2}-p\2}  \nn\\
&\qquad 
-m\up\2
\; \frac{b_{p'}-b\up}{p^{\prime 2}-p\2}
-q\2 \widetilde{ A}_{6}(p^{\prime 2}-m\up\2 ,\,
p\2-m\up\2 ,\, q\2)\,,
\end{align}
\begin{align}
\label{B64}
&\widetilde{ A}_{8}(p^{\prime 2}-m\up\2 , \, p\2-m\up\2 ,\, q\2)\nn\\
&\quad = A_{10}(p^{\prime 2}-m\up\2 , \, p\2-m\up\2 ,\, q\2)\nn  \\
&\qquad +q\2
\widetilde{ A}_{11}(p^{\prime 2}-m\up\2 , \, p\2-m\up\2\ ,\, q\2)\,. 
\end{align}

\begingroup
\allowdisplaybreaks

Now we shall consider the vertex function 
$\widehat{\Gamma}_{\mu}^{(\gamma pp)}$ \\
between on-shell-proton spinors
\bal{B65}
&\bar{u}(p')\widehat{\Gamma}^{(\gamma pp)\mu}(p',p)u(p)\,,\nn \\
&p^{\prime 2}= p\2=m\up\2\,,\nn\\
&p^{\prime 0}, \,p^{0}>0\,.
\end{align}
From \er{B51} we get
\bal{B66}
&\bar{u}(p')\,  \widehat{\Gamma}^{(\gamma pp)\mu}(p',p)u(p) \nn \\
&\quad =\bar{u}(p')\bigg{\lbrace}(p'+p)^{\mu}\Big{[} A_{1}(0,0,q\2)\nn\\
&\qquad +2m\up  A_{4}(0,0,q\2)+q\2  A_{10}(0,0,q\2)\Big{]}\nn\\
&\qquad +\gamma^{\mu} A_{3}(0,0,q\2)
+i\sigma^{\mu\rho}q_{\rho} A_{9}(0,0,q\2)\bigg{\rbrace}u(p)\,.
\end{align}
Using Gordon's identity
\bel{B67}
\bar{u}(p')\Big {[}(p'+p)^{\mu}-2m\up \gamma^{\mu}+i\sigma^{\mu\nu}(p'-p)_{\nu}\Big{]}u(p)=0
\ee
and the relations \er{B20b}, \er{B61}--\er{B63}, 
we arrive at
\bal{B68}
&\bar{u}(p')\, \widehat{\Gamma}^{(\gamma pp)\mu}(p', p)u(p)\nn\\
&\quad =\bar{u}(p')\Big{[} \gamma^{\mu}F_{1}(q\2)+\frac{i}{2m\up}\sigma^{\mu\nu}q_{\nu}F_{2}(q\2)\Big{]}u(p)\,,
\end{align}
where
\bal{B69}
F_{1}(q\2)&=1-q\2 
\Big{[}2m\up\widetilde{A}_{2}(0,0,q\2)
+4m\up\2 \widetilde{A}_{6}(0,0,q\2)
\nn\\
&\quad 
+A_{7}(0,0,q\2) -2m\up A_{10}(0,0,q\2) \Big{]} \,,
\nn\\ 
\nn\\
F_{2}(q\2)&=2m\up\Big{\lbrace} A_{9}(0,0,q\2)+m\up(a_{0}-b_{0})
\nn\\
&\quad +q\2 \Big{[} \widetilde{A}_{2}(0,0,q\2)+2m\up \widetilde{A}_{6}(0,0,q\2)
\nn\\
&\quad -A_{10}(0,0,q\2)\Big{]}\Big{\rbrace}\,.
\end{align}
Here we have also used the expansions of $a\up$ and $b\up$ from \er{B20b}.

With \er{B68} and \er{B69} we have recovered 
the standard expression for the $\gamma p p$ vertex on shell.
We have the correct normalization 
\bel{B70} 
F_{1}(0)=1\,,
\ee
and for $F_{2}(0)$ we have
\bal{B71}
F_{2}(0)&=2m\up \big{[}A_{9}(0,0,0)+m\up (a_{0}-b_{0})\big{]}\nn\\
&=\frac{\mu\up}{\mu_{N} }-1=1.7928\dots \,,
\end{align}
where $\mu_{N}$ is the nuclear magneton $e/(2m\up)$,
and $\mu\up$ is the magnetic moment of the proton.

Now we consider the vertex function 
$\widehat{\Gamma}^{(\gamma pp)\mu}$ \er{B51} 
for an on-shell-proton momentum $p$ and $p'=p-k$ 
where all components of $k$ 
are supposed to be of order $\omega$. 
We will study the expansion of 
$\widehat{\Gamma}^{(\gamma pp)\mu}$ 
in powers of $\omega$ for $\omega \to 0$. 
We have, thus, 
\bal{B72}
&p\2=m\up\2\,, \qquad p'=p-k\,,\nn\\
&p^{\prime 2 }-p\2 = p^{\prime 2 }-m\up\2 =-2p\cdot k +k\2\,,\nn \\
&p'+p=2p-k\,.
\end{align}
From \er{B51} and \er{B54} we get now
\bal{B73}
&\widehat{\Gamma}^{(\gamma pp)\mu}(p-k, p)
=(2p-k)^{\mu}A\uu
(-2p\cdot k + k\2 , 0, 0)\nn\\
&\quad + \gamma^{\mu} A_{3}
(-2p\cdot k + k\2 , 0, 0)\nn\\
&\quad+ (2p-k)^{\mu}(2p-k)^{\nu}\gamma_{\nu}
A_{4}
(-2p\cdot k + k\2 , 0, 0)\nn\\
&\quad -i\sigma^{\mu\nu}\, 2p_{\nu}\, 2(p\cdot k)\widetilde{A}_{8}(0,0,0)\nn\\
&\quad -i\sigma^{\mu\nu}\, k_{\nu}A_{9}(0,0,0)\nn\\
&\quad -i\sigma_{\rho\sigma}p^{\rho}k^{\sigma}4p^{\mu}A_{10}(0,0,0)\nn\\
&\quad + {\mathcal O}(\omega\2)\,.
\end{align}
With \er{B20b}, \er{B61}--\er{B63} we get
\bal{B74}
&(2p-k)^{\mu}A\uu
(-2p\cdot k + k\2 , 0, 0)\nn\\
&\quad =(2p-k)^{\mu}\bigg{\lbrace} -2m\up a_{0}-2m\up^{3}a\uu
+m\up b_{0} + 2m\up^{3} b\uu\nn\\
&\qquad +(-2p\cdot k)\Big{[}-2m\up a\uu +m\up b\uu -2m\up^{3}a\ud +2m\up^{3}b\ud\Big{]}\bigg{\rbrace}\nn\\
&\qquad + {\mathcal O}(\omega\2)\,,\\
\label{B75}
&\gamma^{\mu}A_{3}(-2p\cdot k ,0,0)\nn\\
&\quad =
\gamma^{\mu}\bigg{\lbrace} 1+2m\up\2(a_{0}-b_{0} )\nn\\
&\qquad +(-2p\cdot k) \Big{[} \frac{1}{2}a_{0} +m\up\2 a\uu -m\up\2 b\uu\Big{]} \bigg{\rbrace}
+ {\mathcal O}(\omega\2)\,,\\
\label{B76}
&(2p-k)^{\mu}(2p-k)^{\nu}\gamma_{\nu}A_{4}(-2p\cdot k ,0,0)\nn\\
&\quad =\Big{[}4p^{\mu}\!\! \bp -2k^{\mu}\!\! \bp -2p^{\mu}\!\! \bk\Big {]}
\bigg{\lbrace}\frac{1}{2}a_{0}+m\up\2 a\uu-m\up\2 b\uu\nn\\
&\qquad +(-2p\cdot k)\Big{[}\frac{1}{2}a_{1}+m\up\2 a_{2}-m\up\2 b\ud\Big{]}\bigg{\rbrace}
+ {\mathcal O}(\omega\2)\,.
\end{align}
Putting everything together we find from \er{B73}--\er{B76}
\bal{B77}
&\widehat{\Gamma}^{(\gamma pp)\mu}(p-k, p)\, u(p)\nn\\
&\quad=\bigg{\lbrace}
(2p-k)^{\mu}(-m\up a_{0} +m\up b_{0})\nn\\
&\qquad +\gamma^{\mu}(1+2m\up\2 a_{0}-2m\up\2 b_{0})\nn\\
&\qquad -2p^{\mu}\!\! \bk\big{(}\frac{1}{2}a_{0}+m\up\2 a\uu -m\up\2 b\uu\big{)}\nn\\
&\qquad -i\sigma^{\mu\nu} k_{\nu}A_{9}(0,0,0)\nn\\
&\qquad +2p^{\mu}(\bp\bk -\bk\bp)A_{10}(0,0,0)\nn\\
&\qquad +(-2p\cdot k)\Big{[} (2p-k)^{\mu}(-m\up a\uu +m\up b\uu)\nn\\
&\qquad +\gamma^{\mu}\big{(}\frac{1}{2}a_{0}+m\up\2 a\uu-m\up\2 b\uu\big{)}\nn\\
&\qquad -(\gamma^{\mu}\!\! \bp\;-\bp\gamma^{\mu})\widetilde{A}_{8}(0,0,0)\Big{]}\bigg{\rbrace}u(p)
+ {\mathcal O}(\omega\2)\,.
\end{align}

Now we shall derive the expansion of 
\bel{B78}
S_{F}(p-k)\, \widehat{\Gamma}^{(\gamma pp)\mu}(p-k, p)\,  u(p)\ee
for $\omega\to 0$, where $p$ and $k$ are as in \er{B72}.
Here we use $S_{F}$ from \er{B21}, 
expanded in $\omega$ with the help of \er{B20b}, 
and multiply with 
$\widehat{\Gamma}^{(\gamma pp)\mu}(p-k, p)\, u(p)$ 
from~\er{B77}.
The~result is
\bal{B79}
&S_{F}(p-k)\widehat{\Gamma}^{(\gamma pp)\mu}(p-k, p) \, u(p)\nn\\
&\quad =
\frac{\bp+m\up-\bk}{-2p\cdot k+k\2+i\varepsilon}\Big{[}\gamma^{\mu}-i\sigma^{\mu\nu}k_{\nu}\Big{(}A_{9}(0,0,0)\nn\\
&\qquad +m\up(a_{0}-b_{0})\Big{)}\Big{]} u(p)
+{\mathcal O}(\omega)\nn\\
&\quad=\frac{\bp+m\up-\bk}{-2p\cdot k+k\2+i\varepsilon}\Big{[}\gamma^{\mu}-i\sigma^{\mu\nu}k_{\nu}\frac{1}{2m\up}F_{2}(0)\Big{]}u(p) \nn\\
&\qquad +{\mathcal O}(\omega)\,,
\end{align}
where in the last step we use \er{B71}.

We find it convenient to go from \er{B79} to an equivalent matrix relation. With $\lambda\in \lbrace 1/2 , \, -1/2\rbrace$, the proton's spin indices, we have
\bal{B80}
&S_{F}(p-k)\, \widehat{\Gamma}^{(\gamma pp)\mu}(p-k, p) (\bp+m\up)\nn\\
&\quad = \sum_{\lambda}S_{F}(p-k)\, \widehat{\Gamma}^{(\gamma pp)\mu}(p-k, p)u(p,\lambda)\bar{u}(p,\lambda)\,,\nn\\
&S_{F}(p-k)\, \widehat{\Gamma}^{(\gamma pp)\mu}(p-k, p) u(p,\lambda)\nn\\
&\quad =\frac{1}{2m\up}S_{F}(p-k)\, \widehat{\Gamma}^{(\gamma pp)\mu}(p-k, p)(\bp+m\up)u(p,\lambda)\,.
\end{align}
The result \er{B79} is then equivalent to 
\bal{B81}
&S_{F}(p-k)\, \widehat{\Gamma}^{(\gamma pp)\mu}(p-k, p)(\bp+m\up)\nn\\
&\quad =\frac{\bp+m\up-\bk}{-2p\cdot k+k\2+i\varepsilon}\Big{[}\gamma^{\mu}-\frac{i}{2m\up}\sigma^{\mu\nu}k_{\nu}F\ud(0)\Big{]} ( \bp+m\up)\nn\\
&\qquad
+{\mathcal O}(\omega)\,.
\end{align}

Finally we apply the $T$ transformation to \er{B80}. From \er{B15} and \er{B50} we get
\bal{B82}
&S_{F}(p-k)\, \widehat{\Gamma}^{(\gamma pp)\mu}(p-k, p)(\bp+m\up)\nn\\
&\quad =S(T)S^{\top}_{F}[\cP (p-k)] S^{-1}(T)\nn\\
&\qquad \times \cP^{\mu}{}_{\nu}S(T)\Big{(}\widehat{\Gamma}^{(\gamma pp)\nu}\big{(}\cP p, \cP (p-k)\big{)}\Big{)}^{\top}S^{-1}(T)\nn\\
&\qquad \times S(T)(\cP^{\rho}{}_{\lambda}p^{\lambda}\gamma_{\rho}+m\up)^{\top}S^{-1}(T)\,,
\end{align}
where $S(T)$ and $\cP$ are given in \er{B5} and \er{B8}, respectively. 
Next we make in \er{B82} the replacements
\bal{B83}
p\to\cP p'\,,\quad k\to -\cP k\,, \quad p-k\to \cP (p'+k)\,,
\end{align}
where with $p$ also $p'$ is an on-shell-proton momentum. 
From \er{B81} and \er{B82} we get then
\bal{B84}
&\cP^{\mu}{}_{\nu} S(T)\Big{[} (\bp' +m\up)\widehat{\Gamma}^{(\gamma pp)\nu}(p', p'+k)S_{F}(p'+k)\Big{]}^{\top}S^{-1}(T)\nn\\
&\quad =\frac{\big{(}\cP(p'+k)\big{)}^{\rho}\gamma_{\rho}+m\up}{2p'\cdot k +k\2+i\varepsilon}\Big{[}\gamma^{\mu}+\frac{i}{2m\up}\sigma^{\mu\nu}(\cP k )_{\nu} F\ud (0)\Big{]}\nn\\
&\qquad \times \Big{[} (\cP p')^{\sigma}\gamma_{\sigma}+m\up\Big{]}+{\mathcal O}(\omega)\,,\\
\nn\\
\label{B85}
&(\bp' +m\up)\widehat{\Gamma}^{(\gamma pp)\mu} (p', p'+k)S_{F}(p'+k)\nn\\
&\quad = (\bp'+m\up)\Big{[}\gamma^{\mu}-\frac{i}{2m\up}\sigma^{\mu\nu}k_{\nu} F\ud (0)\Big{]}\nn\\
&\qquad \times \frac{\bp'+m\up +\bk}{2p'\cdot k +k\2+i\varepsilon}
+{\mathcal O}(\omega)\,.
\end{align}
We also recall the following useful relations 
from \er{B1} and \er{B2} of \cite{Lebiedowicz:2022nnn}:
\bal{B86}
&\frac{\bp + m\up -\!\! \bk}{(p-k)\2 -m\up\2 +i\varepsilon}\Big{[} \gamma^{\mu}-\frac{i}{2m\up}\sigma^{\mu\nu}k_{\nu} F\ud (0)\Big{]}(\bp+m\up)\nn\\
&\quad =\frac{1}{-2p\cdot k +k\2 +i\varepsilon}\bigg{\lbrace}2p^{\mu}-k^{\mu}
+\big{(}1+F_{2}(0)\big{)}(k^{\mu}-\!\bk \gamma^{\mu})\nn\\
&\qquad + \frac{F\ud(0)}{2m\up}\Big{[}2\big{(}p^{\mu}\!\!\bk -(p \cdot k)\gamma^{\mu}\big{)}
-(\bk k^{\mu}-k\2\gamma^{\mu})\Big{]}
\bigg{\rbrace} \nn\\
&\qquad \times
(\bp+m\up)\,,\\
\nn\\
\label{B87}
&(\bp'+m\up)
\Big{[}\gamma^{\mu}-\frac{i}{2m\up}\sigma^{\mu\nu}k_{\nu} F\ud (0)\Big{]}\frac{\bp' + m\up +\!\! \bk}{(p'+k)\2 -m\up\2 +i\varepsilon}\nn\\
&\quad =(\bp'+m\up)\, \frac{1}{2p'\cdot k +k\2 +i\varepsilon}
\bigg{\lbrace}2p^{\prime\mu}+k^{\mu}\nn\\
&\qquad -\big{(}1+F\ud(0)\big{)}
(k^{ \mu}-\gamma^{\mu}\!\!\bk)\nn\\
&\qquad -\frac{F\ud(0)}{2m\up}\Big{[}
2\big{(}p^{\prime\mu}\!\! \bk - (p'\cdot k) \gamma^{\mu}\big{)}
+(k^{\mu} \!\!\bk -k\2\gamma^{\mu})\Big{]}\bigg{\rbrace}\,.
\end{align}
In \er{B86} and \er{B87} $p$ and $p'$ 
are on-shell-proton momenta and $k$ is arbitrary.

\section{Definition and details of the calculations for the $\ppm p \to \ppm p \gamma$ amplitudes}
\label{app:C}

Here we consider the reactions
\bel{C1}
\ppm\,(p\ua)+ p\,(p\ub)\to \ppm\,(p'\uu)+p\,(p'\ud)+\gamma(k)
\ee
for $\ppm$ and the proton on shell and $\gamma$ on or off shell. The matrix amplitudes $\cN\ul\equiv\cN^{-}\ul$ from \er{3.16} and the analogous $\cN\ul^{+}$ are defined as follows:
\bal{C2}
\cN\hpm\ul &= 
\int \dv^{4}x\uu\, \dv^{4}x\ud\, \dv^{4}x\ua\, \dv^{4}x\ub \nn\\
&\quad
\times (\slash{p}'\ud+m\up)e^{ip'\ud x\ud}
(-i\! \slash{\rightsidep}{x_{2}}+m\up)\nn\\
&\quad \times
e^{ip'\uu x\uu}(\stackrel{\rightarrow}{\square}_{x\uu}+m_{\pi}\2)\nn\\
&\quad \times 
\braket{0|{\rm T}(\varphi\hpm(x\uu)\varphi\hmp(x\ua)\psi(x\ud)\overline{\psi}(x\ub)(-\mathcal{J}\ul (0)))|0}_{c}\nn\\
&\quad \times (\stackrel{\leftarrow}{\square}_{x\ua}+m_{\pi}\2)e^{-ip\ua x\ua}\nn\\
&\quad \times
(i\! \slash{\leftsidep}{x_{b}} + m\up)e^{-ip\ub x\ub}
(\slash{p}\ub +m\up)\,.
\end{align}
Here the fields and the current are as in \eqref{B1} and \er{B6}, respectively, 
$c$ stands for the connected part,
and we use the reduction formula; 
see, e.g.,~\cite{Bjorken:1965}.

Now we discuss the details of the calculation for 
\mbox{$\cN\ul^{(a)}$, $\dots$, $\cN\ul^{(e)}$} 
corresponding to the diagrams of 
Fig.~\ref{fig:3}(a), $\dots$, Fig.~\ref{fig:3}(e).
We have then from \er{3.18}
\bel{C3}
\cN^{-}\ul\equiv \cN\ul = \cN\ul^{(a)}+\cN\ul^{(b)}+\cN\ul^{(c)}+\cN\ul^{(d)}+\cN\ul^{(e)}\,.
\ee
In the following calculations we shall treat explicitly the $\pi^{-}p\to\pi^{-} p \gamma$ scattering and omit, for brevity, the minus superscript in the expressions.
In Fig.~\ref{fig:3}(a)--(d) 
the off-shell $\pi^{-}p\to \pi^{-} p$ amplitude 
is occurring which is given in \er{2.16}, \er{2.17}. 
We shall need the eight scalar amplitudes 
$\cM\uu, \dots ,\cM_{8}$ ($\cM_{j} \equiv \cM_{j}^{-}$) 
on shell and their partial derivatives
\bal{C4}
&\cM_{j}^{(\text{on})}=\cM_{j}
(s,t,\,m_{\pi}\2,\,m\up\2,\,m_{\pi}\2,\,m\up\2)\,,\nn\\
&\cM_{j},_{s}=\frac{\partial}{\partial s}\cM_{j}
(s,t,\,m_{\pi}\2,\,m\up\2,\,m_{\pi}\2,\,m\up\2)\,,\nn\\
&\cM_{j},_{t}=\frac{\partial}{\partial t}\cM_{j}
(s,t,\,m_{\pi}\2,\,m\up\2,\,m_{\pi}\2,\,m\up\2)\,,\nn\\
&\cM_{j},_{m\uu\2}=\frac{\partial}{\partial m\uu\2}\cM_{j}
(s,t,\,m_{1}\2,\,m\up\2,\,m_{\pi}\2,\,m\up\2)\Big{|}_{m\uu\2 = m_{\pi}\2} \,,\nn\\
&\cM_{j},_{m\ud\2}=\frac{\partial}{\partial m\ud\2}\cM_{j}
(s,t,\,m_{\pi}\2,\,m\ud\2,\,m_{\pi}\2,\,m\up\2)\Big{|}_{m\ud\2 = m_{p}\2} \,,\nn\\
&\cM_{j},_{m\ua\2}=\frac{\partial}{\partial m\ua\2}\cM_{j}
(s,t,\,m_{\pi}\2,\,m\up\2,\,m\ua\2,\,m\up\2)\Big{|}_{m\ua\2 = m_{\pi}\2} \,,\nn\\
&\cM_{j},_{m\ub\2}=\frac{\partial}{\partial m\ub\2}\cM_{j}
(s,t,\,m_{\pi}\2,\,m\up\2,\,m_{\pi}\2,\,m\ub\2)\Big{|}_{m\ub\2 = m_{p}\2} \,,\nn\\
&(j=1, \dots, 8)\,.
\end{align}

From \er{2.24} and \er{2.25} we find then
\bal{C5}
A,_{s}^{\!\!(\text{on})}&=\cM_{1},_{s}+m\up\cM_{2},_{s}-m\up\cM_{4},_{s}\nn\\
&\quad +(-s+m\up\2 +m_{\pi}\2)\cM_{5},_{s}\nn\\
&\quad +(s+t-m\up\2 -m_{\pi}\2)\cM_{7},_{s}\nn\\
&\quad -m\up (2s+t-2m\up\2-2m\upp\2)\cM_{8},_{s}\nn\\
&\quad -\cM_{5}^{(\text{on})}+\cM_{7}^{(\text{on})}-2m\up \cM_{8}^{(\text{on})}\,,
\\
\nn\\
\label{C6}
A,_{t}^{\!\!(\text{on})}&=\cM_{1},_{t}+m\up\cM_{2},_{t}-m\up\cM_{4},_{t}\nn\\
&\quad +(-s+m\up\2 +m_{\pi}\2)\cM_{5},_{t}\nn\\
&\quad +(s+t-m\up\2 -m_{\pi}\2)\cM_{7},_{t}\nn\\
&\quad -m\up (2s+t-2m\up\2-2m\upp\2)\cM_{8},_{t}\nn\\
&\quad +\cM_{7}^{(\text{on})}-m\up \cM_{8}^{(\text{on})}\,,\\
\nn\\
\label{C7}
B,_{s}^{\!\!(\text{on})}&=\cM_{2},_{s}+\cM_{4},_{s}+2m\up\cM_{5},_{s}\nn\\
&\quad -2m\up\cM_{7},_{s}+(4m\up\2 -t)\cM_{8},_{s}\,,\\
\nn\\
\label{C8}
B,_{t}^{\!\!(\text{on})}&=\cM_{2},_{t}+\cM_{4},_{t}+2m\up\cM_{5},_{t}\nn\\
&\quad -2m\up\cM_{7},_{t}+(4m\up\2 -t)\cM_{8},_{t}
-\cM_{8}^{(\text{on})}\,.
\end{align}

In the following calculations the momenta 
$p\ua$, $p\ub$, $p\uu$, $p\ud$, $p_{s}$, $p_{t}$, $p_{u}$, 
as well as $s$, $t$, 
refer to the $\pi^{-}p \to \pi^{-}p$ on-shell reaction; 
see \eqref{3.2} and \eqref{3.37}.

\subsection{The term $(a+b+e1)$}
\label{subsec:C1}

Now we discuss $\cN\ul^{(a)}$ corresponding to 
the diagram of Fig.~\ref{fig:3}(a). 
From \er{3.20}, \er{3.33}, and \er{3.34} we get
\bel{C9}
\cN\ul^{(a)}=-e\, \cN^{(0,a)}\left[\frac{(2p\ua -k)\ul}{-2p\ua \cdot k+k\2}+{\mathcal O}(\omega)\right]\,,
\ee
where $\cN^{(0,a)}$ is given in \eqref{3.39}. In order to simplify the expression for $\cN^{(0,a)}$ we use the following results which are easily checked.
\bal{C10}
&(\bp'\ud +m\up)(\bp_{s}-\bk)(\bp\ub +m\up) \nn\\
&\quad =(\bp'\ud +m\up)\Big{[} m\up+\frac{1}{2}(\bp\ua +\bp'\uu-\bk)\Big{]} (\bp\ub+m\up)\,,
\nn\\ 
\nn\\
&(\bp'\ud +m\up)(\bp_{t}-\!\not{l}\ud)(\bp\ub +m\up)=0\,,
\nn\\
\nn\\
&(\bp'\ud +m\up)(\bp_{u}-\!\not{l}\uu)(\bp\ub +m\up)\nn\\
&\quad =(\bp'\ud +m\up)\Big{[}- m\up+\frac{1}{2}(\bp\ua +\bp'\uu-\bk)\Big{]} (\bp\ub+m\up)\,,
\nn\\
\nn\\
&(\bp'\ud +m\up)i\sigma_{\mu\nu}(p_{s}-k)^{\mu}(p\ut -l\ud)^{\nu}(\bp\ub +m\up)\nn\\
&\quad 
 =(\bp'\ud +m\up)\Big{[}- \frac{1}{2}(p\ua +p'\uu-k\, ,\, p\ub+p'\ud)\nn\\
&\qquad
-(p\ub\cdot p'\ud)+m\up\2 +m\up (\bp\ua+\bp'\uu-\bk)\Big{]} (\bp\ub+m\up)\,,
\nn\\
\nn\\
&(\bp'\ud +m\up)i\sigma_{\mu\nu}(p_{s}-k)^{\mu}(p_{u} -l\uu)^{\nu}(\bp\ub +m\up)\nn\\
&\quad 
 =(\bp'\ud +m\up)\frac{1}{2}(p\ua +p'\uu-k\, ,\, p'\ud-p\ub) (\bp\ub+m\up)\,,
 \nn\\
\nn\\
  &(\bp'\ud +m\up)i\sigma_{\mu\nu}(p_{t}-l\ud)^{\mu}(p_{u}-l\uu)^{\nu}(\bp\ub +m\up)
 \nn\\
 &\quad =(\bp'\ud +m\up)\Big{[}\frac{1}{2}(p\ua +p'\uu-k\, ,\, p\ub+p'\ud)\nn\\
&\qquad
-(p\ub\cdot p'\ud)+m\up\2 -m\up (\bp\ua+\bp'\uu-\bk)\Big{]} (\bp\ub+m\up)\,,
\nn\\
&(\bp'\ud +m\up)i  \gamma_{\mu}\gamma_{5}\varepsilon^{\mu\nu\rho\sigma}(p_{s}-k)_{\nu}(p_{t} -l\ud)_{\rho}
(p_{u}-l\uu)_{\sigma} \nn \\
&\times (\bp\ub +m\up)
=
(\bp'\ud +m\up)\Big{[}-m\up(p\ua +p'\uu-k\, ,\, p\ub+p'\ud)\nn\\
 &\qquad 
 +(\bp\ua+\bp'\uu -\bk)\big{(}m\up\2 +(p'\ud \cdot p\ub)\big{)}\Big{]}
  (\bp\ub+m\up)\,.
  \end{align}
Inserting \er{C10} into \er{3.39} we obtain with $\cM_{j}^{(a)}$ as in \er{3.40}
\bal{C11}
&\cN^{(0,a)}=(\bp'\ud+m\up)\bigg{\lbrace}\cM\uu^{(a)}+m\up\cM\ud^{(a)}-m\up\cM_{4}^{(a)}\nn\\
&\quad +\Big{[} - \frac{1}{2}(p\ua +p'\uu-k\, ,\, p\ub+p'\ud)-(p\ub\cdot p'\ud)+m\up\2\Big{]}\cM_{5}^{(a)}\nn\\
&\quad +\frac{1}{2}(p\ua +p'\uu-k\, ,\, p'\ud-p\ub)\cM_{6}^{(a)}\nn\\
&\quad +\Big{[}  \frac{1}{2}(p\ua +p'\uu-k\, ,\, p\ub+p'\ud)-(p\ub\cdot p'\ud)+m\up\2\Big{]}\cM_{7}^{(a)}\nn\\
&\quad -m\up(p\ua +p'\uu-k\, ,\, p\ub+p'\ud)\cM_{8}^{(a)}\nn\\
&\quad +\frac{1}{2}(\bp\ua +\bp'\uu -\bk)\Big{[}\cM_{2}^{(a)}+\cM_{4}^{(a)}+2m\up\cM_{5}^{(a)}\nn\\
&\quad -2m\up\cM_{7}^{(a)}+\big{(}2m\up\2+2(p'\ud\cdot p\ub)\big{)}\cM_{8}^{(a)}\Big{]}\bigg{\rbrace}
(\bp\ub+m\up)
\nn\\
& \quad
+{\mathcal O}(\omega\2)\,.
\end{align}

Our next topic is to discuss $\cN\ul^{(b)}$ 
corresponding to the diagram of Fig.~\ref{fig:3}(b). 
Here we have from \er{3.21} and \er{B45}
\bel{C12}
\cN\ul^{(b)}=-e\, \frac{(2p'\uu+k)\ul}{2p'\uu\cdot k + k\2} \cN^{(0,b)}+{\mathcal O}(\omega)\,,
\ee
where
\bal{C13}
\cN^{(0,b)}&=(\bp'\ud +m\up)\cM^{(0,b)}(\bp\ub +m\up)\,,
\nn\\ 
\cM^{(0,b)}&=\cM^{(0)}
(p'\uu+k\, ,\, p'\ud \, ,\,  p\ua\, ,\,  p\ub)\,;
\end{align}
see \er{3.25} and \er{3.29}.
Here $\cM^{(0,b)}$ is the off-shell amplitude for $\pi^{-} p\to\pi^{-} p$ as in \er{2.16}, \er{2.17} with the appropriate momenta inserted.

This gives
\bal{C14}
&\cN^{(0,b)}=(\bp'\ud+m\up)\Big{\lbrace}\cM_{1}^{(b)}+\bp\us\cM_{2}^{(b)}\nn\\
&\quad +(\bp\ut-\!\not{l}\ud)\cM_{3}^{(b)}+
(\bp_{u}+\!\not{l}\ud)\cM_{4}^{(b)}\nn\\
&\quad +i\sigma_{\mu\nu}p\us{}^{\mu}(p\ut -l\ud)^{\nu}\cM_{5}^{(b)}\nn\\
&\quad +i\sigma_{\mu\nu}p\us{}^{\mu}(p_{u}+l\ud)^{\nu}\cM_{6}^{(b)}\nn\\
&\quad +i\sigma_{\mu\nu}(p\ut -l\ud)^{\mu}(p_{u}+l\ud)^{\nu}\cM_{7}^{(b)}\nn\\
&\quad +i  \gamma_{\mu}\gamma_{5}\varepsilon^{\mu\nu\rho\sigma}p_{s\nu}(p\ut -l\ud)_{\rho}(p_{u} +l\ud)_{\sigma}\cM_{8}^{(b)}\Big{\rbrace}(\bp\ub+m\up)\nn\\
&\quad +{\mathcal O}(\omega\2)\,,
\end{align}
where
\bal{C15}
\cM_{j}^{(b)}&
=\cM_{j}(s,t-2p\ut\cdot l\ud , m\upp\2 +2p\uu\cdot k , m\up\2 , m\upp\2 , m\up\2)  \nn\\
&\quad +{\mathcal O}(\omega\2) \nn\\
&=\cM_{j}^{(\text{on})}-2(p\ut\cdot l\ud)\cM_{j},_{t}+2(p\uu\cdot k)\cM_{j},_{m\uu\2}\nn\\
&\quad +{\mathcal O}(\omega\2)\,, \nn\\
(j&=1, \dots, 8)\,.
\end{align}
We have the following relations:
\bal{C16}
&(\bp'\ud +m\up)\bp\us (\bp\ub +m\up)\nn\\
&\quad =
 (\bp'\ud +m\up)\Big{[} m\up + 
 \frac{1}{2}(\bp\ua + \bp'\uu +\bk)\Big{]} (\bp\ub +m\up)\,,\nn\\
 \nn\\
&(\bp'\ud +m\up)(\bp\ut -\!\not{l}\ud) (\bp\ub +m\up)=0\,, \nn\\
\nn\\
&(\bp'\ud +m\up)(\bp_{u} +\!\not{l}\ud) (\bp\ub +m\up) 
\nn\\
&\quad=
 (\bp'\ud +m\up)\Big{[}- m\up + \frac{1}{2}(\bp\ua +\bp'\uu +\bk)\Big{]} (\bp\ub +m\up)\,,\nn\\
\nn\\
& (\bp'\ud +m\up) i\sigma_{\mu\nu}p_{s}{}^{\mu}(p\ut -l\ud)^{\nu}(\bp\ub +m\up)
\nn\\
&\quad=
(\bp'\ud +m\up)\Big{[}- \frac{1}{2}(p\ua +p'\uu+k\, ,\,
 p\ub+p'\ud) \nn\\
&\qquad
-(p\ub\cdot p'\ud)+m\up\2 +m\up (\bp\ua+\bp'\uu+\bk)\Big{]} (\bp\ub+m\up)\,,\nn\\
\nn\\
& (\bp'\ud +m\up) i\sigma_{\mu\nu}p_{s}{}^{\mu}(p_{u} +l\ud)^{\nu}(\bp\ub +m\up)
\nn\\
&\quad=
(\bp'\ud +m\up)\frac{1}{2}(p\ua +p'\uu+k\, ,\, p'\ud-p\ub) 
 (\bp\ub+m\up)\,,\nn\\ 
\nn\\
& (\bp'\ud +m\up) i\sigma_{\mu\nu}(p\ut -l\ud)^{\mu}(p_{u}+l\ud)^{\nu}(\bp\ub +m\up)
\nn\\
&\quad=
(\bp'\ud +m\up)
\Big{[} \frac{1}{2}(p\ua +p'\uu+k\, ,\, p\ub+p'\ud) \nn\\
&\qquad
-(p\ub\cdot p'\ud)+m\up\2 -m\up (\bp\ua+\bp'\uu+\bk)\Big{]} (\bp\ub+m\up)\,,
\nn\\
&(\bp'\ud +m\up) i \gamma_{\mu}\gamma_{5}\varepsilon^{\mu\nu\rho\sigma}p_{s\nu}(p\ut -l\ud)_{\rho}(p_{u} +l\ud)_{\sigma}(\bp\ub+m\up)
\nn\\
&\quad =
(\bp'\ud +m\up)\Big{[} -m\up (p\ua +p'\uu+k\, ,\, p\ub+p'\ud) \nn\\
&\qquad 
+ (\bp\ua+\bp'\uu+\bk)\big{(} m\up\2+(p'\ud\cdot p\ub)\big{)}\Big{]}
(\bp\ub+m\up)\,. 
\end{align}
Inserting \er{C16} into \er{C14} we get
\bal{C17}
&\cN^{(0,b)}=(\bp'\ud+m\up)\bigg{\lbrace}
\cM_{1}^{(b)}+m\up\cM_{2}^{(b)}-m\up\cM_{4}^{(b)}\nn\\
&\quad +\Big{[}-\frac{1}{2}(p\ua +p'\uu+k\, ,\, p\ub+p'\ud)-(p\ub\cdot p'\ud)+m\up\2 \Big{]} \cM_{5}^{(b)}\nn\\
&\quad +\frac{1}{2}(p\ua +p'\uu+k\, ,\, p'\ud-p\ub) \cM_{6}^{(b)}\nn\\
&\quad+\Big{[}\frac{1}{2}(p\ua +p'\uu+k\, ,\, p\ub+p'\ud)-(p\ub\cdot p'\ud)+m\up\2 \Big{]} \cM_{7}^{(b)}\nn\\
&\quad -m\up(p\ua +p'\uu+k\, ,\, p\ub+p'\ud)\cM_{8}^{(b)}\nn\\
&\quad+ \frac{1}{2}(\bp\ua +\bp'\uu +\bk)\Big{[} \cM_{2}^{(b)}+ \cM_{4}^{(b)}+2m\up \cM_{5}^{(b)}\nn\\
&\quad -2m\up\cM_{7}^{(b)}
+\big{(} 2m\up\2+2(p'\ud\cdot p\ub)\big{)}\cM_{8}^{(b)}\Big{]}\bigg{\rbrace}(\bp\ub+m\up)\nn\\
&\quad+{\mathcal O}(\omega\2)\,.
\end{align}

Now we define an amplitude $\cN\ul^{(e1)}$ which shall be regular for $\omega=0$ and give the complement needed by gauge invariance for $\cN\ul^{(a)}+\cN\ul^{(b)}$.
That is, we require
\bel{C18}
k^{\lambda}\left(\cN\ul^{(a)}+\cN\ul^{(b)}+\cN\ul^{(e1)}\right) 
= 0\,.
\ee
We shall find that \er{C18} gives a unique solution for $\cN\ul^{(e1)}$ to order $\omega^{0}$.

From \er{3.28} and \er{3.29} we get
\bel{C19}
k^{\lambda}\cN\ul^{(e1)}
=-e\left( \cN^{(0,a)}- \cN^{(0,b)}\right)\,.
\ee
Inserting here $\cN^{(0,a)}$ and $\cN^{(0,b)}$ from \er{C11} and \er{C17}, respectively, we find
\bal{C20}
&k^{\lambda}\cN\ul^{(e1)}=-e(\bp'\ud+m\up)\bigg{\lbrace}\nn\\
&-2(p\us\cdot k)\cM_{1},_{s}-2(p\ua\cdot k)\cM_{1},_{m\ua\2}-2(p\uu\cdot k)\cM_{1},_{m\uu\2}\nn\\
&+m\up\Big{[}-2(p\us\cdot k)\cM_{2},_{s}-2(p\ua\cdot k)\cM_{2},_{m\ua\2}-2(p\uu\cdot k)\cM_{2},_{m\uu\2}\Big{]}\nn\\
&-m\up\Big{[}-2(p\us\cdot k)\cM_{4},_{s}-2(p\ua\cdot k)\cM_{4},_{m\ua\2}-2(p\uu\cdot k)\cM_{4},_{m\uu\2}\Big{]}\nn\\
&+\Big{[} -\frac{1}{2}(p\ua +p'\uu\, ,\, p\ub+p'\ud)-(p\ub\cdot p'\ud)+m\up\2\Big{]}\nn\\
&\quad \times \Big{[}-2(p\us\cdot k)\cM_{5},_{s}-2(p\ua\cdot k)\cM_{5},_{m\ua\2}-2(p\uu\cdot k)\cM_{5},_{m\uu\2}\Big{]}\nn\\
&+(k\, ,\, p\ub +p'\ud ) \cM_{5}^{\text{(on)}}\nn\\
&+\frac{1}{2}(p\ua+p'\uu\, ,\, p'\ud -p\ub)\Big{[}-2(p\ua\cdot k)\cM_{6},_{m\ua\2}-2(p\uu\cdot k)\cM_{6},_{m\uu\2}\Big{]}\nn\\
&+\Big{[}\frac{1}{2}(p\ua +p'\uu\, ,\, p\ub+p'\ud)-(p\ub\cdot p'\ud)+m\up\2\Big{]}\nn\\
&\quad \times \Big{[}-2(p\us\cdot k)\cM_{7},_{s}-2(p\ua\cdot k)\cM_{7},_{m\ua\2}-2(p\uu\cdot k)\cM_{7},_{m\uu\2}\Big{]}\nn\\
&-(k\, ,\,p\ub +p'\ud ) \cM_{7}^{\text{(on)}}
-m\up(p\ua+p'\uu\, ,\, p\ub+p'\ud) \nn\\
& \quad \times
\Big{[}-2(p\us\cdot k)\cM_{8},_{s}%
-2(p\ua\cdot k)\cM_{8},_{m\ua\2}
-2(p\uu\cdot k)\cM_{8},_{m\uu\2}\Big{]}\nn\\
&+2m\up(k\, ,\, p\ub +p'\ud ) \cM_{8}^{\text{(on)}}\nn\\
&+\frac{1}{2}(\bp\ua+\bp'\uu)
\Big{[}-2(p\us\cdot k)\cM_{2},_{s}-2(p\ua\cdot k)\cM_{2},_{m\ua\2}
\nn\\
&\quad -2(p\uu\cdot k)\cM_{2},_{m\uu\2}-2(p\us\cdot k)\cM_{4},_{s}
\nn\\
&\quad -2(p\ua\cdot k)\cM_{4},_{m\ua\2}-2(p\uu\cdot k)\cM_{4},_{m\uu\2} 
\nn\\
&+2m\up\big{(}-2(p\us\cdot k)\cM_{5},_{s}-2(p\ua\cdot k)\cM_{5},_{m\ua\2}-2(p\uu\cdot k)\cM_{5},_{m\uu\2}\big{)}\nn\\
&-2m\up\big{(}-2(p\us\cdot k)\cM_{7},_{s}-2(p\ua\cdot k)\cM_{7},_{m\ua\2}-2(p\uu\cdot k)\cM_{7},_{m\uu\2}\big{)}\nn\\
&+\big{(}2m\up\2+2(p'\ud\cdot p\ub)\big{)}\big{(}-2(p\us\cdot k)\cM_{8},_{s}-2(p\ua\cdot k)\cM_{8},_{m\ua\2}\nn\\
&\quad-2(p\uu\cdot k)\cM_{8},_{m\uu\2}\big{)} \Big{]}\nn\\
&-\bk\Big{[} \cM_{2}^{\text{(on)}}+ \cM_{4}^{\text{(on)}}+2m\up \cM_{5}^{\text{(on)}}-2m\up \cM_{7}^{\text{(on)}}\nn\\
&\quad+\big{(}2m\up\2+2(p'\ud\cdot p\ub)\big{)}\cM_{8}^{\text{(on)}}\Big{]}\bigg{\rbrace}(\bp\ub+m\up)
+{\mathcal O}(\omega\2)\,.
\end{align}
The r.h.s of \er{C20} is, to order $\omega$, a homogeneous linear function of $k\, $. Therefore, we get from \er{C20}, up to the order $\omega^{0}$, a unique solution for $\cN\ul^{(e1)}$:
\vspace{-0.3cm}
\bal{C21}
&\cN\ul^{(e1)}=-e(\bp\ud+m\up)\bigg{\lbrace}\nn\\
&-2p_{s\lambda}\cM_{1},_{s}-2p_{a\lambda}\cM_{1},_{m\ua\2}-2p_{1\lambda}\cM_{1},_{m\uu\2}\nn\\
&+m\up\Big{[}-2p_{s\lambda}\cM_{2},_{s}-2p_{a\lambda}\cM_{2},_{m\ua\2}-2p_{1\lambda}\cM_{2},_{m\uu\2}\Big{]}\nn\\
&-m\up\Big{[}-2p_{s\lambda}\cM_{4},_{s}-2p_{a\lambda}\cM_{4},_{m\ua\2}-2p_{1\lambda}\cM_{4},_{m\uu\2}\Big{]}\nn\\
&+\Big{[} -\frac{1}{2}(p\ua +p\uu\, ,\, p\ub+p\ud)-(p\ub\cdot p\ud)+m\up\2\Big{]}\nn\\
&\quad \times \Big{[}-2p_{s\lambda}\cM_{5},_{s}-2p_{a\lambda}\cM_{5},_{m\ua\2}-2p_{1\lambda}\cM_{5},_{m\uu\2}\Big{]}\nn\\
&+(p\ub +p\ud )_{\lambda}\cM_{5}^{\text{(on)}}\nn\\
&+\frac{1}{2}(p\ua +p\uu\, ,\, p\ud-p\ub)\Big{[}-2p_{a\lambda}\cM_{6},_{m\ua\2}-2p_{1\lambda}\cM_{6},_{m\uu\2}\Big{]}\nn\\
&+\Big{[}\frac{1}{2}(p\ua +p\uu\, ,\, p\ub+p\ud)-(p\ub\cdot p\ud)+m\up\2\Big{]}\nn\\
&\quad \times \Big{[}-2p_{s\lambda}\cM_{7},_{s}-2p_{a\lambda}\cM_{7},_{m\ua\2}-2p_{1\lambda}\cM_{7},_{m\uu\2}\Big{]}\nn\\
&-(p\ub +p\ud )_{\lambda}\cM_{7}^{\text{(on)}}
-m\up(p\ua+p\uu\, ,\, p\ub+p\ud) \nn\\
&\quad \times
\Big{[}-2p_{s\lambda}\cM_{8},_{s}-2p_{a\lambda}\cM_{8},_{m\ua\2}
-2p_{1\lambda}\cM_{8},_{m\uu\2}\Big{]}
\nn\\
&+2m\up (p\ub +p\ud )_{\lambda}\cM_{8}^{\text{(on)}}
\nn\\
&+\frac{1}{2}(\bp\ua+\bp\uu)
\Big{[}-2p_{s\lambda}\cM_{2},_{s}-2p_{a\lambda}\cM_{2},_{m\ua\2}
-2p_{1\lambda}\cM_{2},_{m\uu\2}
\nn\\
&\quad -2p_{s\lambda}\cM_{4},_{s}-2p_{a\lambda}\cM_{4},_{m\ua\2}
-2p_{1\lambda}\cM_{4},_{m\uu\2} 
\nn\\
&\quad +2m\up\big{(}-2p_{s\lambda}\cM_{5},_{s}-2p_{a\lambda}\cM_{5},_{m\ua\2}-2p_{1\lambda}\cM_{5},_{m\uu\2}\big{)}\nn\\
&\quad -2m\up\big{(}-2p_{s\lambda}\cM_{7},_{s}-2p_{a\lambda}\cM_{7},_{m\ua\2}-2p_{1\lambda}\cM_{7},_{m\uu\2}\big{)}\nn\\
&\quad +\big{(} 2m\up\2+2(p\ud\cdot p\ub)\big{)} \big{(}-2p_{s\lambda}\cM_{8},_{s}-2p_{a\lambda}\cM_{8},_{m\ua\2}\nn\\
&\quad -2p_{1\lambda}\cM_{8},_{m\uu\2}\big{)}\Big{]}\nn\\
&-\gamma\ul\Big{[} \cM_{2}^{\text{(on)}}+ \cM_{4}^{\text{(on)}}+2m\up \cM_{5}^{\text{(on)}}-2m\up \cM_{7}^{\text{(on)}}\nn\\
&\quad +\big{(}2m\up\2+2(p\ud\cdot p\ub)\big{)}\cM_{8}^{\text{(on)}}\Big{]}
\bigg{\rbrace}(\bp\ub+m\up)
+{\mathcal O}(\omega)\,.
\end{align}

Our next task is to derive the expansion
in $\omega$ for $\cN\ul^{(a)}$ 
\er{C9} using \er{C11} and \er{3.40}. 
Similarly we treat $\cN\ul^{(b)}$ \er{C12} 
using \er{C13}--\er{C15}, and \er{C17}. 
The results are as follows.
\bal{C22}
&\cN\ul^{(a)}=-e(\bp'\ud +m\up)\bigg{\lbrace} 
\cM_{1}^{\text{(on)}}+m\up \cM_{2}^{\text{(on)}}-m\up \cM_{4}^{\text{(on)}}\nn\\
&+\Big{[}-\frac{1}{2}(p\ua +p'\uu\, ,\, p\ub+p'\ud)-(p\ub\cdot p'\ud)+m\up\2\Big{]}\cM_{5}^{\text{(on)}}\nn\\
&+\Big{[}\frac{1}{2}(p\ua +p'\uu\, ,\, p\ub+p'\ud)-(p\ub\cdot p'\ud)+m\up\2\Big{]}\cM_{7}^{\text{(on)}}\nn\\
&-m\up (p\ua +p'\uu\, ,\, p\ub+p'\ud)\cM_{8}^{\text{(on)}}\nn\\
&+\frac{1}{2}(\bp\ua+\bp'\uu)\Big{[}\cM_{2}^{\text{(on)}}+\cM_{4}^{\text{(on)}}
+2m\up\cM_{5}^{\text{(on)}}-2m\up \cM_{7}^{\text{(on)}}\nn\\
&+\left( 2m\up\2+2(p'\ud\cdot p\ub)\right) 
\cM_{8}^{\text{(on)}}\Big{]}\bigg{\rbrace}
(\bp\ub+m\up)\frac{(2p\ua -k)\ul}{-2p\ua\cdot k +k\2}
\nn\\
&-e(\bp\ud+m\up)
\bigg{\lbrace}-2(p\us\cdot k)\cM_{1},_{s}-2(p\ut\cdot l\ud)\cM_{1},_{t}\nn\\
& \quad  -2(p\ua\cdot k)\cM_{1},_{m\ua\2}\nn\\
&+m\up\big{(}-2(p\us\cdot k)\cM_{2},_{s}-2(p\ut\cdot l\ud)\cM_{2},_{t}-2(p\ua\cdot k)\cM_{2},_{m\ua\2}\big{)}\nn\\
&-m\up\big{(}-2(p\us\cdot k)\cM_{4},_{s}-2(p\ut\cdot l\ud)\cM_{4},_{t}-2(p\ua\cdot k)\cM_{4},_{m\ua\2}\big{)}\nn\\
&+\Big{[}-\frac{1}{2}(p\ua +p\uu\, ,\, p\ub+p\ud)-(p\ub\cdot p\ud)+m\up\2\Big{]}\nn\\
&\quad \times \Big{[}-2(p\us\cdot k)\cM_{5},_{s}-2(p\ut\cdot l\ud)\cM_{5},_{t}-2(p\ua\cdot k)\cM_{5},_{m\ua\2}\Big{]}\nn\\
&+\frac{1}{2}(k\, ,\, p\ub+ p\ud)\cM_{5}^{\text{(on)}}
\nn\\
&+\frac{1}{2}(p\ua +p\uu\,, \,p\ud-p\ub)\big{(}-2(p\ua \cdot k)\cM_{6},_{m\ua\2}\big{)}\nn\\
&+\Big{[}\frac{1}{2}(p\ua +p\uu\, ,\, p\ub+p\ud)-(p\ub\cdot p\ud)+m\up\2\Big{]}\nn\\
&\quad \times \Big{[}-2(p\us \cdot k)\cM_{7},_{s}-2(p\ut\cdot l\ud)\cM_{7},_{t}-2(p\ua\cdot k)\cM_{7},_{m\ua\2}\Big{]}\nn\\
&-\frac{1}{2}(k\, ,\, p\ub+ p\ud)\cM_{7}^{\text{(on)}}
-m\up(p\ua +p\uu\, ,\, p\ub+p\ud)
\nn\\
&\quad \times \Big{[}-2(p\us\cdot k)\cM_{8},_{s}-2(p\ut\cdot l\ud)\cM_{8},_{t}-2(p\ua\cdot k)\cM_{8},_{m\ua\2}\Big{]}\nn\\
&+m\up (k\, ,\, p\ub+ p\ud)\cM_{8}^{\text{(on)}}
+\frac{1}{2}(\bp\ua +\bp \uu) \nn \\
& \quad \times
\Big{[}
-2(p\us\cdot k)\cM_{2},_{s}-2(p\ut\cdot l\ud)\cM_{2},_{t}-2(p\ua\cdot k)\cM_{2},_{m\ua\2}\nn\\
&\qquad \;\;\, 
-2(p\us\cdot k)\cM_{4},_{s}-2(p\ut\cdot l\ud)\cM_{4},_{t}-2(p\ua\cdot k) \cM_{4},_{m\ua\2}
\nn\\
&+2m\up \big{(}-2(p\us\cdot k)\cM_{5},_{s}-2(p\ut\cdot l\ud)\cM_{5},_{t}-2(p\ua\cdot k)\cM_{5},_{m\ua\2}\big{)}\nn\\
&-2m\up \big{(}-2(p\us\cdot k)\cM_{7},_{s}-2(p\ut\cdot l\ud)\cM_{7},_{t}-2(p\ua\cdot k)\cM_{7},_{m\ua\2}\big{)}\nn\\
&
+\big{(}2m\up\2+2(p\ud\cdot p\ub)\big{)}\nn\\
&\quad \times \big{(}-2(p\us\cdot k)\cM_{8},_{s}-2(p\ut\cdot l\ud)\cM_{8},_{t}-2(p\ua\cdot k)\cM_{8},_{m\ua\2}\big{)}
\Big{]}
\nn\\
&-\frac{1}{2}\bk \Big{[}\cM_{2}^{\text{(on)}}+\cM_{4}^{\text{(on)}}+2m\up\cM_{5}^{\text{(on)}}-2m\up \cM_{7}^{\text{(on)}}\nn\\
&\quad +\big{(}2m\up\2+2(p\ud\cdot p\ub)\big{)}\cM_{8}^{\text{(on)}}\Big{]}\bigg{\rbrace}
(\bp\ub+m\up)\frac{2p_{a\lambda}}{(-2p\ua\cdot k )}
\nn\\
&+{\mathcal O}(\omega)
\,.\\
\nn\\
\label{C23}
&\cN\ul^{(b)}=-e(\bp'\ud +m\up)\bigg{\lbrace}
\cM_{1}^{\text{(on)}}+m\up \cM_{2}^{\text{(on) }}-m\up \cM_{4}^{\text{(on)}}
\nn\\
&+\Big{[}-\frac{1}{2}(p\ua +p'\uu\, ,\, p\ub+p'\ud)-(p\ub\cdot p'\ud)+m\up\2\Big{]}\cM_{5}^{\text{(on)}}\nn\\
&+\Big{[}\frac{1}{2}(p\ua +p'\uu\, ,\, p\ub+p'\ud)-(p\ub\cdot p'\ud)+m\up\2\Big{]}\cM_{7}^{\text{(on)}}\nn\\
&-m\up (p\ua +p'\uu\, ,\, p\ub+p'\ud)\cM_{8}^{\text{(on)}}\nn\\
&+\frac{1}{2}(\bp\ua+\bp'\uu)\Big{[}\cM_{2}^{\text{(on)}}+\cM_{4}^{\text{(on)}}+2m\up\cM_{5}^{\text{(on)}}
-2m\up \cM_{7}^{\text{(on)}}\nn\\
&\quad
+\big{(} 2m\up\2+2(p'\ud\cdot p\ub)\big{)} 
\cM_{8}^{\text{(on)}}\Big{]}\bigg{\rbrace}
(\bp\ub+m\up)\frac{(2p'\uu+k)\ul}{2p'\uu\cdot k +k\2}
\nn\\
&-e(\bp\ud+m\up)
\bigg{\lbrace}-2(p\ut\cdot l\ud)\cM_{1},_{t}+2(p\uu\cdot k)\cM_{1},_{m\uu\2}\nn\\
&+m\up\Big{(}-2(p\ut\cdot l\ud)\cM_{2},_{t}+2(p\uu\cdot k)\cM_{2},_{m\uu\2}\Big{)}\nn\\
&-m\up\Big{(}-2(p\ut\cdot l\ud)\cM_{4},_{t}+2(p\uu\cdot k)\cM_{4},_{m\uu\2}\Big{)}\nn\\
&+\Big{[}-\frac{1}{2}(p\ua +p\uu\, ,\, p\ub+p\ud)-(p\ub\cdot p\ud)+m\up\2\Big{]}\nn\\
&\quad \times \Big{[}-2(p\ut\cdot l\ud)\cM_{5},_{t}+2(p\uu\cdot k)\cM_{5},_{m\uu\2}\Big{]}\nn\\
&-\frac{1}{2}(k\, , \, p\ub+ p\ud)\cM_{5}^{\text{(on)}}\nn\\
&+\frac{1}{2}(p\ua +p\uu\, ,\, p\ud-p\ub)\,  2(p\uu \cdot k)\cM_{6},_{m\uu\2}\nn\\
&+\Big{[}\frac{1}{2}(p\ua +p\uu\, ,\, p\ub+p\ud)-(p\ub\cdot p\ud)+m\up\2\Big{]}\nn\\
&\quad \times \Big{[}-2(p\ut\cdot l\ud)\cM_{7},_{t}+2(p\uu\cdot k)\cM_{7},_{m\uu\2}\Big{]}\nn\\
&+\frac{1}{2}(k\, , \,  p\ub+ p\ud)\cM_{7}^{\text{(on)}}
-m\up(p\ua +p\uu\, ,\, p\ub+p\ud)\nn\\
&\quad \times \Big{[}-2(p\ut\cdot l\ud)\cM_{8},_{t}+2(p\uu\cdot k)\cM_{8},_{m\uu\2}\Big{]}\nn\\
&-m\up (k\, , \,  p\ub+ p\ud)\cM_{8}^{\text{(on)}}\nn\\
&+\frac{1}{2}(\bp\ua +\bp \uu)
\Big{[}-2(p\ut\cdot l\ud)\cM_{2},_{t}+2(p\uu\cdot k )\cM_{2},_{m\uu\2}\nn\\
&-2(p\ut\cdot l\ud)\cM_{4},_{t}+2(p\uu\cdot k)\cM_{4},_{m\uu\2}\nn\\
&+2m\up \Big{(}-2(p\ut\cdot l\ud)\cM_{5},_{t}+2(p\uu\cdot k)\cM_{5},_{m\uu\2}\Big{)}\nn\\
&-2m\up \Big{(}-2(p\ut\cdot l\ud)\cM_{7},_{t}+2(p\uu\cdot k)\cM_{7},_{m\uu\2}\Big{)}\nn\\
&
+\big{(}2m\up\2+2(p\ud\cdot p\ub)\big{)}\Big{(}-2(p\ut\cdot l\ud)\cM_{8},_{t}+2(p\uu\cdot k)\cM_{8},_{m\uu\2}\Big{)}\Big{]}\nn\\
&+\frac{1}{2}\bk \Big{[}\cM_{2}^{\text{(on)}}+\cM_{4}^{\text{(on)}}+2m\up\cM_{5}^{\text{(on)}}-2m\up \cM_{7}^{\text{(on)}}\nn\\
&\quad+\big{(}2m\up\2+2(p\ud\cdot p\ub)\big{)}\cM_{8}^{\text{(on)}}\Big{]}\bigg{\rbrace}
(\bp\ub+m\up)\frac{2p_{1\lambda}}{2p\uu\cdot k }
+{\mathcal O}(\omega)\,.
\end{align}
The expressions of \er{C21}--\er{C23} look frightening. 
But it turns out that the sum 
\bel{C24}
\cN\ul^{(a+b+e1)-}=\cN\ul^{(a)}+\cN\ul^{(b)}+\cN\ul^{(e1)}
\ee
is rather simple. We note first that in the sum \er{C24} the terms with
 $\cM_{j},_{m\uu\2}$ and  $\cM_{j},_{m\ua\2}$, that is, the derivatives in off-shell directions drop out. The remaining terms in the sum \er{C24} can be greatly simplified using \er{A10} and the following relations which are valid to the order $\omega$:
\begin{widetext}
\vspace{-0.5cm}
\bal{C25}
&\frac{1}{2}(l\uu \, ,\, p\ub +p\ud)+\frac{1}{2}(l\ud\, ,\,  p\ua +p\uu)+(l\ud\cdot p\ub)+\frac{1}{2}(k\, ,\, p\ub +p\ud)-2(p\us\cdot k)=-(k\cdot p\ua)\,, \nn\\
&-\frac{1}{2}(l\uu\, ,\, p\ub 
+p\ud)-\frac{1}{2}(l\ud\, ,\, p\ua +p\uu)
+(l\ud\cdot p\ub)
+2(k \, ,\, p\ua +p\ub)
+2(l\ud \, ,\, p\ud - p\ub)
-\frac{1}{2}(k \, ,\, p\ub+p\ud)=(k\cdot p\ua)\,, \nn\\
&(l\uu \, ,\, p\ub +p\ud)+(l\ud\, ,\, p\ua +p\uu)-2(l\ud\cdot p\ut)-4(k\cdot p\us)+ (k\, ,\,p\ub+p\ud)=-2(k\cdot p\ua)\,.
\end{align}
In the proof of \er{C25} we make frequent use of \er{3.9}. 
With the help of \er{2.24}, \er{2.25}, \er{C5}--\er{C8}, and \er{C25} we can now greatly simplify the sum \er{C24} of the frightening expressions \er{C21}--\er{C23}. 
The result is given in \er{3.43}.
\end{widetext}

\subsection{The term $(c+d+e2)$}
\label{subsec:C2}

The term $\cN\ul^{(c)}$ corresponding to 
the diagram of Fig.~\ref{fig:3}(c) is defined in \er{3.22}. 
From \er{3.6}, \er{3.26}, and \er{3.30} we have
\begin{align}
\label{C26}
k^{\lambda}\cN\ul^{(c)}&=-e\,\cN^{(0,c)}\nn\\
& =-e (\bp'\ud +m\up )\cM^{(0,c)}(\bp\ub +m\up )\,, \\
\label{C27}
\cM^{(0,c)}&=\cM^{(0)}(p'\uu\, ,\, p'\ud \, ,\, p\ua\, ,\, p\ub -k)\,,\nn\\
p'\uu &= p\uu-l\uu \,, \nn \\
p'\ud &= p\ud-l\ud \,.
\end{align}
With \er{2.16} and \er{2.17} we find
\bal{C28}
&\cN^{(0, c)}=(\bp'\ud +m\up)\bigg{\lbrace}  
\cM_{1}^{(c)}
+\Big{[}m\up +\frac{1}{2}(\bp\ua +\bp'\uu-\bk)\Big{]}\cM_{2}^{(c)}
\nn\\
&+\bk \cM_{3}^{(c)}
+\Big{[}-m\up +\frac{1}{2}(\bp\ua +\bp'\uu+\bk)\Big{]}\cM_{4}^{(c)}\nn\\
&+\Big{[}-\frac{1}{2}(p\ua +p'\uu -k\,  ,\, p\ub+p'\ud)-(p\ub\cdot p'\ud)+m\up\2\nn\\
&\quad +m\up(\bp\ua +\bp'\uu-\bk)-\frac{1}{2}(p\ub-p'\ud\, ,\, k) \nn\\
&\quad 
-\dfrac{1}{4}\Big{(}(\bp\ua+\bp'\uu)\!\bk \;
-\bk (\bp\ua+\bp'\uu)\Big{)}\Big{]}\cM_{5}^{(c)}\nn\\
&+\Big{[}-\frac{1}{2}(p\ua +p'\uu\, ,\, p\ub-p'\ud)
\nn\\
&\quad 
-\dfrac{1}{4}\Big{(}(\bp\ua+\bp'\uu)\!\bk \;
-\bk (\bp\ua+\bp'\uu)\Big{)}\Big{]}\cM_{6}^{(c)}\nn\\
&+\Big{[}\frac{1}{2}(p\ua +p'\uu +k\,  ,\, p\ub+p'\ud)-(p\ub\cdot p'\ud)+m\up\2\nn\\
&\quad -m\up(\bp\ua +\bp'\uu+\bk)
-\frac{1}{2}(p\ub-p'\ud\, ,\, k) 
\nn\\
&\quad +\dfrac{1}{4}\Big{(}(\bp\ua+\bp'\uu)\!\bk \;
-\bk (\bp\ua+\bp'\uu)\Big{)}\Big{]}\cM_{7}^{(c)}\nn\\
&+\Big{[}-m\up (p\ua +p'\uu\, ,\, p\ub+p'\ud)+(\bp\ua+\bp'\uu)\big{(}m\up\2 +(p'\ud\cdot p\ub)\big{)}\nn\\
&\quad+(p\ua +p'\uu\, ,\, p'\ud)\bk - (p'\ud\cdot k )(\bp\ua +\bp'\uu)\nn\\
&-\dfrac{1}{2}m\up\Big{(}(\bp\ua+\bp'\uu)\!\bk \; -\bk (\bp\ua+\bp'\uu)\Big{)}\Big{]}\cM_{8}^{(c)}\bigg{\rbrace}
(\bp\ub+m\up) \nn \\
&+{\mathcal O}(\omega\2)\,,
\end{align}
where
\bal{C29}
\cM_{j}^{(c)}&=
\cM_{j}\Big{(}s-2(p\us\cdot k),\,
t+2(p\ut \cdot l\uu),\nn\\
&\qquad \qquad m\upp\2, \, m\up\2, \, m\upp\2, \, 
m\up\2-2(p\ub\cdot k)\Big{)}+{\mathcal O}(\omega\2)\nn\\
&=\cM_{j}^{\text{(on)}}-2(p\us\cdot k)\cM_{j},_{s}
+ \,2(p\ut\cdot l\uu)\cM_{j},_{t}\nn\\
&\quad -2(p\ub\cdot k)\cM_{j},_{m\ub\2}
+\,{\mathcal O}(\omega\2)\,,\nn\\
(j&=1, \dots , 8)\,;
\end{align}
see \eqref{C4}.
From \er{B81} and \er{B86} we get
\bal{C30}
&S_{F}(p\ub -k) \widehat{\Gamma}\ul^{(\gamma pp)} (p\ub - k\, ,\,  p\ub)(\bp\ub + m\up)\nn\\
&=\frac{\bp\ub + m\up - \!\!\bk}{-2p\ub\cdot k + k\2 + i\varepsilon}\Big{[}\gamma\ul - i\sigma_{\lambda\nu} k^{\nu}\frac{1}{2m\up}F\ud (0) \Big{]} (\bp\ub + m\up)\nn\\
&\quad +{\mathcal O}(\omega)\nn\\
&=\frac{1}{-2p\ub\cdot k + k\2 + i\varepsilon}\bigg{\lbrace}2p_{b\lambda}-k\ul
+(k\ul - \! \bk \gamma\ul)\big{[}1+F\ud (0)\big{]}\nn\\
&\quad + \frac{F\ud(0)}{m \up}\Big{[}p_{b\lambda} \!\! \bk - (p\ub \cdot k)\gamma\ul\Big{]}\bigg{\rbrace}(\bp\ub + m\up)\nn\\
&\quad +{\mathcal O}(\omega)\,.
\end{align}
Inserting this in \er{3.22} we obtain
\bel{C31}
\cN\ul^{(c)}=\cN\ul^{(c1)}+\cN\ul^{(c2)}+\cN\ul^{(c3)}+{\mathcal O}(\omega)\,,
\ee
where
\bal{C32}
\cN\ul^{(c1)}&=e(\bp'\ud + m\up)
\cM^{(0,c)}
(\bp\ub +m\up)\frac{(2p\ub -k)\ul}{-2p\ub\cdot k + k\2}\,,\\
\label{C33}
\cN\ul^{(c2)}&=e(\bp\ud + m\up)
\cM^{(0)}(p_{1},p_{2},p_{a},p_{b})
\nn \\
&\quad \times 
(k\ul -\! \bk\gamma\ul)(\bp\ub +m\up)
\frac{1}{-2p\ub\cdot k}\big{[} 1+F\ud (0)\big{]}\,,\\
\label{C34}
\cN\ul^{(c3)}&=e(\bp\ud + m\up)
\cM^{(0)}(p_{1},p_{2},p_{a},p_{b})\nn\\
&\quad \times 
\big{[}p_{b\lambda}\! \bk - (p\ub \cdot k)\gamma\ul\big{]} (\bp\ub + m\up)
\frac{1}{-2p\ub\cdot k}\frac{F\ud (0)}{m\up}\,.
\end{align}
Here $\cM^{(0,c)}$ is given in (\ref{C27})
and 
$\cM^{(0)}(p_{1},p_{2},p_{a},p_{b}) \equiv \cM^{(0)-}$
is from (\ref{2.16})
for on-shell momenta with
$\cM_{j}^{-} = \cM_{j}^{(\rm on)-} \equiv \cM_{j}^{(\rm on)}$
from (\ref{2.21}).
Inserting \er{2.16}, \er{2.17} in
(\ref{C32})--(\ref{C34})
we get
\bal{C35}
\cN\ul^{(c1)}&=e \, \cN^{(0, c)}\frac{(2p\ub -k)\ul}{-2p\ub\cdot k + k\2}\,,\\
\label{C36}
\cN\ul^{(c2)}&=e(\bp\ud +m\up)\Big{\lbrace} \cM\uu^{(\text{on})}+\bp\us\cM\ud^{(\text{on})}
+ \bp_{u} \cM_{4}^{(\text{on})}\nn\\
&\quad +i\sigma_{\mu\nu}p_{s}^{\mu}p\ut^{\nu}\cM_{5}^{(\text{on})}
+i\sigma_{\mu\nu}p_{t}^{\mu}p_{u}^{\nu}\cM_{7}^{(\text{on})}\nn\\
&\quad +i\gamma_{\mu}\gamma_{5}\varepsilon^{\mu\nu\rho\sigma}p_{s\nu}p_{t\rho}p_{u\sigma}\cM_{8}^{(\text{on})}\Big{\rbrace}\nn\\
&\quad \times (k\ul -\bk \gamma\ul)(\bp\ub + m\up)\frac{1+F\ud(0)}{-2p\ub \cdot k}\,,\\
\label{C37}
\cN\ul^{(c3)}&=e(\bp\ud +m\up)\Big{\lbrace} \cM\uu^{(\text{on})}+\bp\us\cM\ud^{(\text{on})}
+ \bp_{u} \cM_{4}^{(\text{on})}\nn\\
&\quad +i\sigma_{\mu\nu}p_{s}^{\mu}p\ut^{\nu}\cM_{5}^{(\text{on})}
+i\sigma_{\mu\nu}p_{t}^{\mu}p_{u}^{\nu}\cM_{7}^{(\text{on})}\nn\\
&\quad +i\gamma_{\mu}\gamma_{5}\varepsilon^{\mu\nu\rho\sigma}p_{s\nu}p_{t\rho}p_{u\sigma}\cM_{8}^{(\text{on})}\Big{\rbrace}\nn\\
&\quad \times \big{[}p_{b\lambda}\bk - (p\ub \cdot k)\gamma\ul\big{]}
 (\bp\ub + m\up)\frac{F\ud (0)}{m\up}\frac{1}{(-2p\ub\cdot k)}\,.
\end{align}

Now we discuss $\cN\ul^{(d)}$, corresponding to 
the diagram of Fig.~\ref{fig:3}(d), 
as defined in \er{3.23}. 
From \er{3.6}, \er{3.27}, and \er{3.31} we find

\bal{C38}
k^{\lambda}\cN\ul^{(d)}&=e\, \cN^{(0,d)}\nn\\
&=e(\bp'\ud +m\up)\cM^{(0,d)}(\bp\ub +m\up)\,,
\\
\label{C39}
\cM^{(0,d)}&=\cM^{(0)}(p'\uu\,,\, p'\ud + k\,,\, p\ua\,,\, p\ub )\,,
\nn\\
p'\uu&=p\uu-l\uu\,, \nn \\
p'\ud&= p\ud-l\ud\,.
\end{align}
With \er{2.16}, \er{2.17}, and \er{C4}, we find here
\bal{C40}
&\cN^{(0,d)}=
(\bp'\ud + m\up) \Big{\lbrace}\cM\uu^{(d)}
+\Big{[} 
m\up +\frac{1}{2}(\bp\ua +\bp'\uu+\bk)\Big{]}\cM_{2}^{(d)}\nn\\
&+\bk \cM_{3}^{(d)}
+\Big{[} 
-m\up +\frac{1}{2}(\bp\ua +\bp'\uu-\bk)\Big{]}\cM_{4}^{(d)}\nn\\
&+\Big{[}-\frac{1}{2}(p\ua +p'\uu +k\,  ,\, p\ub+p'\ud)-(p\ub\cdot p'\ud)+m\up\2\nn\\
&\quad +m\up(\bp\ua +\bp'\uu+\bk)-\frac{1}{2}(p\ub-p'\ud\, ,\, k) \nn\\
&\quad -\dfrac{1}{4}\Big{(}(\bp\ua+\bp'\uu)\!\bk \;-\bk (\bp\ua+\bp'\uu)\Big{)}\Big{]}\cM_{5}^{(d)}\nn\\
&+\Big{[}-\frac{1}{2}(p\ua +p'\uu\, ,\, p\ub-p'\ud)\nn\\
&\quad +\dfrac{1}{4}
\Big{(}(\bp\ua+\bp'\uu)\!\bk \;-\bk (\bp\ua+\bp'\uu)\Big{)}\Big{]}\cM_{6}^{(d)}
\nn\\
&+\Big{[}\frac{1}{2}(p\ua +p'\uu -k\,  ,\, p\ub+p'\ud)-(p\ub\cdot p'\ud)+m\up\2
\nn\\
&\quad -m\up(\bp\ua +\bp'\uu-\bk)-\frac{1}{2}(p\ub-p'\ud\, ,\, k) \nn\\
&\quad +\dfrac{1}{4}\Big{(}(\bp\ua+\bp'\uu)\!\bk \; -\bk (\bp\ua+\bp'\uu)\Big{)}\Big{]}\cM_{7}^{(d)}\nn\\
&+\Big{[}-m\up (p\ua +p'\uu\, ,\, p\ub+p'\ud)+(\bp\ua+\bp'\uu)\big{(}m\up\2 +(p'\ud\cdot p\ub)\big{)}\nn\\
&\quad + (p\ub\cdot k )(\bp\ua +\bp'\uu)-(p\ua +p'\uu\, ,\, p\ub)\bk\nn\\
&\quad -\dfrac{1}{2}m\up\Big{(}(\bp\ua+\bp'\uu)\!\bk \; -\bk (\bp\ua+\bp'\uu)\Big{)}\Big{]}\cM_{8}^{(d)}\bigg{\rbrace}(\bp\ub+m\up)
\nn\\
&+{\mathcal O}(\omega\2)\,,
\end{align}
where
\bal{C41}
\cM_{j}^{(d)}&=\cM_{j}\big{(}s, \, t+2p\ut \cdot l\uu , \, 
  m\upp\2 ,\,  m\up\2+2 p\ud\cdot k,
  m\upp\2 ,\,  m\up\2 \big{)}
  \nn\\
&\quad +{\mathcal O}(\omega\2)\nn\\
&=\cM_{j}^{\text{(on)}}+2(p\ut\cdot l\uu)\cM_{j},_{t}\ +2(p\ud\cdot k)\cM_{j},_{m\ud\2}\nn\\
&\quad +{\mathcal O}(\omega\2)\,.
\end{align}
From \er{B85} and \er{B87} we find
\bal{C42}
&(\bp'\ud+m\up)\, \widehat{\Gamma}
\ul^{(\gamma pp)}(p'\ud\, ,\, p'\ud +k)S_{F}(p'\ud + k)\nn\\
&=(\bp'\ud +m\up)\Big{[}\gamma\ul -\frac{i}{2m\up}\sigma_{\lambda\nu} k^{\nu}F\ud(0)\Big{]}
\frac{\bp'\ud +m\up + \! \bk}{2p'\ud\cdot k +k\2 +i\varepsilon} \nn\\
&\quad +{\mathcal O}(\omega)\nn\\
&=\frac{1}{2p'\ud\cdot k +k\2 +i\varepsilon}(\bp'\ud +m\up)\bigg{\lbrace}2p'_{2\lambda}+k\ul\nn\\
&\quad -(k\ul -\gamma\ul \!\bk) \big{[}1+F_{2}(0)\big{]}\nn\\
&\quad -\frac{F_{2}(0)}{m\up} \big{(}p'_{2\lambda}\!\bk - (p'\ud\cdot k) \gamma\ul\big{)}\bigg{\rbrace}+{\mathcal O}(\omega)\,.
\end{align}
Inserting \er{C42} into \er{3.23} we obtain
\bal{C43}
\cN\ul^{(d)}&=\cN\ul^{(d1)}+\cN\ul^{(d2)}+\cN\ul^{(d3)}+{\mathcal O}(\omega)\,,
\end{align}
where
\bal{C44}
\cN\ul^{(d1)}=&\;e\frac{2p'_{2\lambda}+k\ul}{2p'\ud\cdot k +k\2 } (\bp'\ud +m\up) \cM^{(0,d)}(\bp\ub +m\up)\,,\\
\label{C45}
\cN\ul^{(d2)}=&-e\big{[}1+F\ud(0)\big{]}\frac{1}{2 p\ud \cdot k} (\bp\ud +m\up)(k\ul-\gamma\ul \!\bk)
\nn \\
&\times 
\cM^{(0)}(p_{1},p_{2},p_{a},p_{b})
(\bp\ub +m\up)\,,\\
\label{C46}
\cN\ul^{(d3)}=&-e\frac{F\ud(0)}{m\up}\, \frac{1}{2p\ud\cdot k}
(\bp\ud+m\up) \big{(}p_{2\lambda} \! \bk - (p\ud\cdot k) \gamma\ul\big{)}
\nn\\
&\times 
\cM^{(0)}(p_{1},p_{2},p_{a},p_{b})
(\bp\ub +m\up) \,.
\end{align}
Here $\cM^{(0,d)}$ is given in (\ref{C39})
and 
$\cM^{(0)}(p_{1},p_{2},p_{a},p_{b}) \equiv \cM^{(0)-}$
is from (\ref{2.16})
for on-shell momenta with
$\cM_{j}^{-} = \cM_{j}^{(\rm on)-} \equiv \cM_{j}^{(\rm on)}$
from (\ref{2.21}).
With \er{2.16}, \er{2.17}, \er{C4}, \er{C38}--\er{C41} 
we find
\bal{C47}
\cN\ul^{(d1)}=&\;
e\frac{(2p'\ud +k)\ul}{2p'\ud\cdot k +k\2 }\; 
\cN^{(0,d)}\,,\\
\label{C48}
\cN\ul^{(d2)}=&-e\big{[}1+F\ud(0)\big{]}\, \frac{1}{2p\ud\cdot k}(\bp\ud+m\up) (k\ul-\gamma\ul \bk)\nn\\
&\times \Big{\lbrace}\cM\uu^{(\text{on})}
+\bp\us\cM\ud^{(\text{on})}
+\bp_{u} \cM_{4}^{(\text{on})}\nn\\&
+i\sigma_{\mu\nu}p_{s}^{\mu}p_{t}^{\nu}\cM_{5}^{(\text{on})}
+i\sigma_{\mu\nu}p_{t}^{\mu}p_{u}^{\nu}\cM_{7}^{(\text{on})}\nn\\
&+i\gamma_{\mu}\gamma_{5}\varepsilon^{\mu\nu\rho\sigma}p_{s\nu}p_{t\rho}p_{u\sigma}\cM_{8}^{(\text{on})}\Big{\rbrace} (\bp\ub + m\up)
\,,\\
\label{C49}
\cN\ul^{(d3)}=&-e\frac{F\ud(0)}{m\up}\, \frac{1}{2p\ud\cdot k}(\bp\ud+m\up) \big{[}p_{2\lambda}\bk - (p\ud\cdot k) \gamma\ul\big{]}\nn\\
&\times \Big{\lbrace}\cM\uu^{(\text{on})}
+\bp\us\cM\ud^{(\text{on})}
+\bp_{u} \cM_{4}^{(\text{on})}\nn\\&
+i\sigma_{\mu\nu}p_{s}^{\mu}p_{t}^{\nu}\cM_{5}^{(\text{on})}
+i\sigma_{\mu\nu}p_{t}^{\mu}p_{u}^{\nu}\cM_{7}^{(\text{on})}\nn\\
&+i\gamma_{\mu}\gamma_{5}\varepsilon^{\mu\nu\rho\sigma}p_{s\nu}p_{t\rho}p_{u\sigma}\cM_{8}^{(\text{on})}\Big{\rbrace} (\bp\ub + m\up)\,.
\end{align}

Now we come to the determination of $\cN\ul^{(e)}$ which corresponds to the diagram of Fig.~\ref{fig:3}(e). 
We have from \er{3.32}
\bal{C50}
k^{\lambda}\cN\ul^{(e)}
=-e\Big{(}\cN^{(0,a)}-\cN^{(0,b)}\Big{)}
+e\Big{(}\cN^{(0,c)}-\cN^{(0,d)}\Big{)}\,.
 \end{align}
Here
$-e\Big{(}\cN^{(0,a)}-\cN^{(0,b)}\Big{)}$ 
is given in \er{C19}, \er{C20}, 
and to order $\omega$ was found 
to be a homogeneous linear function of $k$.
The same is true for $e\Big{(}\cN^{(0,c)}-\cN^{(0,d)}\Big{)}$ for which we find from \er{C28} and \er{C40}
\bal{C51}
&e\Big{[}\cN^{(0,c)}-\cN^{(0,d)}\Big{]}=e(\bp'\ud +m\up)\bigg{\lbrace}
\cM\uu^{(c)}-\cM\uu^{(d)}\nn\\
&+\Big{[}m\up+\frac{1}{2}(\bp\ua +\bp'\uu)\Big{]}\Big{[}\cM\ud^{(c)}-\cM\ud^{(d)}\Big{ ]}\nn\\
&-\bk\cM\ud^{(\text{on})}
+\Big{[}-m\up+\frac{1}{2}(\bp\ua +\bp'\uu)\Big{]}\Big{[}\cM_{4}^{(c)}-\cM_{4}^{(d)}\Big{ ]}\nn\\
&+\bk\cM_{4}^{(\text{on})}
+\Big{[}-\frac{1}{2}(p\ua +p'\uu\,  ,\, p\ub+p'\ud)-(p\ub\cdot p'\ud)\nn\\
&\quad +m\up\2
+m\up(\bp\ua +\bp'\uu)\Big{]}\Big{[}\cM_{5}^{(c)}-\cM_{5}^{(d)}\Big{]}\nn\\
 &+\Big{[}( k\,  ,\, p\ub+p'\ud)-2m\up\bk\Big{]}
 \cM_{5}^{(\text{on})}\nn\\
 &-\frac{1}{2}(p\ua +p'\uu\,  ,\, p\ub+p'\ud)\Big{[}\cM_{6}^{(c)}-\cM_{6}^{(d)}\Big{]}\nn\\
 &
 +\Big{[}\frac{1}{2}(p\ua +p'\uu \,  ,\, p\ub+p'\ud)-(p\ub\cdot p'\ud)\nn\\
&\quad +m\up\2
-m\up(\bp\ua +\bp'\uu)\Big{]}\Big{[}\cM_{7}^{(c)}-\cM_{7}^{(d)}\Big{]}\nn\\
&+\Big{[}(k\, , \, p\ub +p'\ud )-2m\up \bk \Big{]}
\cM_{7}^{(\text{on})}\nn\\
& +\Big{[}-m\up(p\ua +p'\uu \,  ,\, p\ub+p'\ud)
+(\bp\ua+\bp'\uu)
\nn\\
&\quad \times \big{(}m\up\2+(p'\ud\cdot p\ub)\big{)}\Big {]}\Big{[}\cM_{8}^{(c)}-\cM_{8}^{(d)}\Big{]}\nn\\
& +\Big{[}(p\ua +p'\uu \,  ,\, p\ub+p'\ud)\bk
-(\bp\ua+\bp'\uu)(k\, , \, p\ub +p'\ud )\Big{]}\cM_{8}^{(\text{on})}\bigg{\rbrace}\nn\\
&\quad \times (\bp\ub +m\up) 
+ {\mathcal O}(\omega\2)\,.
\end{align}
From \er{C29} and \er{C41} we have
\bal{C52}
&\cM_{j}^{(c)}-\cM_{j}^{(d)}
=-2(p\us\cdot k)\cM_{j},_{s}\nn\\
&\quad -2(p\ub\cdot k)\cM_{j},_{m\ub\2}-2(p\ud\cdot k)\cM_{j},_{m\ud\2}+{\mathcal O}(\omega\2)\ ,\nn\\
&(j=1, \dots , 8)\,.
\end{align} 
Therefore, correct up to order $\omega$, we can everywhere in \er{C51} replace $p'\uu$ by $p\uu$ and $p'\ud$ by $p\ud$. This, indeed, gives for $\Big{[}\cN^{(0,c)}-\cN^{(0,d)}\Big{]}$ a homogeneous linear function in $k$ with corrections starting at order $\omega\2$.
Inserting \er{C19}, \er{C20}, and \er{C51} into \er{C50} we find for $\cN\ul^{(e)}$, up to order $\omega^{0}$, a unique solution which is given by
 \bel{C53}
 \cN\ul^{(e)}=\cN\ul^{(e1)}+\cN\ul^{(e2)}\,.
 \ee
Here $\cN\ul^{(e1)}$ is as in \er{C21} and
 \bal{C54}
&\cN\ul^{(e2)}=e(\bp\ud+m\up)\bigg{\lbrace}
-2p_{s\lambda}\cM_{1},_{s}-2p_{b\lambda}\cM_{1},_{m\ub\2}\nn\\
&-2p_{2\lambda}\cM_{1},_{m\ud\2}+\Big{[}m\up+\frac{1}{2}(\bp\ua +\bp\uu)\Big{]} \nn \\
&\quad \times
\Big{[}-2p_{s\lambda}\cM_{2},_{s}-2p_{b\lambda}\cM_{2},_{m\ub\2}-2p_{2\lambda}\cM_{2},_{m\ud\2}\Big{]}\nn\\
&-\gamma\ul\cM\ud^{(\text{on})}\nn\\
&+\Big{[}-m\up+\frac{1}{2}(\bp\ua +\bp\uu)\Big{]}\nn\\
&\quad\times\Big{[}-2p_{s\lambda}\cM_{4},_{s}-2p_{b\lambda}\cM_{4},_{m\ub\2}-2p_{2\lambda}\cM_{4},_{m\ud\2}\Big{]}\nn\\
&+\gamma\ul\cM_{4}^{(\text{on})}\nn\\
&+\Big{[}-\frac{1}{2}(p\ua +p\uu\,  ,\, p\ub+p\ud)-(p\ub\cdot p\ud)+m\up\2+m\up(\bp\ua +\bp\uu)\Big{]}\nn\\
&\quad \times \Big{[}-2p_{s\lambda}\cM_{5},_{s}-2p_{b\lambda}\cM_{5},_{m\ub\2}-2p_{2\lambda}\cM_{5},_{m\ud\2}\Big{]}\nn\\
&+\Big{[}(p\ub +p\ud)\ul -2m\up\gamma\ul\Big{]}\cM_{5}^{(\text{on})}\nn\\
&-\frac{1}{2}(p\ua +p\uu\,  ,\, p\ub-p\ud)
\Big{[}-2p_{b\lambda}\cM_{6},_{m\ub\2}-2p_{2\lambda}\cM_{6},_{m\ud\2}\Big{]}\nn\\
&+\Big{[}\frac{1}{2}(p\ua +p\uu\,  ,\, p\ub+p\ud)-(p\ub\cdot p\ud)+m\up\2-m\up(\bp\ua +\bp\uu)\Big{]}\nn\\
&\quad \times \Big{[}-2p_{s\lambda}\cM_{7},_{s}-2p_{b\lambda}\cM_{7},_{m\ub\2}-2p_{2\lambda}\cM_{7},_{m\ud\2}\Big{]}\nn\\
&+\Big{[}(p\ub +p\ud)\ul -2m\up\gamma\ul\Big{]}\cM_{7}^{(\text{on})}\nn\\
&+\Big{[}-m\up(p\ua +p\uu \,  ,\, p\ub+p\ud)+(\bp\ua +\bp\uu)\big{(}m\up\2 +(p\ud\cdot p\ub)\big{)}\Big{]}\nn\\
&\quad \times \Big{[}-2p_{s\lambda}\cM_{8},_{s}-2p_{b\lambda}\cM_{8},_{m\ub\2}-2p_{2\lambda}\cM_{8},_{m\ud\2}\Big{]}\nn\\
&+\Big{[}(p\ua +p\uu \,  ,\, p\ub+p\ud)\gamma\ul -(\bp\ua +\bp\uu)(p\ub +p\ud)\ul\Big{]}\cM_{8}^{(\text{on})}\bigg{\rbrace}\nn\\
&\quad\times(\bp\ub +m\up )+{\mathcal O}(\omega)\,.
\end{align}
 
Now the task is to determine $\cN\ul^{(c)}$,
\er{C31}--\er{C37}.
After some straightforward but lengthy calculations we obtain the following results.
\bel{C55}
\cN\ul^{(c1)}=\widetilde{\cN}\ul^{(c1)}+\nldt{c1}\,,
\ee
where, using \er{2.24}, \er{2.25}, and \er{C4}--\er{C8}, we find
 \bal{C56}
&\widetilde{\cN}\ul^{(c1)}=e(\bp'\ud+m\up)\Big{[}A^{(\text{on})}
+\frac{1}{2}(\bp\ua +\bp'\uu)B^{(\text{on})}\Big{]}\nn\\
& \quad \times (\bp\ub +m\up)\frac{(2p\ub-k)\ul}{-2p\ub\cdot k +k\2 }\nn\\
& +e(\bp\ud+m\up)\bigg{\lbrace}\bk \Big{[} -\frac{1}{2}\cM_{2}^{(\text{on})}+ \frac{1}{2}\cM_{4}^{(\text{on})}\nn\\
&\quad -m\up\cM_{5}^{(\text{on})}-m\up \cM_{7}^{(\text{on})}+(p\ua+p\uu \, ,\, p\ud)\cM_{8}^{(\text{on})}\Big{]}\nn\\
&\quad -(\bp\ua +\bp\uu)(p\ub \cdot l\ud + p\ud \cdot k) \cM_{8}^{(\text{on})}\nn\\
 &\quad+\Big{[}\frac{1}{2}(p\ub+p\ud \, , \, l\uu + k)+\frac{1}{2}(p\ua +p\uu \, , \, l\ud) + (p\ub \cdot l\ud)\nn\\
 &\qquad -\frac{1}{2}(p\ub - p\ud \, , \, k )\Big{]}\cM_{5}^{(\text{on})}\nn\\
 &\quad+\Big{[}\frac{1}{2}(p\ub+p\ud \, , \, l\ud)-\frac{1}{2}(p\ua +p\uu \, , \, l\ud) + (p\ub \cdot l\ud)\nn\\
 &\qquad -\frac{1}{2}(p\ub - p\ud \, , \, k )\Big{]}\cM_{7}^{(\text{on})}\nn\\
  &\quad+\Big{[}(p\ub+p\ud \, , \, l\uu )+(p\ua+p\uu\, , \, l\ud )\Big{]}m\up \cM_{8}^{(\text{on})}\nn\\
 &\quad+\Big{[}(\bp\ua +\bp\uu)\!\bk \;- \bk (\bp\ua + \bp\uu)\Big{]}\Big{[}-\frac{1}{4}\cM_{5}^{(\text{on})}+\frac{1}{4}\cM_{7}^{(\text{on})}\nn\\
&\qquad -\frac{1}{2}m\up \cM_{8}^{(\text{on})}\Big{]}\bigg{\rbrace} (\bp\ub +m\up )\frac{2p_{b\lambda}}{(-2p\ub \cdot k)}\nn\\
&+e(\bp\ud +m\up)\bigg{\lbrace}-2(p\us \cdot k)
\Big{[}A,_{s}^{(\text{on})}+\cM_{5}^{(\text{on})}
-\cM_{7}^{(\text{on})}\nn\\
&\quad +2m\up\cM_{8}^{(\text{on})}+\frac{1}{2}(\bp\ua + 
\bp\uu)B,_{s}^{\!\!(\text{on})}\Big{]}\nn\\
&\quad +2(p\ut \cdot l\uu)
\Big{[}A,_{t}^{(\text{on})}-\cM_{7}^{(\text{on})}+m\up \cM_{8}^{(\text{on})}\nn\\
&\quad +\frac{1}{2}(\bp\ua + \bp\uu)( B,_{t}^{\!\!(\text{on})}+\cM_{8}^{(\text{on})})\Big{]}\bigg{\rbrace}
(\bp\ub +m\up )\frac{2p_{b\lambda}}{(-2p\ub \cdot k)}\nn\\
&+{\mathcal O}(\omega)\,,
\end{align}

\bal{C57}
&\overset{\approx}{\cN}\ul^{(c1)}=
 e(\bp\ud+m\up)2p_{b\lambda}\bigg{\lbrace}
\cM_{1, m\ub\2}\nn\\
 &+\Big{[}m\up +\frac{1}{2}(\bp\ua + \bp\uu)\Big{]}\cM_{2, m\ub\2}\nn\\
  &+\Big{[}-m\up +\frac{1}{2}(\bp\ua + \bp\uu)\Big{]}\cM_{4, m\ub\2}\nn\\
 &+\Big{[}-\frac{1}{2}(p\ua+p\uu \, , \, p\ub+p\ud)-(p\ub \cdot p\ud)+m\up\2\nn\\
 &\quad +m\up (\bp\ua + \bp\uu)\Big{]}\cM_{5, m\ub\2}\nn\\
 &-\frac{1}{2}(p\ua+p\uu \, , \, p\ub-p\ud)\cM_{6, m\ub\2}\nn\\
&+\Big{[}\frac{1}{2}(p\ua+p\uu \, , \, p\ub+p\ud)-(p\ub \cdot p\ud)+m\up\2\nn\\
&\quad - m\up (\bp\ua + \bp\uu)\Big{]}\cM_{7, m\ub\2}\nn
  \\ 
&+\Big{[}-m\up(p\ua+p\uu \, , \, p\ub+p\ud)+(\bp\ua + \bp\uu) (m\up\2 +p\ud \cdot p\ub)\Big{]}
\nn\\
&
\quad 
\times \cM_{8, m\ub\2}\bigg{\rbrace}(\bp\ub+m\up)+{\mathcal O}(\omega)\,,\\
\nn\\
\label{C58}
 &\cN\ul^{(c2)}=e(\bp\ud+m\up)\bigg{\lbrace}
 \Big{[}A^{(\text{on})}+\frac{1}{2}(\bp\ua +\bp\uu)B^{(\text{on})}\Big{]}\nn\\
 &\times
 (k\ul -\!\bk\gamma\ul)\nn\\
 &+ \Big{[}\cM_{2}^{(\text{on})}-\cM_{4}^{(\text{on})}+2m\up \cM_{5}^{(\text{on})} +2m\up \cM_{7}^{(\text{on})}\nn\\
 &\quad -(2s+t-2m\up\2 -2m\upp\2)\cM_{8}^{(\text{on})}\nn\\
 &\quad
 + (\bp\ua +\bp\uu)\Big{(}\cM_{5}^{(\text{on})}-\cM_{7}^{(\text{on})}+2m\up \cM_{8}^{(\text{on})} \Big{)}
 \Big{]}\nn\\
 &\times \big{(}\!\bk \,p_{b\lambda}-(p\ub \cdot k)\gamma\ul\big{)}\bigg{\rbrace}(\bp\ub +m\up)\frac{1+F\ud (0)}{(-2p\ub \cdot k)}\,,
\\ \nn\\
\label{C59}
&\cN\ul^{(c3)}=e(\bp\ud+m\up)\bigg{\lbrace}\cM_{1}^{(\text{on})}+(-s-m\up\2 +m\upp\2)\cM_{5}^{(\text{on})}\nn\\
 &+(s+t-3m\up\2 -m\upp\2)\cM_{7}^{(\text{on})}\nn\\
 &+\frac{1}{2}(\bp\ua + \bp\uu)\Big{(}\cM_{2}^{(\text{on})}+\cM_{4}^{(\text{on})}-t\cM_{8}^{(\text{on})}\Big{)}\bigg{\rbrace}\nn\\
&\times \Big{[} p_{b\lambda}\!\bk 
-(p\ub \cdot k)\gamma\ul\Big{]}(\bp\ub+m\up)
\frac{F\ud (0)}{m\up}\frac{1}{(-2p\ub \cdot k)} \,.
 \end{align}
Putting everything together 
we find for $\cN\ul^{(c)}$ \er{C31} the following:
 \bal{C60}
&\cN\ul^{(c)}=e(\bp'\ud+m\up)\Big{[}A^{(\text{on})}+\frac{1}{2}(\bp\ua +\bp'\uu)B^{(\text{on})}\Big{]}\nn\\
 &\quad \times (\bp\ub +m\up)\frac{(2p\ub -k)\ul}{-2p\ub \cdot k + k\2}\nn\\
 &+e(\bp\ud+m\up)\bigg{\lbrace}
 \Big{[}A^{(\text{on})}+\frac{1}{2}(\bp\ua +\bp\uu)B^{(\text{on})}\Big{]}(k\ul -\bk\gamma\ul)\nn\\       
 &\quad -2(p\us \cdot k)\Big{[} A,_{s}^{\!\!(\text{on})} +\frac{1}{2}(\bp\ua +\bp\uu)    
 B,_{s}^{\!\!(\text{on})}\Big{]} 2p_{b\lambda}\nn\\
  &\quad +2(p\ut \cdot l\uu)\Big{[} A,_{t}^{\!\!(\text{on})} +\frac{1}{2}(\bp\ua +\bp\uu)    
 B,_{t}^{\!\!(\text{on})}\Big{]} 2p_{b\lambda}\bigg{\rbrace} \nn\\
 &\quad \times (\bp\ub +m\up)\frac{1}{(-2p\ub \cdot k )}\nn\\
&+ e(\bp\ud+m\up)\Big{[}A^{(\text{on})}+\frac{1}{2}(\bp\ua +\bp\uu)B^{(\text{on})}\Big{]}\nn\\
&\quad \times \Big{[}m\up (k\ul -\bk\gamma\ul)+ \big{(} p_{b\lambda}\bk -(p\ub \cdot k)\gamma\ul\big{)}\Big{]}\nn\\
&\quad \times (\bp\ub+m\up)\frac{F\ud (0)}{m\up}\frac{1}{(-2p\ub \cdot k)}\nn\\
&+e(\bp\ud+m\up)\bigg{\lbrace}\Big{[}-2 \cM_{7}^{(\text{on})} + 2m\up \cM_{8}^{(\text{on})} \nn \\
&\quad
+ (\bp\ua +\bp\uu)\cM_{8}^{(\text{on})} \Big{]}
p_{b\lambda}\nn\\
&\quad +\Big{[}\cM_{2}^{(\text{on})}-\cM_{4}^{(\text{on})}+2m\up \cM_{5}^{(\text{on})} +2m\up \cM_{7}^{(\text{on})}\nn     \\
 &\quad  -( 2s+t-2m\up\2 -2m\upp\2)\cM_{8}^{(\text{on})}\nn\\
  &\quad +(\bp\ua +\bp\uu)  (\cM_{5}^{(\text{on})}-\cM_{7}^{(\text{on})}+2m\up \cM_{8}^{(\text{on})})\Big{]}\frac{1}{2}\gamma\ul\bigg{\rbrace}\nn\\
 &\quad \times(\bp\ub +m\up)\nn\\
 &+\nldt{c1}+{\mathcal O}(\omega)\,.
 \end{align}
Note that $\nldt{c1}$ \er{C57} contains exclusively the derivatives of the amplitudes $\cM_{j}$ with respect to $m\ub\2$ which lead off the mass shell.

The next task is to determine 
$\cN\ul^{(d)}$ from \er{C43}--\er{C49}.
Again, after a straightforward but lengthy calculation we get the following.
\bel{C61}
\cN\ul^{(d1)}=\widetilde{\cN}\ul^{(d1)}+\nldt{d1}\,,
\ee
where
\bal{C62}
&\widetilde{\cN}\ul^{(d1)}=e\frac{(2p'\ud +k)\ul}{2p'\ud \cdot k +k\2 }(\bp'\ud +m\up)
\bigg{\lbrace} \cM_{1}^{(\text{on})} +2(p\ut \cdot l\uu)\cM_{1},_{t}\nn\\
&+\Big{[}m\up + \frac{1}{2}(\bp\ua +\bp'\uu +\bk)\Big{]}\Big{[}\cM_{2}^{(\text{on})} +2(p\ut \cdot l\uu)\cM_{2},_{t}\Big{]}\nn\\
&+\Big{[}-m\up + \frac{1}{2}(\bp\ua +\bp'\uu -\bk)\Big{]}\Big{[}\cM_{4}^{(\text{on})} +2(p\ut \cdot l\uu)\cM_{4},_{t}\Big{]}\nn\\
 &+\Big{[}-\frac{1}{2}(p\ua+p'\uu+k \, , \, p\ub+ p'\ud)-(p\ub \cdot p'\ud)+m\up\2\nn\\
 &\quad+ m\up (\bp\ua + \bp'\uu+\bk)-\frac{1}{2}(p\ub-p'\ud \, , \, k)\nn\\
 &\quad -\frac{1}{4}\Big{(} (\bp\ua +\bp'\uu)\!\bk \;-\bk (\bp\ua +\bp'\uu)\Big{)} \Big{]} \nn \\
 &\quad \times 
    \Big{[}\cM_{5}^{(\text{on})} +2(p\ut \cdot l\uu)\cM_{5},_{t}\Big{]}\nn\\
          &  +\Big{[}\frac{1}{2}(p\ua+p'\uu-k \, , \, p\ub+p'\ud)-(p\ub \cdot p'\ud)+m\up\2\nn\\
  &\quad - m\up (\bp\ua + \bp'\uu-\bk)-\frac{1}{2}(p\ub-p'\ud \, , \, k)\nn\\
    &\quad +\frac{1}{4}\Big{(} (\bp\ua +\bp'\uu)\!\bk \;-\bk (\bp\ua +\bp'\uu)\Big{)}
    \Big{]}\nn\\
    &\quad \times \Big{[}\cM_{7}^{(\text{on})} +2(p\ut \cdot l\uu)\cM_{7},_{t}\Big{]}\nn\\
      &  +\Big{[}-m\up (p\ua+ p'\uu \, , \, p\ub+p'\ud)+ (\bp\ua + \bp'\uu)(m\up\2 + p'\ud\cdot p\ub)\nn\\
  &\quad +(p\ub \cdot k)(\bp\ua + \bp'\uu)-(p\ua+ p'\uu \, , \, p\ub)\bk\nn\\
  &\quad -\frac{1}{2}m\up \Big{(}(\bp\ua + \bp'\uu)\!\bk \;-\bk (\bp\ua + \bp'\uu)\Big{)}\Big{]}\nn\\
    &\quad \times \Big{[}\cM_{8}^{(\text{on})} +2(p\ut \cdot l\uu)\cM_{8},_{t}\Big{]} \bigg{\rbrace}  (\bp\ub + m\up)+{\mathcal O}(\omega)\,,
    \end{align}
    
\bal{C63}
&\nldt{d1}=e\,2p_{2\lambda}(\bp\ud +m\up)\bigg{\lbrace}\cM_{1, m\ud\2}\nn\\
&+\Big{[}m\up + \frac{1}{2}(\bp\ua +\bp\uu )\Big{]}\cM_{2, m\ud\2}\nn\\
&+\Big{[}-m\up + \frac{1}{2}(\bp\ua +\bp\uu )\Big{]}\cM_{4, m\ud\2}\nn\\
 &+\Big{[}-\frac{1}{2}(p\ua+p\uu \, , \, p\ub+p\ud)-(p\ub \cdot p\ud)+m\up\2\nn\\
 &\quad 
 +m\up (\bp\ua + \bp\uu)\Big{]}\cM_{5, m\ud\2}\nn\\ 
 &+\Big{[}-\frac{1}{2}(p\ua+p\uu \, , \, p\ub-p\ud)\Big{]}\cM_{6, m\ud\2}\nn\\
 &+\Big{[}\frac{1}{2}(p\ua+p\uu \, , \, p\ub+p\ud)-(p\ub \cdot p\ud)+m\up\2- m\up (\bp\ua + \bp\uu)\Big{]}\nn\\
  &\quad \times \cM_{7, m\ud\2}\nn\\
 &  +\Big{[}-m\up(p\ua+p\uu \, , \, p\ub+p\ud)+(\bp\ua + \bp\uu) \big{(}m\up\2 +(p\ud \cdot p\ub)\big{)}\Big{]}\nn\\
 &\quad \times
 \cM_{8, m\ud\2}\bigg{\rbrace}
 (\bp\ub+m\up) + {\mathcal O}(\omega)\,.
 \end{align}
Note that $\nldt{d1}$ contains exclusively 
the derivatives $\cM_{j},_{m\ud\2}$ $(j=1, \dots , 8)$
which lead off the mass shell.
With \er{2.24}, \er{2.25}, and \er{C5}--\er{C8} 
we obtain
\bal{C64}
&\widetilde{\cN}\ul^{(d1)}=e\frac{(2p'\ud +k)\ul}{2p'\ud\cdot  k +k\2 }(\bp'\ud +m\up)\nn\\
   &\quad \times \Big{[} A^{(\text{on})} 
   +\frac{1}{2}(\bp\ua +\bp'\uu) B^{(\text{on})}\Big{]}(\bp\ub +m\up)\nn\\
 &+e\frac{p_{2\lambda}}{p\ud \cdot k}\,2(p\ut \cdot l\uu)(\bp\ud +m\up)\nn\\
 &\quad \times \Big{[} A,_{t}^{\!\!(\text{on})} 
  +\frac{1}{2}(\bp\ua +\bp\uu) B,_{t}^{\!\!(\text{on})}\Big{]}(\bp\ub +m\up)\nn\\
 &+e\frac{p_{2\lambda}}{p\ud\cdot k}(\bp\ud +m\up)
 \!\bk
 \nn\\
 &\quad \times \Big{[}\frac{1}{2}\cM_{2}^{(\text{on})}-\frac{1}{2}\cM_{4}^{(\text{on})}+m\up \cM_{5}^{(\text{on})} +m\up \cM_{7}^{(\text{on})}\nn\\
 &\qquad -(p\ua +p\uu\, , \, p\ub)\cM_{8}^{(\text{on})}\nn\\
 &\qquad +\frac{1}{2}(\bp\ua +\bp\uu )(\cM_{5}^{(\text{on})}-\cM_{7}^{(\text{on})}+2m\up \cM_{8}^{(\text{on})})\Big{]} \nn\\
&\qquad \times (\bp\ub +m\up)\nn\\
 &+e \,p_{2\lambda}(\bp\ud +m\up)\Big{[}-2\cM_{7}^{(\text{on})}+2m\up \cM_{8}^{(\text{on})}\nn\\
& \quad +(\bp\ua +\bp\uu) \cM_{8}^{(\text{on})}\Big{]}(\bp\ub +m\up)+{\mathcal O}(\omega)\,,
 \end{align}
 \bal{C65}
&\cN\ul^{(d2)}=-e\big{[}1+F\ud (0)\big{]}\frac{1}{2p\ud \cdot k}(\bp\ud +m\up)\nn\\
 & \times \bigg{\lbrace}(k\ul -\gamma\ul \! \bk)\Big{[}A^{(\text{on})}+\frac{1}{2}(\bp\ua +\bp\uu)B^{(\text{on})}\Big{]}\nn\\
 &\quad +\Big{[} p_{2\lambda}\bk -(p\ud \cdot k)\gamma\ul\Big{]} \Big{[}\cM_{2}^{(\text{on})}-\cM_{4}^{(\text{on})}\nn\\
 &\qquad +\cM_{5}^{(\text{on})}\big{(}2m\up +(\bp\ua +\bp\uu)\big{)}\nn\\
  &\qquad +\cM_{7}^{(\text{on})}\big{(}2m\up -(\bp\ua +\bp\uu)\big{)}\nn\\
  & \qquad -\cM_{8}^{(\text{on})}\big{(}2s + t -2m\up\2 -2m\upp\2 -2m\up(\bp\ua +\bp\uu)\big{)}\Big{]} \bigg{\rbrace}\nn\\
  &\quad\times (\bp\ub +m\up)+{\mathcal O}(\omega)\,,
  \end{align}
 \bal{C66}
&\cN\ul^{(d3)}=-e\frac{F\ud(0)}{m\up}\frac{1}{2p\ud\cdot k}(\bp\ud +m\up)\Big{[} p_{2\lambda}\!\bk -(p\ud \cdot k)\gamma\ul\Big{]} \nn\\
&\times \bigg{\lbrace} \cM_{1}^{(\text{on})}+\cM_{2}^{(\text{on})}\frac{1}{2}  (\bp\ua +\bp\uu) +\cM_{4}^{(\text{on})}    \frac{1}{2}  (\bp\ua +\bp\uu) \nn\\
&\quad+\cM_{5}^{(\text{on})}(-s +m\up\2  +m\upp\2 -2m\up\2)\nn\\
&\quad+\cM_{7}^{(\text{on})}(s +t-m\up\2  -m\upp\2 -2m\up\2)\nn\\
&\quad+\cM_{8}^{(\text{on})}(-\frac{1}{2}t)(\bp\ua +\bp\uu) \bigg{\rbrace}(\bp\ub +m\up)+{\mathcal O}(\omega)\,.\end{align}
\mbox{Collecting things together we get from 
\er{C43}, \er{C61}--\er{C66}}
\bal{C67}
&\cN\ul^{(d)}=\widetilde{\cN}\ul^{(d1)}+ \cN\ul^{(d2)}+ \cN\ul^{(d3)}
+\nldt{d1}+{\mathcal O}(\omega)\ \nn\\
&=e \frac{(2p'\ud+k)\ul}{2p'\ud\cdot k +k\2}(\bp'\ud +m\up)\nn\\
 &\quad \times \Big{[}A^{(\text{on})}+\frac{1}{2}(\bp\ua +\bp'\uu)B^{(\text{on})}\Big{]}(\bp\ub +m\up)\nn\\
 &+e\frac{p_{2\lambda}}{p\ud \cdot k} 2(p\ut\cdot l\uu)(\bp\ud +m\up)\nn\\
 &\quad \times \Big{[} A,_{t}^{\!\!(\text{on})} +\frac{1}{2}(\bp\ua +\bp\uu) B,_{t}^{\!\!(\text{on})}\Big{]} (\bp\ub +m\up)\nn\\
 &-e\frac{1}{2p\ud\cdot k}(\bp\ud +m\up) (k\ul -\gamma\ul \!\bk)\nn\\
 &\quad \times \Big{[}A^{(\text{on})}+\frac{1}{2}(\bp\ua +\bp\uu)B^{(\text{on})}\Big{]} (\bp\ub +m\up)\nn\\
 &-e\frac{F\ud (0)}{m\up}\frac{1}{2p\ud \cdot k}(\bp\ud +m\up) \nn\\
 &\quad \times \Big{[}m\up (k\ul -\gamma\ul \!\bk)+\big{(} p_{2\lambda}\bk -(p\ud \cdot k)\gamma\ul\big{)}\Big{]}\nn\\
 &\quad \times \Big{[}A^{(\text{on})}+\frac{1}{2}(\bp\ua +\bp\uu)B^{(\text{on})}\Big{]} (\bp\ub +m\up)\nn\\
 &+e \,p_{2\lambda}(\bp\ud +m\up)\Big{[} -2\cM_{7}^{(\text{on})}+2m\up\cM_{8}^{(\text{on})}\nn\\
 &\quad+(\bp\ua +\bp\uu) \cM_{8}^{(\text{on})}\Big{]}(\bp\ub +m\up)\nn\\
 &+e(\bp\ud +m\up)\gamma\ul \Big{[}\frac{1}{2}\cM_{2}^{(\text{on})} -\frac{1}{2}\cM_{4}^{(\text{on})}\nn\\
 &\quad+\cM_{5}^{(\text{on})}\big{(}m\up+ \frac{1}{2}(\bp\ua +\bp\uu)\big{)}
 +\cM_{7}^{(\text{on})}\big{(}m\up-\frac{1}{2}(\bp\ua +\bp\uu)\big{)}\nn\\
 &\quad +\cM_{8}^{(\text{on})}\big{(}-\frac{1}{2}(2s + t - 2 m\up\2 - 2m\upp\2)+m\up(\bp\ua +\bp\uu)\big{)}\Big{]}\nn\\
&\quad \times (\bp\ub +m\up)\nn\\
&+\nldt{d1} +{\mathcal O}(\omega)\,.
  \end{align}
 
Now we study the term
 \bal{C68}
 &\cN\ul^{(e3)}=\cN\ul^{(e2)}+ \nldt{c1}+ \nldt{d1}\,.
 \end{align}
From  \er{C54}, \er{C57}, and \er{C63} 
we get
\bal{C68A}
&\cN\ul^{(e3)}=e (\bp\ud +m\up)\bigg{\lbrace}-2p_{s\lambda}\Big{[}\cM_{1},_{s}\nn\\
 &+ \Big{(}m\up+ \frac{1}{2}(\bp\ua +\bp\uu )\Big{)}\cM_{2},_{s}\nn\\
 &+ \Big{(}-m\up+ \frac{1}{2}(\bp\ua +\bp\uu )\Big{)}\cM_{4},_{s}\nn\\
 &+ \Big{(}-\frac{1}{2}(p\ua +p\uu\, , \, p\ub + p\ud )-(p\ub \cdot p\ud ) + m\up\2 \nn\\
  &\quad +m\up (\bp\ua +\bp\uu)\Big{)}\cM_{5},_{s}\nn\\
   & + \Big{(}\frac{1}{2}(p\ua +p\uu\, , \, p\ub + p\ud )-(p\ub \cdot p\ud ) + m\up\2 \nn\\
 &\quad  -m\up (\bp\ua +\bp\uu)\Big{)}\cM_{7},_{s}\nn\\
   & + \Big{(}-m\up (p\ua +p\uu\, , \, p\ub + p\ud )+ \big{(}m\up\2 +(p\ud \cdot p\ub )\big{)} (\bp\ua +\bp\uu) \Big{)} \nn\\
 &\quad \times \cM_{8},_{s}\Big{]}\nn\\
 &-\gamma\ul \Big{[}\cM_{2}^{(\text{on})}
 -\cM_{4}^{(\text{on})}+2m\up\cM_{5}^{(\text{on})}
 +2m\up \cM_{7}^{(\text{on})}\nn\\
 &\quad -(p\ua +p\uu\, , \, p\ub + p\ud) 
 \cM_{8}^{(\text{on})}\Big{]}\nn\\
 &+(p\ub + p\ud)\ul \Big{[}\cM_{5}^{(\text{on})}+\cM_{7}^{(\text{on})}
 - (\bp\ua +\bp\uu) \cM_{8}^{(\text{on})}\Big{]}\bigg{\rbrace} \nn\\
& \times (\bp\ub +m\up)
+ {\mathcal O}(\omega)\,,
\end{align}
and with \er{A10}, \er{C5}, and \er{C7}
\bal{C69}
 &\cN\ul^{(e3)}=e (\bp\ud +m\up)\bigg{\lbrace}-2p_{s\lambda}\Big{[}\cM_{1},_{s}
+m\up \cM_{2},_{s}-m\up \cM_{4},_{s} \nn\\
&+(-s+m\up\2 +m\upp\2 )\cM_{5},_{s}
+ (s+t-m\up\2 -m\upp\2 )\cM_{7},_{s}\nn \\
&-m\up  (2s+t-2m\up\2 -2m\upp\2 )\cM_{8},_{s}\nn \\
&+\frac{1}{2}(\bp\ua +\bp\uu)\Big{(}\cM_{2},_{s}+\cM_{4},_{s}+2m\up \cM_{5},_{s} \nn\\
&-2m\up \cM_{7},_{s} +(4m\up\2 -t) \cM_{8},_{s}\Big{)}\Big{]}\nn\\
&-\gamma\ul \Big{[}\cM_{2}^{(\text{on})}-\cM_{4}^{(\text{on})}+2m\up\cM_{5}^{(\text{on})}+2m\up \cM_{7}^{(\text{on})}\nn\\
 &\quad -   (2s+t-2m\up\2 -2m\upp\2 )    \cM_{8}^{(\text{on})}\Big{]}\nn\\
 &+(p\ub + p\ud)\ul \Big{[}\cM_{5}^{(\text{on})}+\cM_{7}^{(\text{on})}- (\bp\ua +\bp\uu)   \cM_{8}^{(\text{on})}\Big{]}\bigg{\rbrace} \nn\\
& \times(\bp\ub +m\up) 
+ {\mathcal O}(\omega)\,,\\
\nn\\
\label{C70}
 &\cN\ul^{(e3)}=e (\bp\ud +m\up)\bigg{\lbrace}-2p_{s\lambda}\Big{[}A,_{s}^{\!\!(\text{on})}
 +\frac{1}{2}(\bp\ua +\bp\uu)
 B,_{s}^{\!\!(\text{on})}\Big{]}\nn\\
  &-\gamma\ul \Big{[}\cM_{2}^{(\text{on})}-\cM_{4}^{(\text{on})}+2m\up\cM_{5}^{(\text{on})}+2m\up \cM_{7}^{(\text{on})}\nn\\
 &\quad -(2s+t-2m\up\2 -2m\upp\2) 
 \cM_{8}^{(\text{on})}\Big{]}\nn \\
 &+(p\ub + p\ud)\ul \Big{[}\cM_{5}^{(\text{on})}+\cM_{7}^{(\text{on})}- (\bp\ua +\bp\uu)   \cM_{8}^{(\text{on})}\Big{]}\nn
\\
  &-2(p\uu + p\ud)\ul \Big{[}\cM_{5}^{(\text{on})}-\cM_{7}^{(\text{on})}+2m\up  \cM_{8}^{(\text{on})}\Big{]}\bigg{\rbrace} \nn\\
&\times (\bp\ub +m\up) + {\mathcal O}(\omega)\,.
\end{align}

Now we can collect everything together and calculate
\bal{C71}
\cN\ul^{(c+d+e2)}&=\cN\ul^{(c)}+\cN\ul^{(d)}+\cN\ul^{(e2)}\nn\\
&=\widetilde{\cN}\ul^{(c1)}   +\cN\ul^{(c2)}+\cN\ul^{(c3)} \nn\\
&\quad +\widetilde{\cN}\ul^{(d1)}+\cN\ul^{(d2)}+\cN\ul^{(d3)}+\cN\ul^{(e3)}\,;
\end{align}
see \er{C31}--\er{C37}, \er{C43}--\er{C49}, \er{C55}, \er{C60}, \er{C61}, \er{C68}, and \er{C70}. 
The result is given in \er{3.44}
\bel{C72}
\cN\ul^{(c+d+e2)-}\equiv \cN\ul^{(c+d+e2)}\,.
\ee
 
\section{Comparison with results from the literature}
\label{app:D}

In this appendix we compare our findings with some results
from the literature. We start with Low's result
for soft-photon production in the scattering of 
a neutral spin-zero boson on a spin-one-half fermion
having non-zero charge and magnetic moment;
see Eq.~(1.8) of \cite{Low:1958sn}.
To be concrete we shall consider $\pi^{0} p$ scattering
without and with soft-photon emission:
\begin{align}
&\pi^{0}\,(p\ua)+p\,(p\ub,\lambda\ub) \to \pi^{0}\,(p\uu)+p\,(p\ud,\lambda\ud)
\,,\nn \\
&p_{a} + p_{b} = p_{1} + p_{2}\,,
\label{D1}
\intertext{and}
&\pi^{0}\,(p\ua)+p\,(p\ub,\lambda\ub) \to \pi^{0}\,(p\uu')+p\,(p\ud',\lambda\ud')
+\gamma\,(k , \varepsilon)\,,\nn \\
&p_{a} + p_{b} = p_{1}' + p_{2}' + k\,.
\label{D2}
\end{align}
All kinematic relations of the reactions (\ref{D1}) and (\ref{D2})
can be taken over from those for charged pion-proton scattering;
see Secs.~\ref{sec:2} and \ref{sec:3}.
As done in \cite{Low:1958sn} we shall in the following
only consider real photon emission.
That is, we set $k^{2} = 0$.

For the reaction (\ref{D1}) we have the amplitude as in (\ref{2.23})
\begin{align}
\label{D3}
&\braket{\pi^{0} (p\uu), \,  p(p\ud ,\lambda\ud)|{\mathcal T}|\pi^{0}(p\ua), \, p(p\ub ,\lambda\ub)} \nn\\
& = \bar{u}(p\ud , \lambda\ud)
\Big{[}A^{(\text{on})}(s,t)+\frac{1}{2}(\slash{p}_{a}+\slash{p}_{1})
B^{(\text{on})}(s,t)\Big{]}u(p\ub,\lambda\ub )\,,
\end{align}
where
\begin{align}
\label{D4}
&s = (p_{a} + p_{b})^{2} = (p_{1} + p_{2})^{2}\,, \nn\\
&t = (p_{a} - p_{1})^{2}\,.
\end{align}
In \cite{Low:1958sn} a different energy variable is used
\begin{align}
\label{D5}
\nu = p_{a} \cdot p_{b} + p_{1} \cdot p_{2} 
    = s - m_{p}^{2} - m_{\pi}^{2}\,.
\end{align}
We have then with $A_{\rm L}$, $B_{\rm L}$ Low's invariant
functions for $\pi^{0} p$ scattering
\begin{align}
\label{D6}
&A_{\rm L}(\nu,t) = A^{(\text{on})}(s,t)
= A^{(\text{on})}(\nu + m_{p}^{2} + m_{\pi}^{2},t)\,, \nn\\
&B_{\rm L}(\nu,t) = B^{(\text{on})}(s,t)
= B^{(\text{on})}(\nu + m_{p}^{2} + m_{\pi}^{2},t)\,.
\end{align}

Low's result for the amplitude of (\ref{D2})
in the soft-photon limit reads [see Eq.~(1.8) of \cite{Low:1958sn}]
\begin{align}
\label{D7}
&\bar{u}(p\ud', \lambda\ud')\,
{\cal M}_{\lambda}^{\rm Low}(p_{1}',p_{2}',k,p_{a},p_{b})\,
u(p\ub,\lambda\ub) \nn \\
&=
e\bar{u}(p\ud', \lambda\ud')
\bigg\lbrace
\Big{[}
\gamma_{\lambda} - 
\frac{i}{2 m_{p}} \sigma_{\lambda \nu} k^{\nu} F_{2}(0)
\Big{]}
\frac{1}{\slash{p}_{2}' + \slash{k} - m_{p}} \nn \\
& \quad \times \Big{[}
A_{\rm L}(\nu_{\rm L},t_{1})
+\frac{1}{2}(\slash{p}_{a}+\slash{p}_{1}')
B_{\rm L}(\nu_{\rm L},t_{1})
\Big{]}
 \nn \\
& \quad +
\Big{[}
A_{\rm L}(\nu_{\rm L},t_{1})
+\frac{1}{2}(\slash{p}_{a}+\slash{p}_{1}')
B_{\rm L}(\nu_{\rm L},t_{1})
\Big{]} \nn \\
& \quad \times
\frac{1}{\slash{p}_{b} - \slash{k} - m_{p}}
\Big{[}
\gamma_{\lambda} - 
\frac{i}{2 m_{p}} \sigma_{\lambda \nu} k^{\nu} F_{2}(0)
\Big{]}\nn \\
& \quad +
\Big{[}
p_{b \lambda} \frac{p_{a} \cdot k}{p_{b} \cdot k} +
p_{2 \lambda}' \frac{p_{1}' \cdot k}{p_{2}' \cdot k} -
p_{a \lambda} - p_{1 \lambda}'
\Big{]} \nn \\
& \quad \times
\Big{[}
\frac{\partial}{\partial \nu_{\rm L}} A_{\rm L}(\nu_{\rm L},t_{1})
+\frac{1}{2}(\slash{p}_{a}+\slash{p}_{1}')
\frac{\partial}{\partial \nu_{\rm L}} B_{\rm L}(\nu_{\rm L},t_{1})
\Big{]}
\bigg\rbrace \nn \\
&\quad \times u(p\ub,\lambda\ub) + {\cal O}(k)\,.
\end{align}
Here we use our metric conventions and notations and
\begin{align}
\label{D8}
\nu_{\rm L} &= p_{a} \cdot p_{b} + p_{1}' \cdot p_{2}'  \nn \\
            &= \frac{1}{2} (p_{a} + p_{b})^{2} 
            - \frac{1}{2} m_{p}^{2} - \frac{1}{2} m_{\pi}^{2} \nn \\
&\quad + \frac{1}{2} (p_{1}' + p_{2}')^{2} 
            - \frac{1}{2} m_{p}^{2} - \frac{1}{2} m_{\pi}^{2} \nn \\
&= \frac{1}{2} (p_{a} + p_{b})^{2} 
 + \frac{1}{2} (p_{a} + p_{b} - k)^{2} 
 - m_{p}^{2} - m_{\pi}^{2} \nn \\
&= s - (p_{a}+p_{b},k) - m_{p}^{2} - m_{\pi}^{2}\,.
\end{align}

We emphasize, that the result (\ref{D7}) represents an approximate expression
for the radiative amplitude which is valid at the given phase-space
configuration $(p_{1}', p_{2}', k)$;
see the extensive discussion concerning this point for the analogous
reaction $\pi^{-} \pi^{0} \to \pi^{-} \pi^{0} \gamma$
in Sec.~V of \cite{Lebiedowicz:2023ell}.
Eq.~(\ref{D7}) does \textit{not} represent
an expansion of the radiative amplitude in $k$ around
some phase-space point.
In fact, $k$ is here fixed by energy-momentum conservation
for given $p_{a}$, $p_{b}$, $p_{1}'$, and $p_{2}'$; see (\ref{D2}).

As in (\ref{3.16}) we will now go over to
\begin{align}
\label{D9}
&\mathcal{N}_{\lambda}^{\rm Low}(p'\uu, p'\ud, k, p\ua, p\ub)
=\sum_{\lambda'\ud , \lambda\ub} u(p'\ud ,\lambda'\ud)\bar{u}(p'\ud ,\lambda'\ud) \nn\\
&\qquad \times \mathcal{M}_{\lambda}^{\rm Low}(p'\uu, p'\ud, k, p\ua ,p\ub) \,
u(p\ub ,\lambda\ub)\bar{u}(p\ub ,\lambda\ub)\,.
\end{align}
Our aim is to derive the relation of Low's result 
(\ref{D7})--(\ref{D9}) and the Laurent expansion of the amplitude
for $\pi^{0} p \to \pi^{0} p \gamma$
which is constructed in the same way as shown for
$\pi^{\pm} p \to \pi^{\pm} p \gamma$ in Sec.~\ref{sec:3}.
That is, we shall construct the Laurent expansion
according to (\ref{3.34a}) of 
$\mathcal{N}_{\lambda}^{\rm Low}(p'\uu, p'\ud, k, p\ua, p\ub)$
around a phase-space point $(p_{1}, p_{2}, k=0)$,
where $(p_{1}, p_{2})$ has to be close to $(p_{1}', p_{2}')$;
see Fig.~\ref{fig:2}.
We set, as in (\ref{3.6}),
\begin{equation}
p'\uu=p\uu-l\uu\,,\quad p'\ud=p\ud-l\ud\,,
\label{D10}
\end{equation}
and assume $l_{1,2}$ to be of order $k^{0} = \omega$.

The next task is to substitute (\ref{D10}) 
in (\ref{D7}) and (\ref{D9}) and to expand in $\omega$.
Using the notation of (\ref{3.37}) and (\ref{3.41})
we find with (\ref{D8}) and $t = (p_{a} - p_{1})^{2}$
\begin{align}
A_{\rm L}(\nu_{\rm L},t_{1}) &=
A^{(\text{on})}(s - p_{s} \cdot k, t + 2 p_{t} \cdot l_{1} + l_{1}^{2}) \nn\\
&= A^{(\text{on})}(s,t) 
- p_{s} \cdot k \,A,_{s}^{\!\!(\text{on})}(s,t)\nn\\
& \quad + 2 p_{t} \cdot l_{1} \,A,_{t}^{\!\!(\text{on})}(s,t)
+ {\cal O}(\omega^{2})\,,
\label{D11}\\
B_{\rm L}(\nu_{\rm L},t_{1}) &=
B^{(\text{on})}(s,t) 
- p_{s} \cdot k \,B,_{s}^{\!\!(\text{on})}(s,t)\nn\\
& \quad + 2 p_{t} \cdot l_{1} \,B,_{t}^{\!\!(\text{on})}(s,t)
+ {\cal O}(\omega^{2})\,,
\label{D12}\\
\frac{\partial A_{\rm L}(\nu_{\rm L},t_{1})}{\partial \nu_{\rm L}} &=
A,_{s}^{\!\!(\text{on})}(s,t) + {\cal O}(\omega)\,,
\label{D13}\\
\frac{\partial B_{\rm L}(\nu_{\rm L},t_{1})}{\partial \nu_{\rm L}} &=
B,_{s}^{\!\!(\text{on})}(s,t) + {\cal O}(\omega)\,.
\label{D14}
\end{align}
In addition we use the relations \er{B86} and \er{B87}.
Putting everything together we get from (\ref{D7}) and (\ref{D9})
the following Laurent expansion of $\mathcal{N}_{\lambda}^{\rm Low}$
\begin{align}
\label{D15}
&\mathcal{N}_{\lambda}^{\rm Low}(p_{1}',p_{2}',k,p_{a},p_{b})\,
\nn \\
&  =
e (\slash{p}_{2}' + m_{p})
\Big{[}
A^{(\text{on})}(s,t)
+\frac{1}{2}(\slash{p}_{a}+\slash{p}_{1}')
B^{(\text{on})}(s,t)
\Big{]}
 \nn \\
& \quad \times
(\slash{p}_{b}+m_{p})
\Big{[}
\frac{2 p_{2 \lambda}' + k_{\lambda}}{2 p_{2 \lambda}' \cdot k}
-
\frac{2 p_{b \lambda} - k_{\lambda}}{2 p_{b \lambda} \cdot k}
\Big{]}\nn \\
& \quad +
e (\slash{p}_{2} + m_{p}) \Big{[}
A,_{s}^{\!\!(\text{on})}(s,t)+\frac{1}{2}(\slash{p}_{a}+\slash{p}_{1})
B,_{s}^{\!\!(\text{on})}(s,t) \Big{]}
\nn\\
& \quad \times
(\slash{p}_{b}+m_{p})
\Big{[}
2 (p_{s} \cdot k) \frac{p_{b \lambda}}{p_{b} \cdot k} - 2 p_{s \lambda} 
\Big{]} \nn \\
& \quad +
e (\slash{p}_{2} + m_{p}) \Big{[}
A,_{t}^{\!\!(\text{on})}(s,t)+\frac{1}{2}(\slash{p}_{a}+\slash{p}_{1})
B,_{t}^{\!\!(\text{on})}(s,t) \Big{]}\nn \\
& \quad \times
(\slash{p}_{b}+m_{p})\,
2 (p_{t} \cdot l_{1}) 
\Big{[}\frac{p_{2 \lambda}}{p_{2} \cdot k} - \frac{p_{b \lambda}}{p_{b} \cdot k}
\Big{]} \nn \\
& \quad +
e (\slash{p}_{2} + m_{p})
\Big{[}
A^{(\text{on})}(s,t)
+\frac{1}{2}(\slash{p}_{a}+\slash{p}_{1})
B^{(\text{on})}(s,t)
\Big{]} 
\nn \\
&\quad \times 
\Big{[}
(k_{\lambda} - \slash{k} \gamma_{\lambda}) (1 + F_{2}(0)) 
+ \frac{F_{2}(0)}{m_{p}}(p_{b\lambda} \slash{k} - p_{b} \cdot k \,\gamma_{\lambda})
\Big{]} \nn \\
&\quad \times 
(\slash{p}_{b}+m_{p}) \frac{1}{(- 2 p_{b} \cdot k)}
\nn \\
& \quad +
e (\slash{p}_{2} + m_{p})
\Big{[}
(k_{\lambda} - \gamma_{\lambda} \slash{k} ) (1 + F_{2}(0)) \nn \\
&\qquad 
+ \frac{F_{2}(0)}{m_{p}}(p_{2\lambda} \slash{k} - p_{2} \cdot k \,\gamma_{\lambda})
\Big{]} 
\nn \\
&\quad \times 
\Big{[}
A^{(\text{on})}(s,t)
+\frac{1}{2}(\slash{p}_{a}+\slash{p}_{1})
B^{(\text{on})}(s,t) \Big{]} 
\nn \\
&\quad \times 
(\slash{p}_{b}+m_{p}) \frac{1}{(- 2 p_{2} \cdot k)}
+ {\cal O}(\omega)\,.
\end{align}
As it should be, this result agrees completely with our result (\ref{3.44})
setting there $k^{2} = 0$ and
interpreting $A^{(\text{on})}$ and $B^{(\text{on})}$
as the $\pi^{0} p \to \pi^{0} p$ invariant amplitudes from (\ref{D3}).
We have thus shown that Low's approximate expression 
which is valid at the given phase-space point $(p_{1}', p_{2}', k)$
gives the correct Laurent expansion around the phase-space point
$(p_{1}, p_{2}, k=0)$ as specified above in (\ref{D10}).

Finally we compare our results to those of \cite{Liou:1978sz}.
The general methodology of \cite{Liou:1978sz} is close to ours:
if one wants to construct an expansion of the radiative amplitude
(\ref{3.1})
in the photon momentum $k$ one has to expand all relevant terms in $k$.
There are, however, also important differences between our work and \cite{Liou:1978sz}.
In \cite{Liou:1978sz} only the coplanar case of (\ref{1.3}) is treated,
whereas we treat the general case.
We discuss the phase-space of (\ref{1.3}) and give the general form
of the expansion of the radiative amplitude for real and virtual photon
emission in all directions of phase space starting from an arbitrary
phase-space point of no radiation.
In \cite{Liou:1978sz} only real photon emission is considered
and only a special expansion point is used for a given radiative amplitude.
The off-shell $\pi p \to \pi p$ amplitudes used in Eq.~(17) of \cite{Liou:1978sz}
contain only two invariant amplitudes whereas we use the most general
off-shell amplitude (\ref{2.16}) containing eight invariant amplitudes.
In \cite{Liou:1978sz} all off-shell corrections to the electromagnetic vertices
are neglected.
In contrast, we have given in Appendix~\ref{app:B} an extensive \textit{derivation}
of the fact that the relevant combinations of propagators and electromagnetic
vertices contain no off-shell effects to the relevant orders
$\omega^{-1}$ and $\omega^{0}$; 
see (\ref{B41}), (\ref{B45}), (\ref{B81}), and (\ref{B85}).
In this way we could make sure that our results represent genuine theorems of QFT.


\bibliography{refs}

\end{document}